\documentclass[11pt,a4paper]{article}

\pdfoutput=1

\usepackage{jheppub}

\usepackage{amsmath} 
\usepackage{amssymb}
\usepackage{hhline}
\usepackage{url}
\usepackage{lscape}
\usepackage{caption}
\usepackage[numbers,sort&compress]{natbib}
\usepackage{epsfig}
\usepackage{multirow} 
\usepackage{color}
\usepackage{verbatim}
\usepackage{nicefrac}
\usepackage{upgreek}
\usepackage{bbm}
\usepackage{setspace}
\usepackage{pstricks}
\usepackage{psfrag}
\usepackage{color}
\usepackage{subfigure}

\newcommand{\GeV}{{\rm GeV}}
\newcommand{\MeV}{{\rm MeV}}
\newcommand{\eV}{{\rm eV}}

\usepackage{hyperref}
\newcommand{\ee}{e}

\renewcommand{\Re}{\text{Re}}
\newcommand{\im}{{\rm Im}}

\renewcommand{\Im}{{\rm Im}}

\newcommand{\ii}{{\rm i}}
\newcommand{\dd}{d}

\newcommand{\bse}{\begin{subequations}}
\newcommand{\ese}{\end{subequations}}

\newcommand{\Ns}{N_{\rm s}}
\newcommand{\Nw}{N_{\rm w}}

\definecolor{green}{rgb}{0,0.5,0}  

\title{Testing the low scale seesaw and leptogenesis}

\author[a]{Marco Drewes}
\author[a]{Bj\"orn~Garbrecht}
\author[a,b,c]{Dario Gueter}
\author[a]{Juraj Klari\'{c}}

\affiliation[a]{Physik Department T70, Technische Universit\"at M\"unchen,\\
James Franck Stra\ss e 1, 85748 Garching, Germany}
\affiliation[b]{Max-Planck-Institut f\"ur Physik (Werner-Heisenberg-Institut),\\
F\"ohringer Ring 6, 80805 M\"unchen, Germany}
\affiliation[c]{Excellence Cluster Universe,\\
Boltzmannstra{\ss}e 2, 85748 Garching, Germany}
\emailAdd{marco.drewes@tum.de}
\emailAdd{garbrecht@tum.de}
\emailAdd{dario.gueter@tum.de}
\emailAdd{juraj.klaric@tum.de}
\abstract{Heavy neutrinos with masses below the electroweak scale can simultaneously generate the light neutrino masses via the seesaw mechanism and the baryon asymmetry of the universe via leptogenesis. The requirement to explain these phenomena imposes constraints on the mass spectrum of the heavy neutrinos, their flavour mixing pattern and their $CP$ properties. We first combine bounds from different experiments in the past to map the viable parameter regions in which the minimal low scale seesaw model can explain the observed neutrino oscillations, while being consistent with the negative results of past searches for physics beyond the Standard Model. We then study which additional predictions for the properties of the heavy neutrinos can be made based on the requirement to explain the observed baryon asymmetry of the universe. Finally, we comment on the perspectives to find traces of heavy neutrinos in future experimental searches at the LHC, NA62, BELLE II, T2K, SHiP or a future high energy collider, such as ILC, CEPC or FCC-ee. If any heavy neutral leptons are discovered in the future, our results can be used to assess whether these particles are indeed the common origin of the light neutrino masses and the baryon asymmetry of the universe. If the magnitude of their couplings to all Standard Model flavours can be measured individually, and if the Dirac phase in the lepton mixing matrix is determined in neutrino oscillation experiments, then all model parameters can in principle be determined from this data. This makes the low scale seesaw a fully testable model of neutrino masses and baryogenesis.}
\arxivnumber{1609.09069}

\begin{document}
\maketitle

\section{Introduction}

\label{sec:introduction}
All fermions in the Standard Model (SM) of particle physics are known to exist with both left handed (LH) and right handed (RH) chirality, with the exception of neutrinos. 
RH neutrinos could, if they exist, generate the light neutrino masses $m_a$ via the type I seesaw mechanism \cite{Minkowski:1977sc,GellMann:1980vs,Mohapatra:1979ia,Yanagida:1980xy,Schechter:1980gr,Schechter:1981cv}. 
Traditionally it is assumed that the Majorana masses $M_i$ of the RH neutrinos are much larger than the masses of any known particles. 
However, a \emph{low scale seesaw} with Majorana masses below the electroweak scale is in perfect agreement with all known experimental and cosmological constraints \cite{Atre:2009rg,Ibarra:2011xn,Asaka:2013jfa,Abada:2013aba,Drewes:2013gca,Abada:2014vea,Hernandez:2014fha,Antusch:2014woa,Gorbunov:2014ypa,Drewes:2015iva,deGouvea:2015euy,Abada:2015oba,Fernandez-Martinez:2016lgt,Abada:2016awd,Blennow:2016jkn,Ge:2016xya}. 
A strong theoretical motivation for this choice is provided by the observed values of the Higgs boson and top quark masses, which lie in the region in which the SM could be a viable effective field theory valid up to the Planck scale.
While the existence superheavy RH neutrinos would destabilise the Higgs mass \cite{Vissani:1997ys}, this problem is alleviated if the RH neutrinos have masses below the electroweak scale \cite{Bezrukov:2012sa}.
Hence, in absence of any other New Physics, a low seesaw scale $M_i$ is required to preserve the technical naturalness of the SM with RH neutrinos.
This scenario is technically natural if the difference between baryon number $B$ and lepton number $L$, which is conserved in the SM due to an accidental symmetry, is an approximately conserved quantity in Nature. 
This allows to explain the smallness of the light neutrino masses even for comparably large values of the neutrino Yukawa couplings $Y$ (slightly smaller than the electron Yukawa coupling for $M_i$ below the electroweak scale).

The probably most studied model that invokes the low scale seesaw is the \emph{Neutrino Minimal Standard Model} ($\nu$MSM) \cite{Asaka:2005pn,Asaka:2005an}, a minimal extension of the SM by three RH neutrinos that aims to address several problems in particle physics and cosmology.\footnote{Detailed descriptions of the $\nu$MSM are e.g. given in refs.~\cite{Boyarsky:2009ix,Canetti:2012kh}.}
Other frameworks that can accommodate a low scale seesaw include the possibility that the scale(s) $M_i$ and the electroweak scale have a common origin \cite{Iso:2009ss,Iso:2012jn,Khoze:2013oga,Khoze:2016zfi}, minimal flavour violation  \cite{Cirigliano:2005ck,Gavela:2009cd} and left-right-symmetric models \cite{Pati:1974yy,Mohapatra:1974hk,Senjanovic:1975rk,Wyler:1982dd} in which the complete breaking of GUT symmetry happens near the TeV scale, and the possibility that $B-L$ is a spontaneously broken symmetry \cite{Chikashige:1980ui,Gelmini:1980re,GonzalezGarcia:1988rw} and/or is approximately conserved \cite{Branco:1988ex,Gluza:2002vs,Shaposhnikov:2006nn,Kersten:2007vk,Gavela:2009cd,Racker:2012vw}.
Two of the most popular classes of scenarios are often referred to as ``inverse seesaw models'' \cite{Mohapatra:1986bd,Mohapatra:1986aw,GonzalezGarcia:1988rw} and ``linear seesaw models'' \cite{Akhmedov:1995vm,Akhmedov:1995ip,Barr:2003nn,Malinsky:2005bi} (see also \cite{Bernabeu:1987gr,Pilaftsis:1991ug,Abada:2007ux,Sierra:2012yy,Fong:2013gaa}).

In addition to the generation of light neutrino masses via the seesaw mechanism, RH neutrinos with masses below the electroweak scale could also generate the baryon asymmetry of the universe (BAU) via low scale leptogenesis.\footnote{A short review of the observational evidence for a matter-antimatter asymmetry in the observable universe and its theoretical implications is given in ref.~\cite{Canetti:2012zc}.}
In contrast to thermal leptogenesis in scenarios with superheavy RH neutrinos \cite{Fukugita:1986hr}, the BAU in these scenarios is not generated in the decay of the heavy neutrinos, but via $CP$ violating oscillations during their production \cite{Akhmedov:1998qx,Asaka:2005pn}.

The minimal number $n_s$ of RH neutrinos that is required to explain the two observed light neutrino mass differences is $n_s=2$. This is the scenario on which we focus in the following. 
The same choice also effectively describes baryogenesis in the $\nu$MSM, where it was first shown that leptogenesis from neutrino oscillations is feasible for $n_s=2$ \cite{Asaka:2005pn}.  
Two of the three RH neutrinos in the $\nu$MSM generate the light neutrino masses and the BAU, while the third one is a Dark Matter (DM) candidate. The constraints on the mass and lifetime of the DM candidate imply that its mixing with ordinary neutrinos must be extremely tiny, see \cite{Adhikari:2016bei} and references therein, so that its effect on the magnitude of the light neutrino masses and the BAU is negligible. Hence, one can effectively describe the seesaw mechanism and baryogenesis in the $\nu$MSM by setting $n_s=2$. An attractive feature of the $\nu$MSM is that it could in principle be an effective field theory up to the Planck scale \cite{Shaposhnikov:2007nj},
i.e. , all known phenomena in particle physics and cosmology may be explained without adding any new particles other than the RH neutrinos to the SM.
From an experimental viewpoint this scenario is very attractive because the new particles (or traces of them) could be found in the laboratory \cite{Shrock:1980ct,Shrock:1981wq,Langacker:1988ur}, and there is a realistic possibility of solving two of the most important puzzles in particle physics and cosmology.

The goal of this work is to derive constraints on the  properties of the RH neutrinos from the requirement to simultaneously explain the light neutrino flavour oscillations and the BAU. 
If any heavy neutral leptons are discovered in future experiments, then it will be possible to use these constraints in order to assess whether these particles can indeed be the common origin of light neutrino masses and the baryonic matter in the universe. 
In the present analysis, we focus on the heavy neutrino mass range $100 \,\MeV < M_i < 40\,\GeV$. 
In the context of the seesaw mechanism, masses below $100 \,\MeV$ are disfavoured by cosmological constraints \cite{Hernandez:2014fha}. For masses larger than $40 \,\GeV$ our treatment of leptogenesis in the early universe
 requires some refinements because the underlying assumption that the heavy neutrinos are fully relativistic while the BAU is generated is not justified.

This paper is organised as follows: In section~\ref{Sec:Osc} we introduce the seesaw mechanism, define the active-sterile mixing and discuss the allowed parameter region that can be explained by neutrino oscillation data. Further constraints emerge by simultaneously imposing that the heavy neutrinos generate the experimentally observed BAU. The discussion of these constraints and corresponding plots are presented in section~\ref{Sec:Lepto}. Future improvements of flavour predictions are listed in section~\ref{Sec:Future}, and
appendix~\ref{App:Mixing} summarises analytic expressions for the heavy neutrino mixing angles.

\section{
Testing the seesaw mechanism
}
\label{Sec:Osc}

\begin{table}[t]
  \centering
  \begin{tabular}{l||c|c|c|c|c|r}
 &$m_1^2$ & $m_2^2$ & $m_3^2$  & $\sin^2\uptheta_{12}$ & $\sin^2\uptheta_{13}$ & $\sin^2\uptheta_{23}$ \\
  \hline\hline
   NH & $0$ & $m^2_{\rm sol}$ & $\Delta m^2_{31}$ & $0.308$ & $0.0219$ & $0.451$\\
      \hline
  IH & $-\Delta m^2_{32}-m^2_{\rm sol}$ & $-\Delta m^2_{32}$ & $0$ & $0.308$ & $0.0219$ & $0.576$\\
  \end{tabular}
    \caption{Best fit values for the active neutrino masses $m_a$ and mixings $\uptheta_{ab}$ for normal hierarchy (NH) in the top row and inverted hierarchy (IH) in the bottom row from ref.~\cite{Gonzalez-Garcia:2014bfa}. 
The smaller measured mass difference, the so-called solar mass difference, is given by $m^2_{\rm sol}=m_2^2-m_1^2=7.49\times 10^{-5}\,\eV^2$. The larger mass differences are given by $\Delta m^2_{31}=m_3^2-m_1^2=2.477\times 10^{-3}\,\eV^2$ for NH and $\Delta m^2_{32}=m_3^2-m_2^2=-2.465\times 10^{-3}\,\eV^2<0$ for IH. In case of $n_s=2$ flavours the lightest neutrino is massless ( $m_1=0$ for NH and $m_3=0$ for IH). The absolute value $|m_3^2-m_1^2|$ is often referred to as the atmospheric mass difference $m^2_{\rm atm}$. Thus, for NH and $n_s=2$ we have $m^2_{\rm atm}=m_3^2$ and for IH $m^2_{\rm atm}=m_1^2= m_2^2+\mathcal{O}(m^2_{\rm sol}/m^2_{\rm atm})$. These values for $m^2_{\rm atm}$ differ slightly for the two hierarchies. However, the errors are of order $m^2_{\rm sol}/m^2_{\rm atm}$.}
     \label{tab:active_bounds}
\end{table}

The requirement to reproduce the observed properties of light SM neutrinos imposes considerable constraints on the properties of heavy RH neutrinos.
With the exception of the Majorana phase, all light neutrino properties can be determined in the foreseeable future.
As we discuss in more detail in section~\ref{Sec:Testability}, independent measurements of the 
couplings of the individual heavy neutrino species to all SM flavours 
then allow to determine all model parameters in the seesaw Lagrangian (\ref{eq:Lagrangian}).
If the splitting $\Delta M$ between the two heavy neutrino masses is too small to be experimentally resolved, which is the case in most of the leptogenesis parameter space, this is not possible.
However, the relative size of the RH neutrinos' couplings to the three SM generations is fixed by the properties of the light neutrinos alone~\cite{Shaposhnikov:2008pf,Asaka:2011pb,Ruchayskiy:2011aa,Hernandez:2016kel}, cf. eqs.~(\ref{eq:mixing_NHsimple},\ref{eq:mixing_IHsimple}) and (\ref{eq:mixing_NH},\ref{eq:mixing_IH}) further below. These relations offer a powerful tool to test the seesaw mechanism as the origin of neutrino mass experimentally, even if the heavy neutrino mass spectrum cannot be fully resolved. 
In the subsequent sections, we investigate which additional constraints can be derived from the requirement to explain the observed BAU and from other experiments.

\subsection{The model parameters}
\label{SubSec:model}
The minimal type-I seesaw model is described by the Lagrangian
\begin{align}
\label{eq:Lagrangian}
{\cal L}=\mathcal{L}_{\rm SM}
+{\rm i} \overline{\nu_{{\rm R} i}}\partial\!\!\!/\nu_{{\rm R} i}
-\frac{1}{2}(\overline{\nu_{{\rm R} i}^c}M_{ij}\nu_{{\rm R} j} +  \overline{\nu_{{\rm R} i}}M_{ji}^*\nu_{{\rm R} j}^c)-Y_{ia}^*\overline{\ell_a}\varepsilon\phi \nu_{{\rm R} i}
-Y_{ia}\overline{\nu_{{\rm R} i}}\phi^\dagger \varepsilon^\dagger \ell_a\,,
\end{align}
where $\mathcal{L}_{\rm SM}$ is the SM Lagrangian. The heavy neutrinos are represented by RH spinors $\nu_{{\rm R} i}$ and the superscript $c$ denotes charge conjugation. 
They interact with the SM solely through their Yukawa interactions $Y$ to the SM lepton doublets $\ell_a$ ($a=e,\mu,\tau$) and the Higgs field $\phi$, where  $\varepsilon$ is the antisymmetric ${\rm SU}(2)$-invariant tensor with $\varepsilon^{12}=1$.

The connection between the Lagrangian (\ref{eq:Lagrangian}) and low energy neutrino oscillation data is well known. Here we only recapitulate the main results that are required for the present analysis and refer the reader to refs.~\cite{Drewes:2013gca,Drewes:2015iva} for a recent summary. 
A convenient way to express the relation at tree level is provided by the Casas-Ibarra parametrisation \cite{Casas:2001sr}\footnote{The relation (\ref{CasasIbarraDef}) between $Y$ and the light neutrino properties holds at tree level and at leading order in the parameters $\theta_{ai}$ defined below.  
A generalisation that holds at one loop level has been introduced in ref.~\cite{Lopez-Pavon:2015cga}. However, the results found in ref.~\cite{Drewes:2016lqo} suggest that the tree level treatment is sufficient in the parameter region where leptogenesis is viable.
}
\begin{align}\label{CasasIbarraDef}
Y^\dagger=\frac{\ii}{v}U_\nu\sqrt{m_\nu^{\rm diag}}\mathcal{R}\sqrt{M^{\rm diag}}\,.
\end{align}
Here $(m_\nu^{\rm diag})_{ab}=\delta_{ab} m_a$ is the light neutrino mass matrix, and $m_a$ are the light neutrino masses.
Note that the number of non-vanishing eigenvalues $m_a$ cannot be larger than $n_s$.
This immediately implies that the lightest neutrino is massless in the scenario with $n_s=2$ we consider here ($m_{\rm lightest}=0$).\footnote{This statement practically also applies to the $\nu$MSM because the coupling of the third RH neutrino is so feeble that its contribution to the generation of light neutrino masses can be neglected.}
The light neutrino mixing matrix $V_\nu$ can be expressed as 
\begin{eqnarray}\label{VnuDef}
V_\nu= \left(\mathbbm{1}-\frac{1}{2}\theta\theta^{\dagger}\right)U_\nu\,,
\end{eqnarray} 
where the unitary matrix $U_\nu$ diagonalises the matrix 
\begin{equation}\label{SeesawRelation}
m_\nu=v^2 Y^\dagger M^{-1} Y^*
\end{equation}
and can be factorised as
\begin{align}
\label{PMNS}
U_\nu=V^{(23)}U_\delta V^{(13)}U_{-\delta}V^{(12)}{\rm diag}(e^{\ii \alpha_1/2},e^{\ii \alpha_2 /2},1)\,,
\end{align}
with $U_{\pm \delta}={\rm diag}(1,e^{\mp \ii \delta/2},e^{\pm \ii \delta /2})$.
The light neutrino mixing parameters shown in table \ref{tab:active_bounds} are extracted from neutrino oscillation data under the assumption that $V_\nu=U_\nu$,  though this assumption is in principle not necessary \cite{Antusch:2006vwa}. 
In the following we ignore the effect of the effect of the $\theta\theta^\dagger$ term in eq.~(\ref{VnuDef}) on light neutrino oscillations and use the data in table \ref{tab:active_bounds} to fix the parameters in $U_\nu$.
This is justified because the best fit parameters for the light neutrinos are not strongly affected by the properties of the heavy neutrinos in the range of $M_i$ we consider, see \cite{Blennow:2016jkn} and references therein for a recent discussion.

The non-vanishing entries of the matrix $V=V^{(23)}V^{(13)}V^{(12)}$ are given by
\begin{eqnarray}
V^{(ab)}_{aa}=V^{(ab)}_{ba}=\cos \uptheta_{ab} \ , \
V^{(ab)}_{ab}=-V^{(ab)}_{ba}=\sin \uptheta_{ab} \ , \
V^{(ab)}_{cc}=1 \quad \text{for $c\neq a,b$}\,.
\end{eqnarray}
The parameters $\uptheta_{ab}$ are the the mixing angles of the light neutrinos, $\delta$ is referred to as the Dirac phase and $\alpha_{1,2}$ as Majorana phases.
Only one combination of the Majorana phases is physical for $n_s=2$. For normal hierarchy this is $\alpha_2$, for inverted hierarchy it is $\alpha_1-\alpha_2$. We can therefore, without loss of generality, set $\alpha_1=0$ and $\alpha_2\equiv \alpha$.
The deviation of $V_\nu$ from unity is due to the mixing between the doublet states $\nu_{\rm L}$ and the singlet state $\nu_{\rm R}$. It is quantified by the parameters
\begin{equation}
\theta=v  Y^\dagger M^{-1}\,.
\end{equation}
Here $v$ is the temperature dependent Higgs expectation value with $v=174\,\GeV$ at temperature $T=0$. 
The complex orthogonal matrix $\mathcal{R}$  fulfils the condition $\mathcal{R}\mathcal{R}^T=1$ and can be expressed as
\begin{align}
\mathcal{R}^{\rm NH}=
\begin{pmatrix}
0 && 0\\
\cos \omega && \sin \omega \\
-\xi \sin \omega && \xi \cos \omega
\end{pmatrix}\,,\quad \quad 
\mathcal{R}^{\rm IH}=
\begin{pmatrix}
\cos \omega && \sin \omega \\
-\xi \sin \omega && \xi \cos \omega \\
0 && 0
\end{pmatrix}
\,,
\end{align}
where $\xi=\pm 1$ and $\omega=\Re\omega+\ii\Im\omega$ is the complex mixing angle.
For large $|\Im\omega|\gg 1$ one obtains the useful expansion
\begin{align}
\label{eq:R_exp}
\mathcal{R}^{\rm NH}_{|\Im\omega| \gg 1}=\frac12 \ee^{\Im\omega}\ee^{-\ii \Re\omega}
\begin{pmatrix}
0 & 0 \\
1 & \ii \\
-\xi \ii& \xi
\end{pmatrix}
\,,\quad\quad
\mathcal{R}^{\rm IH}_{|\Im\omega| \gg 1}=\frac12 \ee^{\Im\omega}\ee^{-\ii \Re\omega}
\begin{pmatrix}
1 & \ii \\
-\xi \ii& \xi \\
0 & 0
\end{pmatrix}
\,.
\end{align}
To second order in $|\theta_{a i}|\ll 1$, the three light neutrino mass eigenstates with masses $m_a$ can be expressed in terms of the Majorana spinors
\begin{equation}\label{LightMassEigenstates}
\upnu_i=\left[ V_\nu^{\dagger}\nu_{\rm L}-U_\nu^{\dagger}\theta\nu_{\rm R}^c + V_\nu^{T}\nu_{\rm L}^c-U_\nu^{T}\theta\nu_{\rm R} \right]_i.
\end{equation}
In addition, there are $n_s$ heavy mass eigenstates
\begin{equation}
N_i=\left[V_N^\dagger\nu_{\rm R}+\Theta^{T}\nu_{\rm L}^{c} +  V_N^T\nu_{\rm R}^c+\Theta^{\dagger}\nu_{\rm L}\right]_i\,.
\end{equation}
The unitary matrix $U_N$ diagonalises the heavy neutrino mass matrix 
\begin{eqnarray}
M_N=M + \frac{1}{2}(\theta^{\dagger} \theta M + M^T \theta^T \theta^{*})\,,\label{MNDef}
\end{eqnarray}
and $V_N= (1-\frac{1}{2}\theta^T\theta^*)U_N$. 
We shall refer to the basis where $M_N$ is diagonal as the \emph{mass basis} of the  heavy neutrinos, while the basis where $Y^\dagger Y$ is diagonal is the \emph{interaction basis}. The \emph{mass states} are the eigenstates of $M_N^\dagger M_N$, with physical mass squares given by the eigenvalues of that matrix.
The \emph{interaction states} are the eigenstates of $Y Y^\dagger$, their interaction strengths are characterised by  the eigenvalues of that matrix.\footnote{That statement has to be taken with some care because the interactions are in principle helicity dependent.}

An experimental confirmation of the seesaw mechanism requires the discovery of the new particles $N_i$. 
They interact with ordinary matter via their quantum mechanical mixing.
If kinematically allowed, they appear in any process that involves ordinary neutrinos, but with an amplitude that is suppressed by the ``mixing angle''
\begin{equation}
\Theta_{a i}=(\theta U_N^*)_{a i}\approx \theta_{a i}\ \,.
\end{equation}
Hence, the branching ratios can be expressed in terms of the quantities 
\begin{eqnarray}
\label{eq:U_theta}
U_{a i}^2 = |\Theta_{a i}|^2\approx |\theta_{a i}|^2\,.
\end{eqnarray}
It is a main goal of this work to make predictions about the flavour mixing pattern of the heavy neutrinos, i.e. the relative size of the $U_{a i}^2$ for the different $a=e,\mu,\tau$.  
Low scale leptogenesis with $n_s=2$ is known to require a mass degeneracy $|\Delta M|\ll \bar{M}$, where
\begin{equation}
\bar{M}=\frac{M_2 + M_1}{2} 
\, , \quad 
\Delta M = \frac{M_2 - M_1}{2}\,. 
\end{equation}
If the mass splitting $\Delta M$ is too small to be resolved experimentally, then experimental constraints should be applied to the quantities 
\begin{eqnarray}
U_a^2=\sum_i U_{a i}^2\,.
\end{eqnarray} 
Finally, the overall coupling strength of the heavy neutrinos can be quantified by
\begin{equation}
U^2=\sum_a U_a^2\,.
\end{equation}
In terms of the Casas-Ibarra parameters, $U^2$ reads
\begin{eqnarray}\label{U2NH}
U^2&=&\frac{M_2-M_1}{2M_1 M_2} (m_2-m_3)\cos(2 {\rm Re}\omega)+\frac{M_1+M_2}{2M_1 M_2}(m_2+m_3)\cosh(2 {\rm Im}\omega) 
\end{eqnarray}
with normal hierarchy and
\begin{eqnarray}\label{U2IH}
U^2&=&\frac{M_2-M_1}{2M_1 M_2} (m_1-m_2)\cos(2 {\rm Re}\omega)+\frac{M_1+M_2}{2M_1 M_2}(m_1+m_2)\cosh(2 {\rm Im}\omega) 
\end{eqnarray}
with inverted hierarchy. For given mass spectrum, the magnitude of $U^2$ is determined by $\omega$ alone. 

\subsection{Constraints from neutrino oscillation data}
If there are no cancellations in the light neutrino mass matrix $m_\nu$, then one expects all $U_{ai}^2$ to be of the order of the \emph{naive seesaw expectation}
\begin{eqnarray}\label{F0} 
U^2 \sim \sqrt{m_{\rm atm}^2 + m_{\rm lightest}^2}/\bar{M}\,.
\end{eqnarray}
Mixing angles much larger than the estimate (\ref{F0}) can only be made consistent with small neutrino masses and non-observation of neutrinoless double $\beta$ ($0\nu\beta\beta$) decay if there are cancellations that keep the eigenvalues of $m_\nu$ small in spite of comparably large Yukawa couplings $Y_{ia}$ \cite{Blennow:2010th,LopezPavon:2012zg,Drewes:2015iva,Lopez-Pavon:2015cga}. 
Such cancellations are e.g. expected in models with an approximate conservation of $B-L$. 
If $B-L$ were an exact symmetry of the Lagrangian (\ref{eq:Lagrangian}), the heavy neutrino mass eigenstates $N_i$ would have exactly the same mass $M_1=M_2=\bar{M}$ and couplings 
\begin{equation}\label{Uequality}
Y_{a2}^\dagger=\ii Y_{a1}^\dagger \, , \quad U_{a 1}^2 = U_{a 2}^2 = U_a^2/2\,,
\end{equation}
and they could be combined into a Dirac spinor 
\begin{equation}
\Psi_N=\frac{1}{\sqrt{2}}(N_1 + \ii N_2)\,.\label{DiracSpinor}
\end{equation}
While this limit cannot be realised within the seesaw approximation with $m_a\neq 0$, it  can be reached approximately in the limit $\Delta M\rightarrow 0$, $|{\rm Im}\omega|\rightarrow \infty$. 
In the experimentally most accessible region we can therefore expand in the small $B-L$ violating parameters\footnote{Note that $\epsilon \gg 1$ also leads to an approximate $B-L$ conservation. However, changing simultaneously the sign of $\xi$, $\Im\omega$ and $\Delta M$, as well as shifting $\Re\omega \to \pi-\Re\omega$ results in a swap of the labels $N_1$ and $N_2$, without any physical consequences. It is therefore sufficient to discuss the case $\epsilon<1$ here.} 
\begin{eqnarray}
\label{mu_eps}
\mu= \Delta M/\bar{M} \, , \quad  \epsilon=e^{-2\Im \omega}\,. 
\end{eqnarray}
In the limit $\mu=0$ and $\epsilon\rightarrow 0$, one approximately recovers the relation (\ref{Uequality}),
\begin{equation}\label{pseudoDiractheta}
\theta_{a 2} \simeq \ii \theta_{a 1}
\end{equation}
and the interaction eigenstates approximately become
\begin{equation}
\Ns\simeq\frac{1}{\sqrt{2}}(N_1 + i N_2) = \Psi_N \, , \quad
\Nw\simeq\frac{1}{\sqrt{2}}(N_1 - i N_2) = \Psi_N^c.
\end{equation}
The interactions of $\Nw$ are of order $\epsilon$, while those of $\Ns$ scale as $1/\epsilon$. More precisely, the larger eigenvalue of $Y Y^\dagger$ scales as $[(m_a-m_b)\bar{M}]/(4v^2\epsilon)$, where $m_a$ and $m_b$ are the two non-zero light neutrino masses. 
As usual, only the right chiral part of the spinors $\Ns$ and $\Nw$ interacts with the SM.

This argument can be generalised to the case $n_s>2$, where leptogenesis does not require a mass degeneracy \cite{Drewes:2012ma}; mixings much larger than (\ref{F0}) always require that the heavy neutrino mass spectrum is organised in pairs of particles with quasi-degenerate masses.
This has important phenomenological implications.
Searches for lepton number violating processes are often considered to be the golden channel to look for heavy Majorana neutrinos. 
However, in the context of the seesaw mechanism, these processes are suppressed by small parameters $\mu$ and $\epsilon$ if $U^2$ is much larger than suggested by (\ref{F0}). Hence, the bounds published by experimental collaborations cannot be directly applied to the seesaw partners $N_i$ if they are based on searches for LNV.

Most realistic future experiments can only access the paramater region with $U^2$ much larger than suggested by eq.~(\ref{F0}),i.e. $\epsilon\ll 1$. 
Consistency with the smallness of the light neutrino masses $m_a$ in this regime implies $\mu\ll 1$. 
In addition, leptogenesis with $n_s=2$ also requires $\mu\lesssim 10^{-2}$. 
Therefore, the remaining sections of this paper, in particular sections~\ref{Sec:Lepto} and~\ref{Sec:Future}, are mostly devoted to study the perspectives of testing the seesaw mechanism and leptogenesis in the limit $\mu,\epsilon\ll1$.
It is instructive to briefly recall the relation between the $U_a^2$ and the properties of light neutrinos in this regime.
In the following we set $\mu=0$ and only keep terms of order $1/\epsilon$. Just for illustrative purposes, we follow refs.~\cite{Shaposhnikov:2008pf,Asaka:2011pb,Ruchayskiy:2011aa,Hernandez:2016kel} and further expand in $m_{\rm sol}/m_{\rm atm}$ and find that 
\bse
\label{eq:mixing_total}
\begin{eqnarray}
U^2&\simeq&\frac{1}{2\bar{M}}m_{\rm atm}\ee^{2 {\rm Im}\omega} \ {\rm for} \ {\rm NH}\,, \\
U^2&\simeq&
\frac{1}{\bar{M}}m_{\rm atm}\ee^{2 {\rm Im}\omega} \ {\rm for} \ {\rm IH}\,.
\end{eqnarray}
\ese
Hence, the overall interaction strength of the $N_i$  in this approximation is fixed by the parameters $\bar{M}$, ${\rm Im}\omega$ and $m_{\rm atm}$ alone. 
For normal hierarchy we furthermore obtain
\bse
\begin{align}
U^2_{e}&\approx \ee^{2\Im\omega}\frac{m_{\rm atm}}{2\bar{M}}
\sin^2\uptheta_{13}
\,,\\
U^2_{\mu}&\approx \ee^{2\Im\omega}\frac{m_{\rm atm}}{2\bar{M}}
\cos^2\uptheta_{13}\sin^2\uptheta_{23}
\,,\\
U^2_{\tau}&\approx \ee^{2\Im\omega}\frac{m_{\rm atm}}{2\bar{M}}
\cos^2\uptheta_{13}\cos^2\uptheta_{23}\,.
\end{align}
\label{eq:mixing_NHsimple}
\ese
The main prediction of eqs.~(\ref{eq:mixing_NHsimple}) is that $U_e^2$ is suppressed with respect to 
$U_\mu^2$ and $U_\tau^2$ due to the relative smallness of $\uptheta_{13}$.
On the other hand, $U_\mu^2$ and $U_\tau^2$ are expected to be comparable in size because $\uptheta_{23}\simeq \pi/4$.
A measurement of any of the $U_a^2$ already determines $\Im \omega$ in this approximation.
Since $\bar{M}$ can be determined kinematically, $\Im \omega$ is the only free parameter, and the system of eqs.~(\ref{eq:mixing_NHsimple}) is overconstrained. Note that this is only true within the (bad) approximation $m_{\rm sol}/m_{\rm atm}\ll 1$. 
For inverted hierarchy we neglect $\uptheta_{13}$ and $\uptheta_{23}- \pi/4$ in order to obtain simple relations:
\bse
\begin{align}
U^2_{e}&\approx 
\ee^{2\Im\omega}
\frac{m_{\rm atm}}{2\bar{M}}
\left[
1 + \xi\sin\left(\frac{\alpha_2-\alpha_1}{2}\right)\sin(2\uptheta_{12})
\right]
,\\
U^2_{\mu}&\approx\ee^{2\Im\omega}
\frac{m_{\rm atm}}{4\bar{M}}
\left[
1 - \xi\sin\left(\frac{\alpha_2-\alpha_1}{2}\right)\sin(2\uptheta_{12})
\right]\label{SimpleResult}
,\\
U^2_{\tau}&\approx U^2_{\mu}\,.
\end{align}
\label{eq:mixing_IHsimple}
\ese
In the approximation above, observing  $U^2_e$ and $U_\mu^2$ would not only allow to measure $\Im\omega$, but also the Majorana phase difference $\alpha_2-\alpha_1$.\footnote{Recall that only the Majorana phase difference $\alpha_2-\alpha_1$ is physical for $n_s=2$.} 
The results from eqs. (\ref{eq:mixing_total})--(\ref{eq:mixing_IHsimple}) are consistent with previous results from ref.~\cite{Shaposhnikov:2008pf,Asaka:2011pb,Ruchayskiy:2011aa,Hernandez:2016kel} as well as the more detailed expressions in appendix~\ref{App:Mixing}.

The relations (\ref{eq:mixing_NHsimple}) and (\ref{eq:mixing_IHsimple}) are quite useful to get a rough understanding of the parametric dependencies and, further, they allow to identify which parameters would be most strongly constrained by measurements of the $U_a^2$. At a quantitative level, the quantities $\uptheta_{23} - \pi/4$, $\uptheta_{13}$ and $m_{\rm sol}/m_{\rm atm}$ are not small enough to be neglected, and the dependence of the $U_a^2$  on the phases $\delta$, $\alpha_1$ and $\alpha_2$ has to be taken into account.\footnote{Note that the leading order correction in $m_{\rm sol}$ is of order $\sqrt{m_{\rm sol}/m_{\rm atm}}\sim 1/2$, which is not a good expansion parameter.} The full expressions in the limit $\mu,\epsilon \ll 1$ can be obtained from the more general expressions for $U_{ai}^2$ in appendix~\ref{App:Mixing}. 
Figure~\ref{fig:regions_NH.pdf} shows the range of $U_a^2/U^2$ that can be realised by varying the phases. 
\begin{figure}
	\centering
		\includegraphics[scale=0.9]{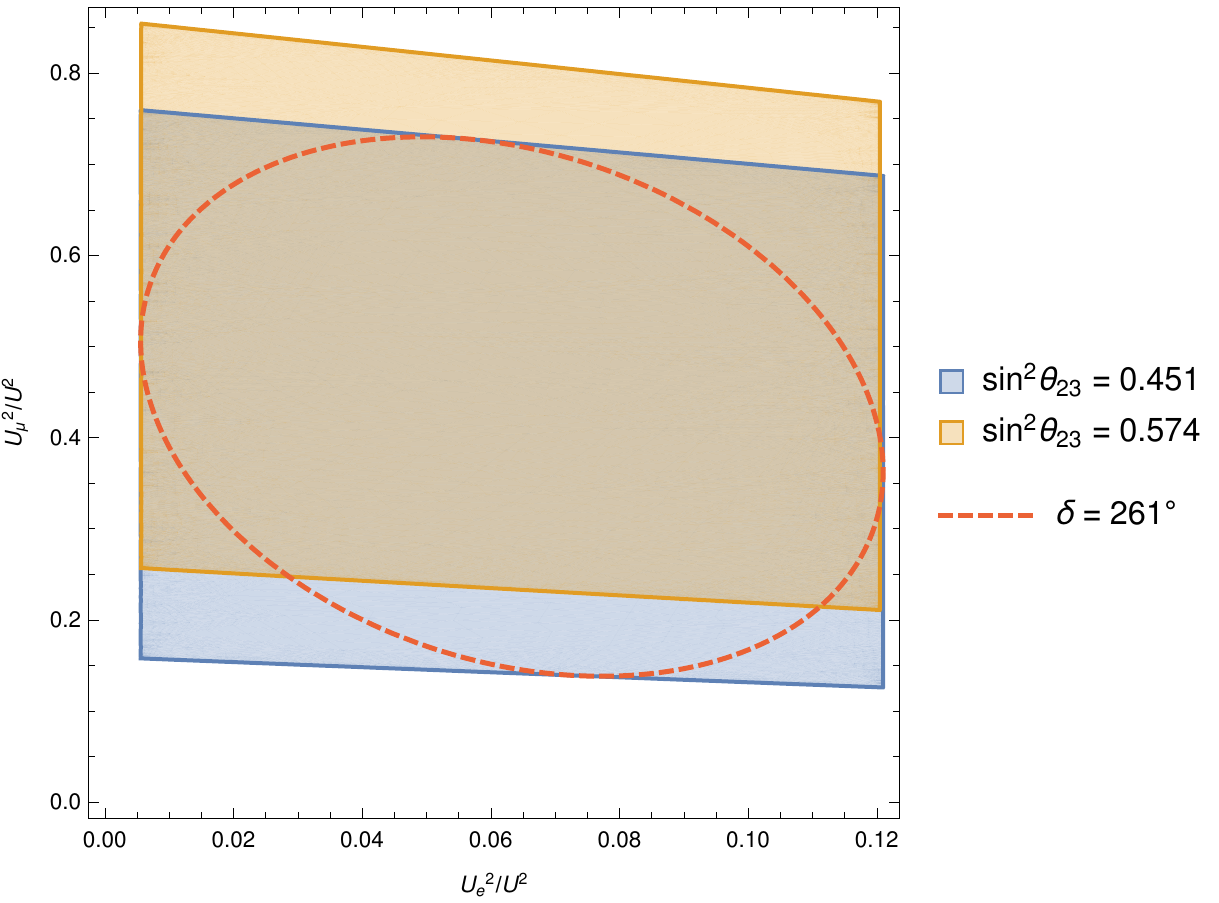}\\
		\includegraphics[scale=0.9]{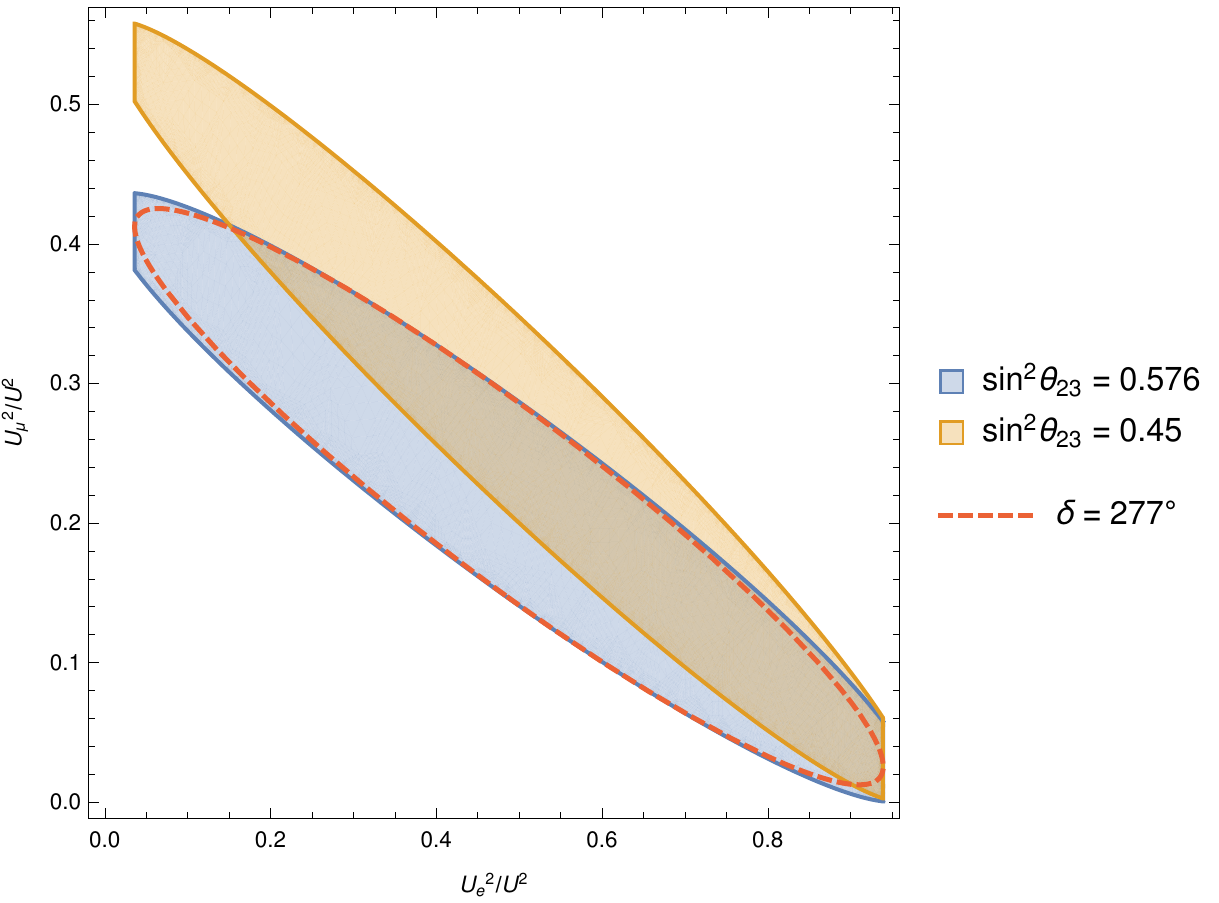}
	\caption{The coloured regions indicate the allowed range of $U_a^2/U^2$ that can be realised by varying the phases in $U_\nu$ after fixing the light neutrino mass splittings and mixing angles to their best fit values given in table \ref{tab:active_bounds}.  $U_\tau^2/U^2$ is fixed by the requirement $\sum_a U_a^2/U^2 = 1$.
	The difference between the orange and blue region illustrates the change of the predictions if one varies these parameters within their experimental uncertainties. The upper panel assumes normal light neutrino mass hierarchy, the lower panel assumes inverted hierarchy.
	If the Dirac phase $\delta$ is measured independently in light neutrino oscillation experiments, the two dimensional regions will reduce to rings in the $U_e^2/U^2$-$U_\mu^2/U^2$ plane as demonstrated by the red dashed lines, where we take the best fit values for $\delta=261^\circ$ for normal hierarchy with $\sin^2\uptheta_{23}=0.451$ and $\delta=277^\circ$ with $\sin^2\uptheta_{23}=0.576$ for inverted hierarchy. Note that we assume here $\epsilon, \mu\ll 1$, while for non-small $\epsilon$ (for smaller $\Im\omega$) the shape and size of the coloured region changes.
	}
	\label{fig:regions_NH.pdf}
\end{figure}
Figure~\ref{fig:Mixing_Angles_NH_IH.pdf} illustrates the dependence of $U_a^2/U^2$ on the phases.

\subsection{Other constraints}\label{other_constraints}
\begin{landscape}
\begin{figure}
\begin{center}
\subfigure{
\includegraphics[width=0.61\textwidth]{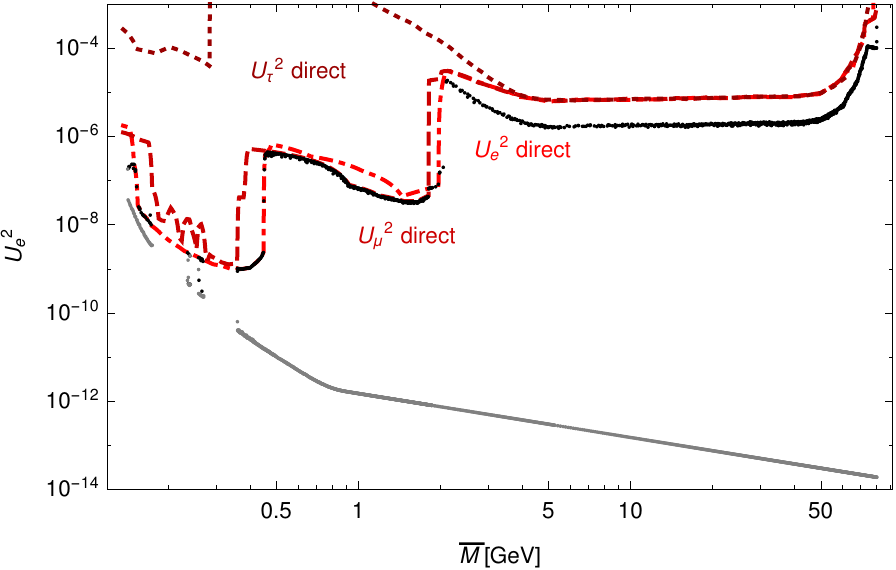}
}
\quad
\subfigure{
\includegraphics[width=0.61\textwidth]{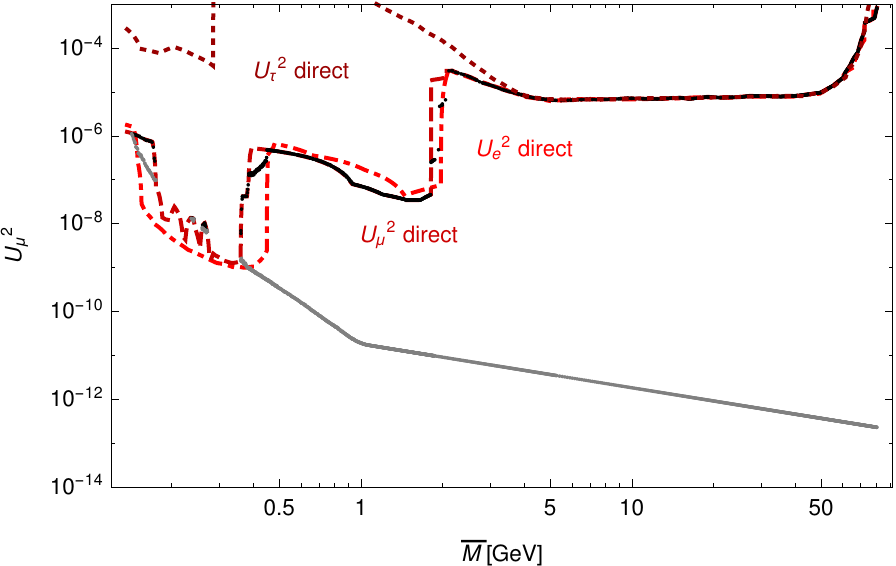}
}
\\
\subfigure{
\includegraphics[width=0.61\textwidth]{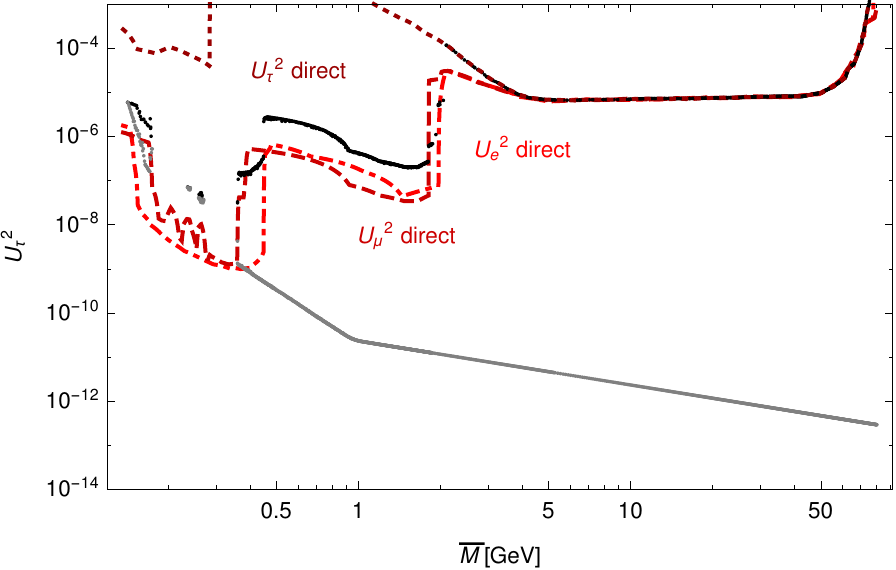}
}
\quad
\subfigure{
\includegraphics[width=0.61\textwidth]{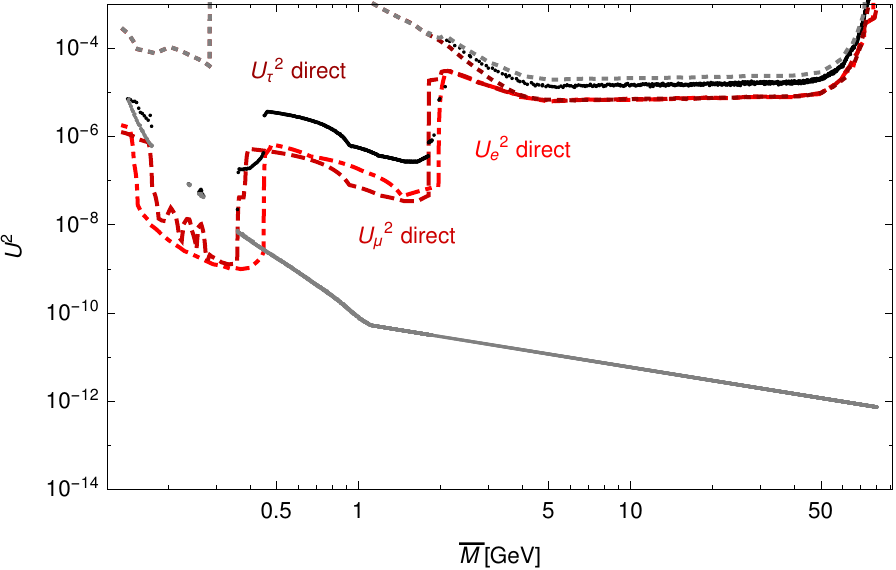}
}
\caption{\label{Fig:Flavor_Plots_NH}
Global constraints from direct and indirect search experiments and BBN for normal hierarchy and $\Delta M<10$ MeV. The black and grey dots mark the largest and smallest value of  $U_a^2$ and $U^2$ we found to be consistent with all constraints  in each mass bin.
For comparison, the red lines show the direct search constraints from collider and fixed target experiments on $U_e^2$ (dashed-dotted), $U_\mu^2$ (dashed) and $U_\tau^2$ (dotted) alone. 
In the plot for $U^2$ we also added their sum as dashed grey line. 
}
\end{center}
\end{figure}
\end{landscape}

\begin{landscape}
\begin{figure}
\begin{center}
\subfigure{
\includegraphics[width=0.61\textwidth]{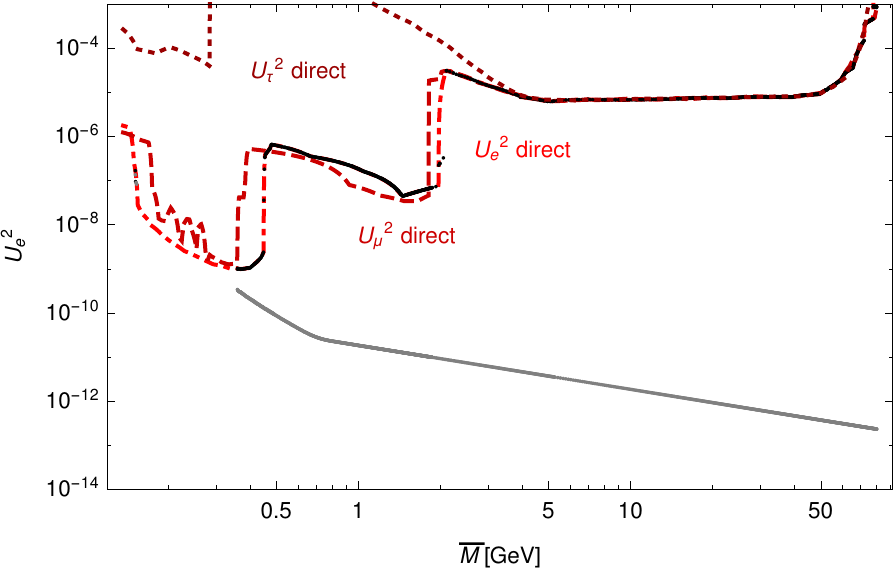}
}
\quad
\subfigure{
\includegraphics[width=0.61\textwidth]{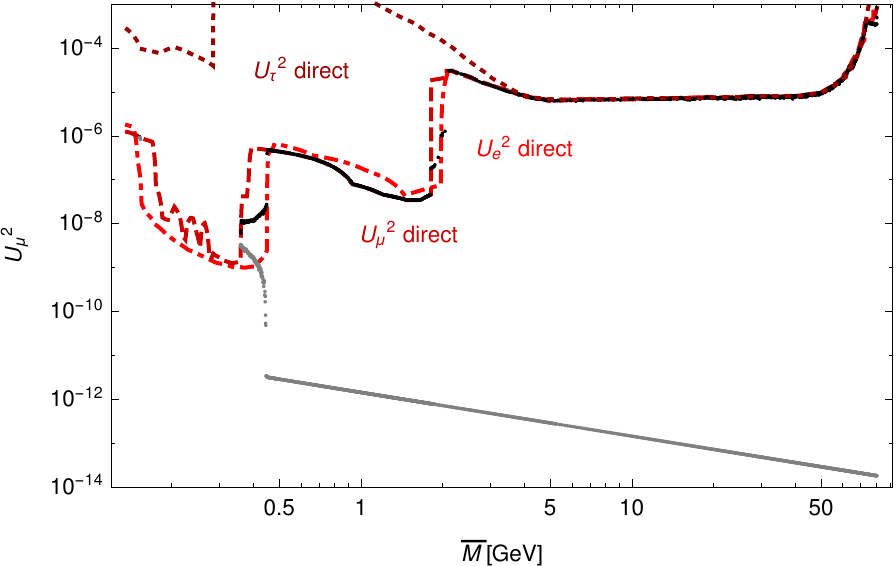}
}
\\
\subfigure{
\includegraphics[width=0.61\textwidth]{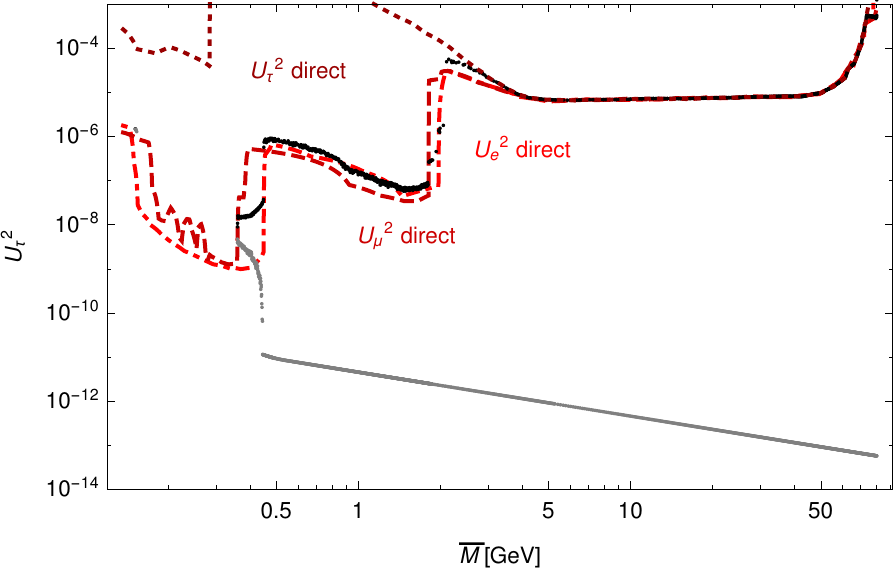}
}
\quad
\subfigure{
\includegraphics[width=0.61\textwidth]{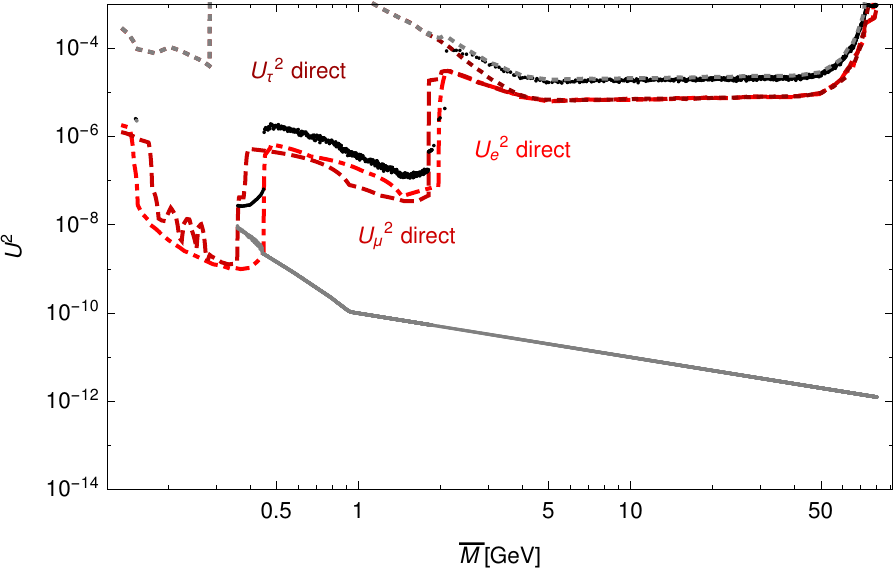}
}
\caption{\label{Fig:Flavor_Plots_IH}
Global constraints from direct and indirect search experiments and BBN for inverted hierarchy and $\Delta M<10$ MeV. The black and grey dots mark the largest and smallest value of  $U_a^2$ and $U^2$  we found to be consistent with all constraints  in each mass bin.
For comparison, the red lines show the direct search constraints from collider and fixed target experiments on $U_e^2$ (dashed-dotted), $U_\mu^2$ (dashed) and $U_\tau^2$ (dotted) alone.
In the plot for $U^2$ we also added their sum as dashed grey line. 
}
\end{center}
\end{figure}
\end{landscape}

Direct searches for heavy neutrinos in fixed target experiments and colliders impose upper bounds on the $U_{a i}^2$ for given $M_i$.
The properties of heavy neutrinos are constrained indirectly by searches for rare processes and precision observables, which can be affected by their existence.
Various collections of constraints have been presented in refs.~\cite{Atre:2009rg,
Blennow:2016jkn,
Kusenko:2009up,
Ibarra:2011xn,
Ruchayskiy:2011aa,
Asaka:2011pb,
Abazajian:2012ys,
Asaka:2013jfa,
Abada:2013aba,
Drewes:2013gca,
Hernandez:2014fha,
Antusch:2014woa,
Asaka:2014kia,
Gorbunov:2014ypa,
Abada:2014vea,
Abada:2014kba,
Abada:2015oba,
Drewes:2015iva,
Deppisch:2015qwa,
Escrihuela:2015wra,
Fernandez-Martinez:2015hxa,
deGouvea:2015euy,
Lopez-Pavon:2015cga,
Fernandez-Martinez:2016lgt,
Rasmussen:2016njh,
Abada:2016awd,
Drewes:2016lqo}.
In order to identify the range of $U_a^2$ that is consistent with the constraints from past experiments for given $\bar{M}$, we perform a numerical scan of the parameter space. We fix all light neutrino parameters to their best fit values, set $\alpha_1=0$ and $\xi=1$.
We then randomise all other parameters, using flat priors for $\delta$, $\alpha_2$, ${\rm Re}\omega$ and ${\rm Im}\omega$, and a logarithmic prior for $\Delta M$.
We restrict  $\Delta M$ to values less than $10$ MeV, which roughly corresponds to the largest mass splitting found to be consistent with leptogenesis for $n_s=2$ in ref.~\cite{Canetti:2010aw,Canetti:2012kh}.\footnote{
More recent studies \cite{Drewes:2016lqo,Hernandez:2016kel} suggest that there exists a small part of the parameter space where leptogenesis is possible with slightly larger $\Delta M$. For $n_s=3$ leptogenesis does not require a mass degeneracy \cite{Drewes:2012ma}.
}
This means that the experiments under consideration could not resolve $N_1$ and $N_2$ kinematically, and constraints apply to $U_a^2$ (rather than the individual $U_{ai}^2$).
From a sample of $10^8$ randomly generated sets of parameters we identify the largest and smallest $U_a^2$ for fixed values $\bar{M}$ between the pion and W boson mass that are consistent with the negative results of past direct and indirect searches. 
We take into account constraints from the direct search experiments 
DELPHI \cite{Abreu:1996pa}, 
L3 \cite{Adriani:1992pq}, 
LHCb \cite{Aaij:2014aba},
ATLAS \cite{Aad:2015xaa},
CMS \cite{Khachatryan:2015gha}, 
BELLE \cite{Liventsev:2013zz}, 
BEBC \cite{CooperSarkar:1985nh}, 
FMMF \cite{Gallas:1994xp}, 
E949 \cite{Artamonov:2014urb},
PIENU \cite{PIENU:2011aa},
NOMAD \cite{Astier:2001ck}, 
TINA \cite{Britton:1992xv},
PS191 \cite{Bernardi:1987ek},
CHARM \cite{Bergsma:1985is,Orloff:2002de},  
CHARMII \cite{Vilain:1994vg},
NuTeV \cite{Vaitaitis:1999wq}, 
NA3 \cite{Badier:1985wg}
and kaon decays \cite{Yamazaki:1984sj,Hayano:1982wu}.
For peak searches below the kaon mass,
 we use the summary given in \cite{Atre:2009rg}, and for PS191 we use the re-interpretation given in \cite{Ruchayskiy:2011aa}.
We combine these with indirect constraints from searches for $0\nu\beta\beta$ decays \cite{Agostini:2013mzu,KamLAND-Zen:2016pfg},
lepton universality \cite{Lazzeroni:2012cx,Czapek:1993kc,Beringer:1900zz},
violation of CKM unitarity \cite{Beringer:1900zz,Antonelli:2010yf,Aoki:2013ldr,Follana:2007uv,Amhis:2012bh,Antusch:2014woa}, LNV decays \cite{Adam:2013mnn,Blankenburg:2012ex}
and with electroweak precision data \cite{Baak:2014ora,ALEPH:2005ab,Group:2012gb,Ferroglia:2012ir}. 
In order to remain consistent with neutrino oscillation data even for large values of ${\rm Im}\omega$, we use the radiatively corrected Casas Ibarra parametrisation introduced in ref.~\cite{Lopez-Pavon:2015cga} instead of eq.~(\ref{CasasIbarraDef}) for the purpose of the scan.
Finally, we impose the constraint that the $N_i$ lifetime is shorter than $0.1$ s, based on the various decay rates given in refs.~\cite{Gorbunov:2007ak,Canetti:2012kh}.
This requirement is motivated by the fact that the heavy neutrinos, which come into thermal equilibrium in the early universe \cite{Hernandez:2013lza}, should have decayed before the formation of light elements in big bang nucleosynthesis (BBN) \cite{Ruchayskiy:2012si}.
The precise implementation of all these conditions is discussed in detail in ref.~\cite{Drewes:2015iva}.
The results are shown in figures~\ref{Fig:Flavor_Plots_NH}-\ref{Fig:Flavor_Plots_IH}.
Potential future improvements are discussed in section~\ref{Sec:Future}. We find that indirect search constraints other than neutrino oscillation data do not significantly affect the range of the allowed $U_a^2$. This is consistent with what has previously been found in refs.~\cite{Gorbunov:2014ypa,Canetti:2013qna}. However, the interplay between direct search bounds, neutrino oscillation data and BBN leads to non trivial combined constraints that are much stronger than one would expect from superimposing the regions excluded by each of these individually in the $\bar{M}-U_a^2$ plane.
\subsubsection{Normal hierarchy}

\paragraph{Mixing $U_e^2$ with electron flavour}\label{NHe}--
The constraints on $U_e^2$ for normal hierarchy are shown in the upper left panel of figure~\ref{Fig:Flavor_Plots_NH}.
For $\bar{M}$ below the kaon mass $m_K$, the global constraints are much stronger than the direct search constraints alone.
The reason is the interplay between the direct search bounds on $U_e^2$ and $U_\mu^2$, neutrino oscillation data and the lower bound on $U_a^2$ from BBN.
The latter requires at least one of the $U_a^2$ to be sufficiently large that the heavy neutrinos decay in less than $0.1$s.
In order to make this consistent with very small $U_e^2$, either $U_\mu^2$ or $U_\tau^2$ have to be sizeable. The direct search constraint on $U_\tau^2$ is rather weak, but neutrino oscillation data does not allow $U_\tau^2$ to be larger than  $U_\mu^2$ by more than a factor $\sim 5-6$, see figure~\ref{fig:regions_NH.pdf}.
 Hence, the rather strong bounds on $U_\mu^2$ from the experiments PS191 and E494 also forbids large values of $U_\tau^2$. 
In combination, this implies that $U_e^2$ must not be too small in order to be consistent with BBN, and a significant fraction of the heavy neutrino mass region below $\bar{M}=m_K$ is in fact already ruled out. 
This is clearly visible by looking and the constraints on $U_\mu^2$ in the upper right panel of figure~\ref{Fig:Flavor_Plots_NH}.
This is not obvious if one simply superimposes the upper bound on $U_e^2$ from direct searches  and the lower bound from BBN, which are separated by an order of magnitude for $\bar{M}$ near the kaon mass.
The lower bound is dominated by BBN for masses below $\sim 1$ GeV. 
For larger masses it is determined by the requirement that at least one of the $N_i$ must mix with electron flavour in order explain the observed light neutrino oscillation data. This constraint is sometimes called the \emph{seesaw constraint}. It does not apply to individual $U_{a i}^2$, but does impose a lower bound on all $U_a^2$.

For $\bar{M}$ between the kaon and D meson mass $m_D$, the combined constraints on $U_e^2$ are essentially identical with the stronger one amongst the direct search constraints on $U_e^2$ and $U_\mu^2$. The reason why the direct search constraint on $U_\mu^2$ effectively also acts as an upper bound on $U_e^2$ is that, for normal hierarchy, $U_e^2$ must be smaller than $U_\mu^2$ to be consistent with neutrino oscillation data, see figure~\ref{fig:regions_NH.pdf}.
There is only a small region slightly above $m_D$ in which the combined constraints are much stronger than the direct constraint on any of the $U_a^2$. This has already been pointed out in ref.~\cite{Drewes:2015iva}. It is related to the fact that the CHARM experiment imposes constraints on combinations of the $U_a^2$ \cite{Ruchayskiy:2011aa}.

For all other masses between $m_D$ and the $W$ boson mass $m_W$, the global constraint can be understood as a combination of the direct search bound from the DELPHI experiment at LEP and neutrino oscillation data. The shape of the upper bound on $U_e^2$ as a function of $\bar{M}$ is determined by the DELPHI bound, which is more or less the same for all $U_a^2$.
However, since neutrino oscillation data requires $U_e^2$ to be an order of magnitude smaller than $U_\mu^2+U_\tau^2$, the combined constraint on $U_e^2$ is about a factor $5$ stronger than the direct search bound and is saturated when $U_\mu^2 \simeq U_\tau^2$.

\paragraph{Mixing $U_\mu^2$ with muon flavour}--
As already explained in the for $U_e^2$, significant fractions of the region $\bar{M}<m_K$ are excluded by the interplay of direct search constraints, neutrino oscillation data and BBN. For almost all values above $m_K$, the combined bound is identical with the direct search bound alone. This is easy to understand: Constraints on $U_e^2$ have almost no effect on the combined constraint on $U_\mu^2$ because neutrino oscillation data in any case implies $U_e^2<U_\mu^2$, and constraints on $U_\tau^ 2$ are never stronger than those on $U_\mu^2$. 
The only exception is the small region near $m_D$ already mentioned in the discussion on $U_e^2$.

\paragraph{Mixing $U_\tau^2$ with tauon flavour} --  The direct search constraints on $U_\tau^2$ for $\bar{M}$ below $m_D$ are much weaker than those on $U_e^2$ and $U_\mu^2$. 
As a result, the combined constraint on $U_\tau^2$ is much stronger than the direct search constraint.
As mentioned above, the strong constraints on $U_\mu^2$ and $U_e^2$ rule out some part of the region $\bar{M}<m_D$. In the allowed mass region below $m_D$, the combined bound on $U_\tau^2$ tracks the direct search constraint on $U_e^2$, but is weaker by the amount that neutrino oscillation data allows $U_\tau^2$ to exceed $U_e^2$ (about two orders of magnitude if $U_\mu^2$ takes the smallest allowed value).
For $m_K<\bar{M}<m_D$ it tracks the direct search bound on $U_\mu^2$ in the same way, but the factor by which $U_\tau^2$ can exceed $U_\mu^2$ is smaller (roughly a factor $5$, see figure~\ref{fig:regions_NH.pdf}).
For masses $\bar{M}>m_D$, the combined constraint essentially equals the direct search constraint from DELPHI.

\subsubsection{Inverted hierarchy}

\paragraph{Mixing $U_e^2$ with electron flavour}--
For inverted hierarchy, most of the mass region $\bar{M}<m_K$ is excluded due to the interplay of constraints described in the following paragraph. 
For larger masses, the global bound is essentially identical to the direct search bound.

\paragraph{Mixing $U_\mu^2$ with muon flavour}--
The combined lower bound on $U_\mu^2$ for $\bar{M}<m_K$ is much stronger than that from BBN or the seesaw constraint alone. The reason is that a small $U_\mu^2$ for inverted hierarchy necessarily comes at the price of a comparably large $U_e^2$, see  figure~\ref{fig:regions_NH.pdf}. This leads to a conflict with the upper bound on $U_e^2$ from the PS191 experiment even for values of $U_\mu^2$ that are two orders of magnitude larger than the lower bound from BBN and/or the seesaw constraint. As a result, almost the entire mass region below $\sim 350$ MeV is ruled out, as this combined lower bound is in conflict with the upper bound from the PS191 experiment.

For $\bar{M}>m_K$, the combined upper bound on $U_\mu^2$ equals the direct search bound in almost the entire mass range we consider, with the exception of the small region above $m_D$ that also exists for normal hierarchy (cf. section~\ref{NHe}).
The lower bound is given by the seesaw constraint.

\paragraph{Mixing $U_\tau^2$ with tauon flavour}--
For $\bar{M}$ below $m_K$, the situation for $U_\tau^2$ is essentially the same as for $U_\mu^2$.  In the regime $m_K<\bar{M}<m_D$, the combined upper bound  tracks the direct search bound on $U_\mu^2$, but is slightly weaker because neutrino oscillation data permits $U_\tau^2$ to be slightly larger than $U_\mu^2$. For masses $\bar{M}>m_D$, the combined constraint essentially equals the direct search constraint from DELPHI.
The lower bound is given by the seesaw constraint.

\subsection{Higgs contribution to heavy neutrino masses}\label{Higgs}
Eq.~(\ref{MNDef}) implies that the neutrino mass matrix has an $\mathcal{O}(\theta^2)$ contribution from the coupling to the Higgs field. 
This contribution is crucial in the $\nu$MSM because it allows to adjust the physical mass splitting in a way that the asymmetry production is resonantly enhanced during both, the $N_i$ production at $T>T_{\rm sph}$ and the $N_i$ decay at $T \ll T_{\rm sph}$ \cite{Shaposhnikov:2008pf}. The latter is needed for resonant DM production \cite{Laine:2008pg} via the Shi-Fuller mechanism \cite{Shi:1998km}.
This adjustment imposes a constraint on the parameter ${\rm Re}\omega$ \cite{Canetti:2012kh}.
The role of the $\mathcal{O}(\theta^2)$ term in eq.~(\ref{MNDef}) has largely been neglected in studies that address the behaviour of $N_i$ in the laboratory. Before moving on to discuss cosmological bounds, we point out that it may be relevant for the interpretation of experimental data in the pseudo-Dirac regime $\epsilon,\mu\ll1$ and affect some of the future analyses proposed in section~\ref{Sec:Future}.
By neglecting terms of $\mathcal{O}(\mu\, m_{i})$, one can approximate
the mass of the right-handed neutrinos by
\begin{align}
\notag
	M_N&=M + v^2\Re[Y Y^\dagger]/\bar{M}\\
	&=
	\begin{cases}
		M + \frac{m_2+m_3}{2} \cosh(\Im \omega) + \frac{m_2-m_3}{2}
		\begin{pmatrix} \cos 2 \Re \omega && \sin 2 \Re \omega \\ \sin 2 \Re \omega && -\cos 2 \Re \omega \end{pmatrix}
		&\text{for NH,}\\
		M + \frac{m_1+m_2}{2} \cosh(\Im \omega) + \frac{m_1-m_2}{2}
		\begin{pmatrix} \cos 2 \Re \omega && \sin 2 \Re \omega \\ \sin 2 \Re \omega && -\cos 2 \Re \omega \end{pmatrix}
		&\text{for IH.}
	\end{cases}
\end{align}
For $\epsilon,\mu\ll1$,
the contribution to the difference between the eigenvalues of $M_N$ from the coupling to the Higgs field can be estimated as
\begin{align}
	\Delta M_{\theta \theta} \approx
	\begin{cases}
		m_2-m_3 \approx 4.11 \times 10^{-11} \,\GeV 
		&\text{for NH,}\\
		m_1-m_2 \approx 7.60 \times 10^{-13} \,\GeV
		&\text{for IH.}
	\end{cases}
\end{align}
As a consequence,  the physical mass splitting cannot be much smaller than the light neutrino mass splitting $\Delta m_{\rm atm}$ unless the parameters are specifically chosen such that there are cancellations between the term involving the mass splitting in $M$ and the $\mathcal{O}(\theta^2)$ term in $M_N$. 
In such a situation, also radiative corrections to the mass splitting have to
be considered~\cite{Roy:2010xq}.
Furthermore, the inclusion of the $\theta^2$ term can change the alignment between the interaction and mass bases: For $\Delta M \lesssim \Delta M_{\theta \theta}$, the mass basis is effectively identical to the interaction basis.
As a result, for $\Delta M \lesssim \Delta M_{\theta \theta}$  the calculations of the $CP$ violation~\cite{Cvetic:2014nla,Cvetic:2015naa}, discussions about the oscillations in the detector~\cite{Cvetic:2015ura,Boyanovsky:2014una} and work on distinguishing the Dirac and Majorana nature of the
right-handed neutrinos from collider signatures~\cite{Dib:2016wge,Dib:2015oka,Anamiati:2016uxp} may need to be revisited.

\section{Constraints from leptogenesis}
\label{Sec:Lepto}
In the previous section we have seen that the requirement to explain the observed light neutrino masses via the seesaw mechanism already imposes considerable constraints on the mass spectrum and flavour mixing pattern of heavy neutrinos with observably large $U^2$. In this section we explore which additional constraints can be obtained from the requirement to explain the observed BAU. 

\subsection{Leptogenesis from neutrino oscillations}
\label{SubSec:Lepto}
The basic idea behind leptogenesis is that a matter-antimatter asymmetry is  generated in the lepton sector and then partly transferred into a baryon number by weak sphaleron processes \cite{Kuzmin:1985mm} that violate $B+L$ and conserve $B-L$. 
$B$ is conserved at temperatures $T$ below the temperature $T_{\rm sph}\simeq  130\,\GeV$ \cite{D'Onofrio:2014kta} of sphaleron freezeout, so that the BAU is determined by the lepton asymmetry $L$ at $T=T_{\rm sph}$.
Traditionally it is assumed that $\bar{M}$ is much larger than the electroweak scale. In this case, the $N_i$ occupation numbers become Maxwell suppressed long before sphaleron freezeout, and the BAU is generated by their $CP$ violating decay~\cite{Fukugita:1986hr}.
For $\bar{M}$ below the electroweak scale, the seesaw relation (\ref{SeesawRelation}) implies that the Yukawa couplings must be rather small, and the $N_i$ production in the early universe is so slow that they may not reach thermal equilibrium before $T=T_{\rm sph}$.\footnote{This implicitly assumes that no $N_i$ are present at the onset of the radiation dominated era.} 
Then the BAU is generated via $CP$ violating flavour oscillations amongst the $N_i$ during their production. This idea has first been proposed in ref.~\cite{Akhmedov:1998qx} and was shown to be feasible for $n_s=2$ in ref.~\cite{Asaka:2005pn}.
We do not repeat the detailed description of this mechanism here, which has meanwhile been studied by various authors \cite{Shaposhnikov:2008pf,Anisimov:2010dk,Anisimov:2010aq,Canetti:2010aw,Garny:2011hg,Garbrecht:2011aw,Canetti:2012vf,Canetti:2012kh,Drewes:2012ma,Canetti:2014dka,Khoze:2013oga,Shuve:2014zua,Garbrecht:2014bfa,Abada:2015rta,Hernandez:2015wna,Kartavtsev:2015vto,Drewes:2016lqo,Hernandez:2016kel,Drewes:2016gmt,Hambye:2016sby}. 
Depending on the model parameters, one can identify two qualitatively different ranges of dynamic behaviour
that we refer to as the \emph{oscillatory regime} and the \emph{overdamped regime}.
These are distinguished by the hierarchy between the heavy neutrino oscillation frequency and the rates at which the heavy neutrino interaction eigenstates approach thermal equilibrium in the symmetric phase of the SM, which are given by the eigenvalues of the matrix $\Gamma_N\simeq Y Y^\dagger \gamma_{\rm av}T$. 
Here $\gamma_{\rm av}$ is a numerical coefficient which we set to $\gamma_{\rm av}=0.012$ for the temperatures of consideration here, corresponding to the value used in ref.~\cite{Garbrecht:2014bfa} based on refs.~\cite{Garbrecht:2013urw}, cf. \ also \cite{Besak:2012qm,Ghisoiu:2014ena,Laine:2013lka}. 

If the equilibration is slow compared to the oscillation frequency of the heavy neutrinos, then baryogenesis happens in the oscillatory regime, which includes two steps. First, $CP$ violating oscillations amongst the $N_i$ generate non-zero lepton numbers $L_a$ in the individual flavours, while the total lepton number $L=\sum_a L_a$ remains negligibly small. Later these asymmetries are washed out, primarily by transferring them into helicity asymmetries in the $N_i$ that are invisible to sphalerons. 
The washout rates $\Gamma_a\simeq (Y^\dagger Y)_{aa} \gamma_{\rm av}T/g_w$ differ for the individual flavours, leading to a net $L\neq 0$ that is partly transferred into $B$ by sphalerons.
Here the factor $g_w=2$ accounts for the fact that $\gamma_{\rm av}$ has been derived in the context of $\Gamma_N$, i.e.  the $N_i$ interact with both components of the SU(2) doublet $\ell_{\rm L}$, while the $L_a$ violating interactions of $\ell_{\rm L}$ only involve the singlet heavy neutrinos. For $U^2$ in the range close to the naive seesaw relation~(\ref{F0}) baryogenesis typically happens in the oscillatory regime.

For large $U^2$ and small $\Delta M$ (i.e.\ $\epsilon,\mu\ll1$), the equilibration time scale of the more strongly coupled interaction state $\Ns$ is much faster than the oscillation time scale. 
This parameter regime can be referred to as the overdamped regime because the BAU is generated within a single overdamped oscillation shortly before sphaleron freezeout. A more detailed description of the two regimes is e.g. given in ref.~\cite{Drewes:2016gmt}.

Quantitatively the SM lepton asymmetries are expressed in terms of the variables $\Delta_a=B/3 - L_a$. The heavy neutrinos can be described in terms of correlation functions in flavour space. Practically we use the quantities $(\delta n_h)_{ij}$, i.e. the deviation of the momentum averaged two point correlator between flavour $i$ and $j$ with helicity $h$ from its equilibrium value. The flavour-diagonal, helicity-odd components of this deviation correspond to the \emph{sterile charges} $q_{Ni}$. At the present level of approximations, this formulation leads to the same results as a description in terms of \emph{density matrices} \cite{Sigl:1992fn} for the heavy neutrinos \cite{Asaka:2005pn}.
We describe the time evolution in terms of the variable $z=T_{\rm ref}/T$, where $T_{\rm ref}$ is  an arbitrarily chosen reference temperature which we take to be the temperature $T_{\rm sph}$ of sphaleron freezeout in the following.
The time evolution is then given by the set of equations~\cite{Drewes:2016gmt}
\bse
\begin{align}
\label{Diff:Sterile}
\frac{\dd}{\dd z}\delta n_{h} &=-\frac{\ii}{2}[H_N^{\rm th}+z^2 H_N^{\rm vac},\delta n_{h}]-\frac{1}{2}\{\Gamma_N,\delta n_{h}\}+\sum_{a,b=e,\mu,\tau}\tilde{\Gamma}_N^a (A_{ab} + C_b/2)\Delta_b\,,\\
\label{Diff:Active}
\frac{\dd \Delta_a}{\dd z} &=
\frac{\gamma_{\rm av}}{g_w} \frac{a_{\rm R}}{T_{\rm ref}}\sum_{i} Y_{ia}Y_{ai}^\dagger
\,\left(\sum_b (A_{ab} + C_b/2)\Delta_b
-q_{Ni}\right)-
2\frac{\gamma_{\rm av}}{g_w} \frac{a_{\rm R}}{T_{\rm ref}}\sum\limits_{\overset{i,j}{i\not=j}}Y^*_{ia}Y_{ja}\,.
\end{align}
\ese
Here $a_{\rm R}=m_{\rm Pl}\sqrt{45/(4\pi^3 g_*)}=T^2/H$ can be interpreted as the comoving temperature in a radiation dominated universe with Hubble parameter $H$, $m_{\rm Pl}$ is the Planck mass and $g_*$ is the number of relativistic degrees of freedom in the primordial plasma. 
The flavour matrix $H_N^{\rm vac}$ can be interpreted as an effective Hamiltonian in vacuum, and $H_N^{\rm th}$ is the hermitian part of the finite temperature correction. 
The contributions involving the matrix $\Gamma_N$ and the vector $\tilde{\Gamma}_N$ correspond to collision terms. They are given by
\bse
\label{RHN:rates}
\begin{align}
H^{\rm vac}_N &=
\frac{\pi^2 }{18 \zeta(3)}\frac{a_{\rm R}}{T_{\rm ref}^3}
\left(\Re[M^\dagger M] + \ii h  \Im[M^\dagger M]\right)\,,
\label{avg:hamiltonian}\\
H^{\rm th}_N&=
\mathfrak{h}_{\rm th}
\frac{a_{\rm R}}{{T_{\rm ref}}}
\left(\Re[Y^* Y^t]-\ii h\Im [Y^* Y^t]\right)\,,\label{avg:hamiltonianth}\\
\label{avg:decayrate}
\Gamma_N &=\gamma_{\rm av} \frac{a_{\rm R}}{{T_{\rm ref}}}
\left(\Re[Y^* Y^t]-\ii h \Im [Y^*Y^t]\right)\,,\\
\label{avg:backreaction}
(\tilde{\Gamma}^a_N)_{ij}&= \frac{h}{2}\gamma_{\rm av} \frac{a_{\rm R}}{T_{\rm ref}}
\left(
\Re [Y^*_{ia}Y^t_{aj}] - \ii h  \Im [Y^*_{ia}Y^t_{aj}]
\right)\,,
\end{align}
\ese
with $\gamma_{\rm av}=0.012$,
and $\mathfrak{h}_{\rm th}\approx 0.23$.\footnote{
Note that eqs.~(\ref{Diff:Sterile},\ref{Diff:Active}) are not flavour covariant because the Majorana condition $N_i=N_i^c$ has been used in their derivation \cite{Drewes:2016gmt}, which (in this form) only holds in the mass basis. 
We have added the imaginary part  in (\ref{avg:hamiltonian})  for formal reasons, it vanishes in the mass basis.
} 
Finally, the matrix $A$ and vector $C$ are given by
\begin{align}
A=
\frac{1}{711}
\left(
\begin{array}{ccc}
-221 & 16 & 16\\
16 & -221  & 16\\
16 & 16 & -221
\end{array}
\right)
\,,
\quad
C=
-\frac{8}{79}
\left(
\begin{array}{ccc}
1 & 1 & 1
\end{array}
\right) \,.
\end{align}
These equations are valid under the assumptions that the $N_i$ are relativistic at all relevant temperatures, and that their mass splitting is kinematically negligible. 
Moreover, they do not include processes that violate the generalised lepton number $\tilde{L}=L+\sum_i (\delta n_{+ ii} - \delta n_{- ii})$.
Finally, we neglect several effects that occur at  $T\sim T_{\rm sph}$. This includes the  kinematic effect of the top quark and gauge boson masses, the correction to $H^{\rm th}_N$ from $\Delta M_{\theta\theta}$ and the temperature dependence of the sphaleron rate (we assume that sphalerons freeze our instantaneously at $T=T_{\rm sph}$).\footnote{This can be taken into account by replacing (\ref{avg:hamiltonianth}) with $H^{\rm th}_N =\frac{a_{\rm R}}{{T_{\rm ref}}} \left(\Re[Y^* Y^t][\mathfrak{h}_{\rm th} + \mathfrak{h}_{\rm EV}(z)] - \ii h\Im [Y^* Y^t]\mathfrak{h}_{\rm th}\right)$, where $\mathfrak{h}_{\rm EV}(z)=\frac{2\pi^2}{18\zeta(3)}\frac{v^2(z)}{T_{\rm ref}^2}z^2$ and $v(z)$ is the temperature dependent Higgs field expectation value.} All of these effects are subdominant in most of the parameter space we study, but may play an important role if the BAU is generated very late (during sphaleron freezeout) and/or the $N_i$ are relatively heavy (with masses comparable to the W boson).

The kinetic equations can be solved analytically in the two limits
\begin{align}\label{conditon1}
\frac{||Y^* Y^t||\gamma_{\rm av}a_{\rm R}^{2/3}}{|M_1^2-M_2^2|^{1/3}}
\begin{cases}
&\ll 1 \quad \text{oscillatory}\\
&\gg 1 \quad \text{overdamped}
\end{cases}
\,,
\end{align}
where $|| Y Y^\dagger ||$ refers to the largest eigenvalue of the matrix.
The largest values for the $U_a^2$ consistent with the observed BAU can be found within the overdamped regime. This regime is particularly interesting for two reasons.
First, it is naturally realised when $B-L$ is an approximate symmetry, such that at high temperatures $T\gg M_i$ the violation of $L$ is suppressed when comparing to a violation of the individual lepton numbers $L_a$. Second, as this regime implies large mixings $U_a^2$, such heavy neutrinos are more likely to be observed in experiments.

\begin{figure}
	\centering
	\subfigure{
	\includegraphics[width=0.65\textwidth]{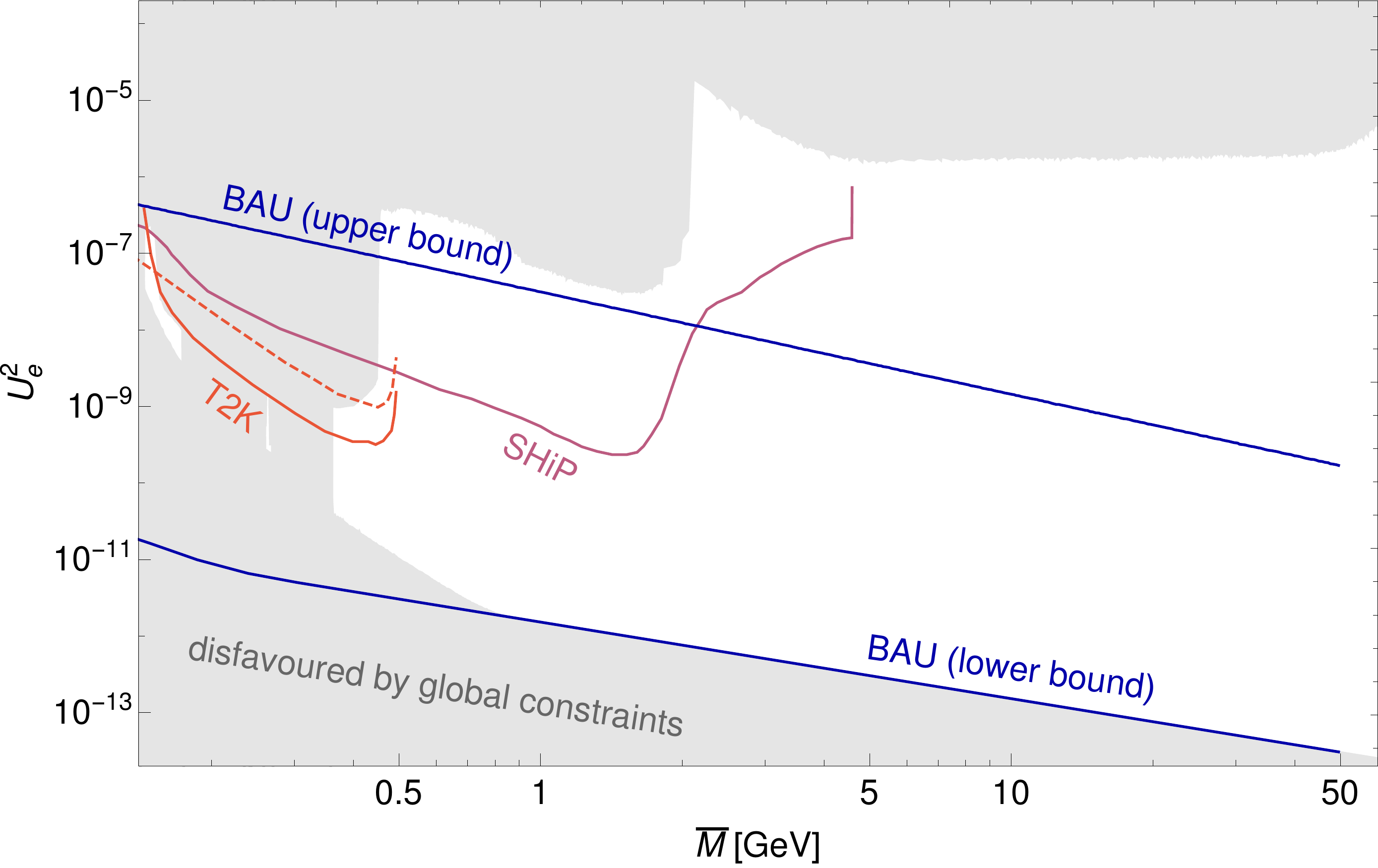}
	}	
	\quad
	\subfigure{
	\includegraphics[width=0.65\textwidth]{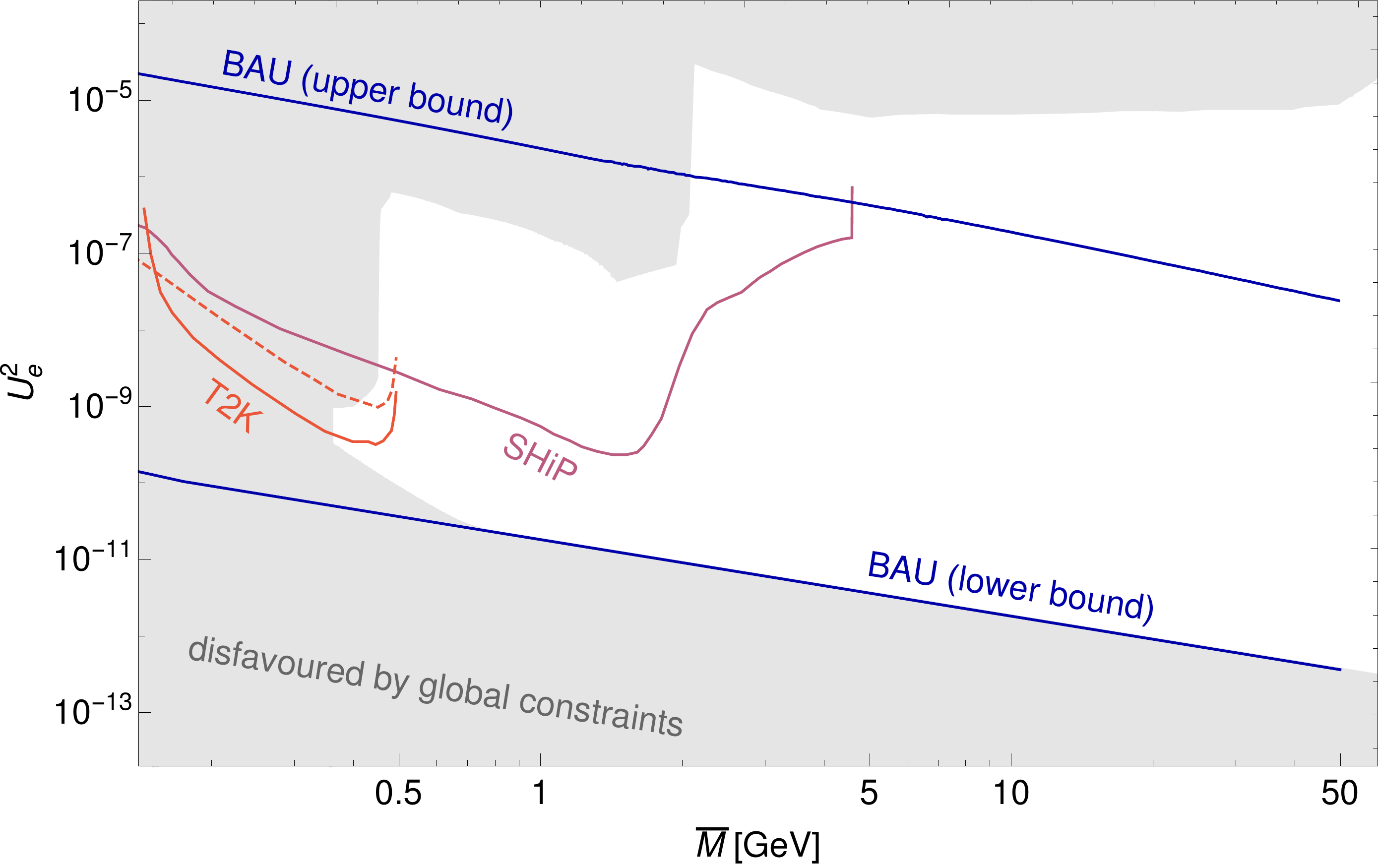}
	}
	\caption{Limits on $U_e^2$ as a function of $\bar{M}=(M_1+M_2)/2$ for normal hierarchy (top panel) and inverted hierarchy (bottom panel). The grey area corresponds to the parameter region that is disfavoured by the combined constraints discussed in subsection~\ref{other_constraints}. The dark blue lines are the upper and lower bound of $U_e^2$ consistent with neutrino oscillation data and the requirement to account for the observed BAU. These are compared to the sensitivity of future experiments: The SHiP line (purple) show the $90\%$ c.l. upper limits assuming $0.1$ background events in  $2\times 10^{20}$ proton target collisions for a ratio of $U_e^2:U_\mu^2:U_\tau^2\sim 52:1:1$~\cite{Anelli:2015pba,Graverini:2214085}. 
The T2K sensitivity has been estimated in ref.~\cite{Asaka:2012bb} for $10^{21}$ protons on target at $90\%$ c.l. with full volume for both the $K^+\to e^+N\to e^+e^-\pi^+$ two-body decays (red, solid) and the $K^+\to e^+N\to e^+e^-e^+\nu_e$ three-body decays (red, dashed)~\cite{Asaka:2012bb}.
Further, $U_e^2$ can e.g. be probed by LBNE/DUNE \cite{Anelli:2015pba,Graverini:2214085}, FCC-ee \cite{Blondel:2014bra,Blondel:2014bra}, CEPC \cite{Antusch:2016vyf} and ILC \cite{Antusch:2016vyf}. However, no experimental sensitivities have been published for benchmark scenarios that would allow to extract the sensitivity to $U_e^2$ in a simple way.
}
		\label{Fig:Flavor_Plots_e}
\end{figure}

\begin{figure}
	\centering
	\subfigure{
	\includegraphics[width=0.65\textwidth]{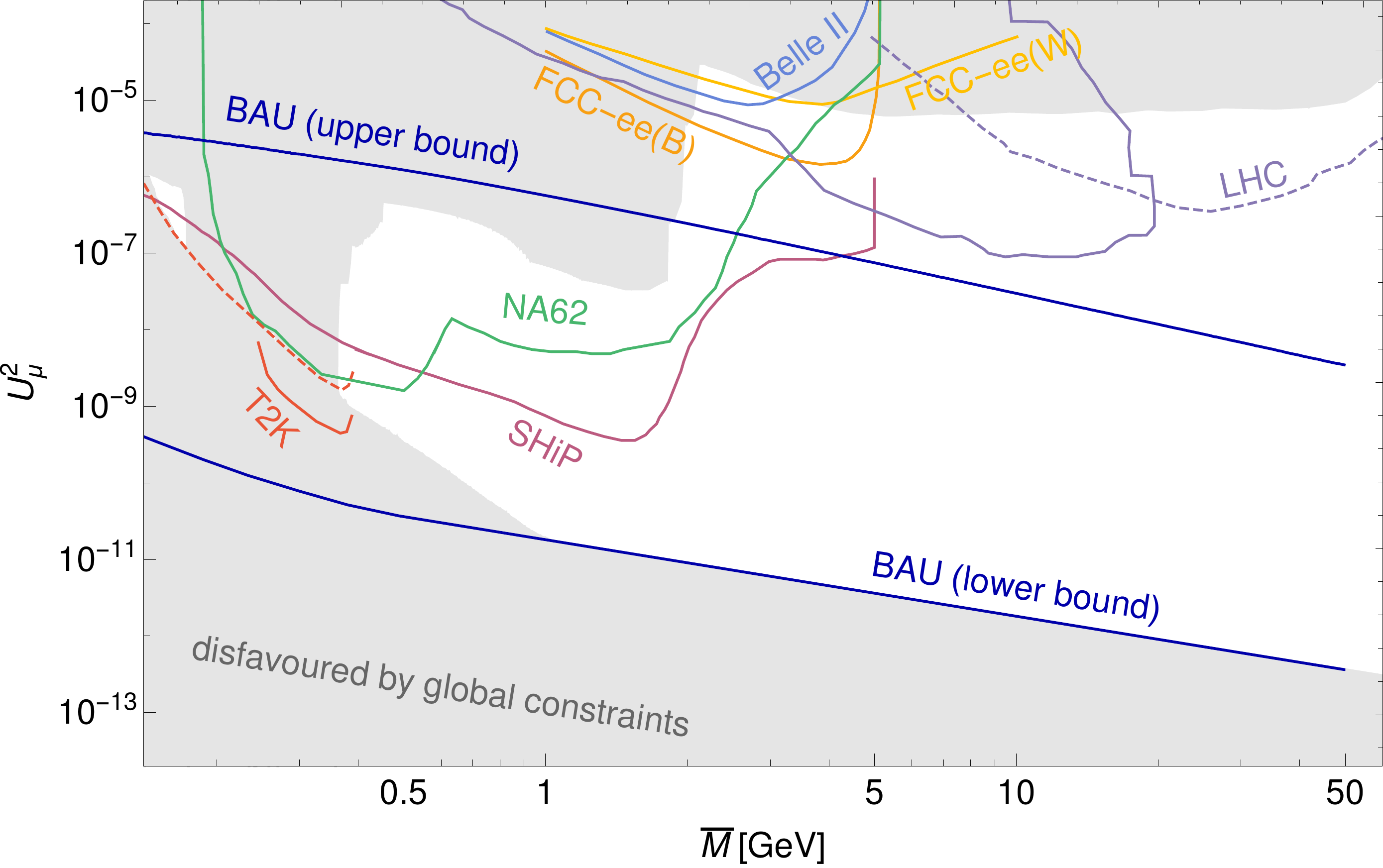}
	}	
	\quad
	\subfigure{
	\includegraphics[width=0.65\textwidth]{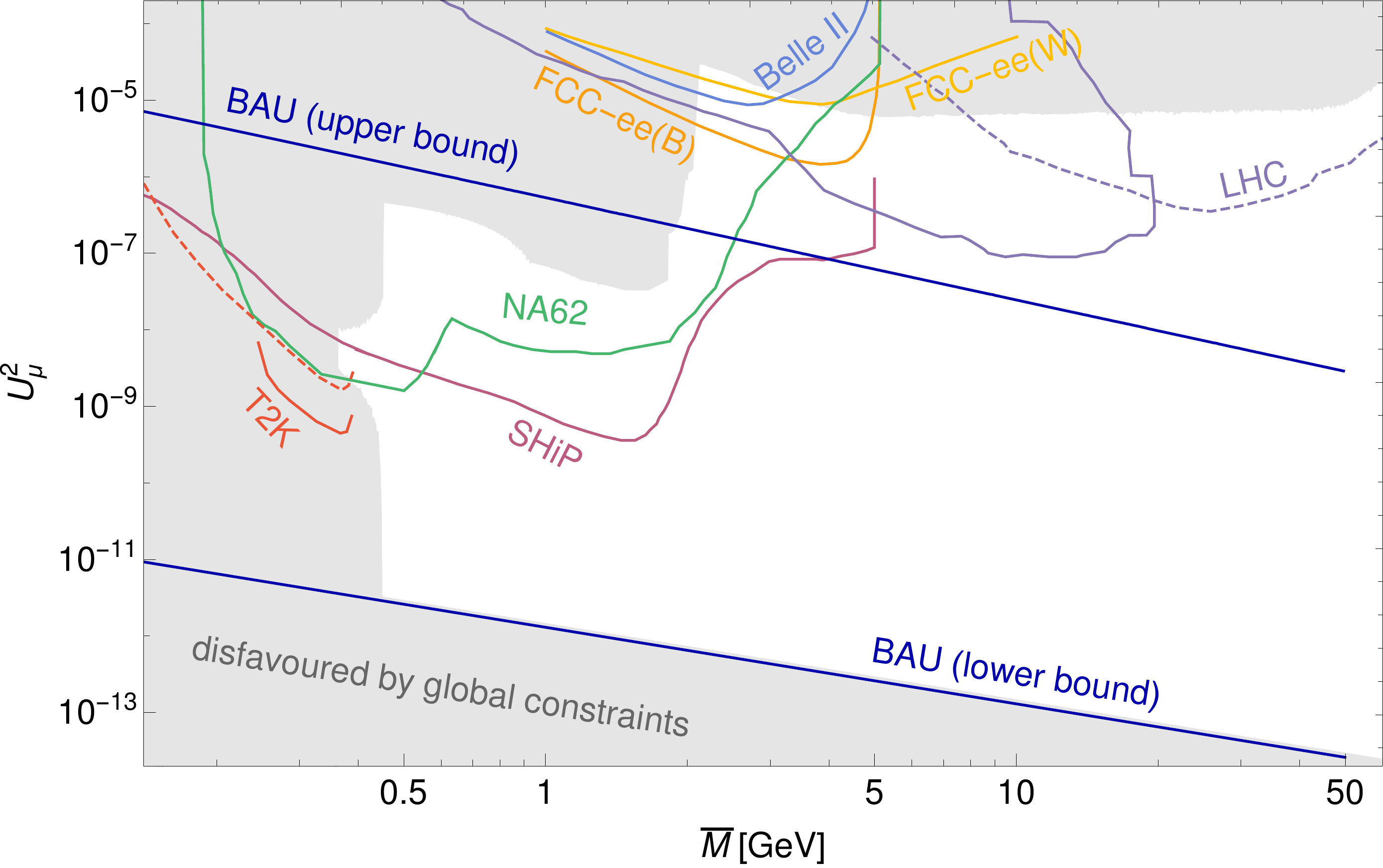}
	}
	\caption{Limits on $U_\mu^2$ as a function of $\bar{M}=(M_1+M_2)/2$ for normal hierarchy (top panel) and inverted hierarchy (bottom panel). The grey area corresponds to the parameter region that is disfavoured by the combined constraints discussed in subsection~\ref{other_constraints}. The dark blue lines are the upper and lower bound of $U_\mu^2$ consistent with neutrino oscillation data and the requirement to account for the observed BAU. These are compared to the sensitivity of future experiments: The SHiP lines (purple) show the $90\%$ c.l. upper limits assuming $0.1$ background events in  $2\times 10^{20}$ proton target collisions for a ratio of $U_e^2:U_\mu^2:U_\tau^2\sim 1:16:3.8$~~\cite{Anelli:2015pba,Graverini:2214085}. 
The NA62 line (turquoise) is the expected limit of the NA62 experiment on $U_\mu^2$ with $2\times 10^{18}$ $400\,\GeV$ protons on target~\cite{Talk:Spadaro}. The T2K sensitivity (red) has been estimated in ref.~\cite{Asaka:2012bb} for $10^{21}$ protons on target at $90\%$ c.l. with full volume for both the $K^+\to \mu^+N\to \mu^+\mu^-\pi^+$ two-body decays (red, solid) and the $K^+\to \mu^+N\to \mu^+\mu^-e^+\nu_e$ three-body decays (red, dashed)~\cite{Asaka:2012bb}. Limits on $U_\mu^2$ can be obtained from LNV decays of $5\times 10^{10}$ $B^+$ mesons at Belle II (blue) and $2\times 10^8$ W bosons (yellow) at the FCC-ee~\cite{Asaka:2016rwd}.
Similarly, the expected sensitivity from $B$ meson decay at the FCC-ee is given by the light orange line~\cite{Asaka:2016rwd}.
The violet lines are the limits on  $U_\mu^2$ from displaced lepton jet (solid) and prompt trilepton (dashed) searches for $\sqrt{s}=13\,{\rm TeV}$ and $300\,{\rm fb}^{-1}$ at the LHC~\cite{Izaguirre:2015pga}. 
Further, $U_\mu^2$ can e.g. be probed by  LBNE/DUNE \cite{Anelli:2015pba,Graverini:2214085}, FCC-ee \cite{Blondel:2014bra,Blondel:2014bra}, CEPC \cite{Antusch:2016vyf} and ILC \cite{Antusch:2016vyf}.
However, no experimental sensitivities have been published for benchmark scenarios that would allow to extract the sensitivity to $U_\mu^2$ in a simple way.}
	\label{Fig:Flavor_Plots_mu}
\end{figure}

\begin{figure}
	\centering
	\subfigure{
	\includegraphics[width=0.65\textwidth]{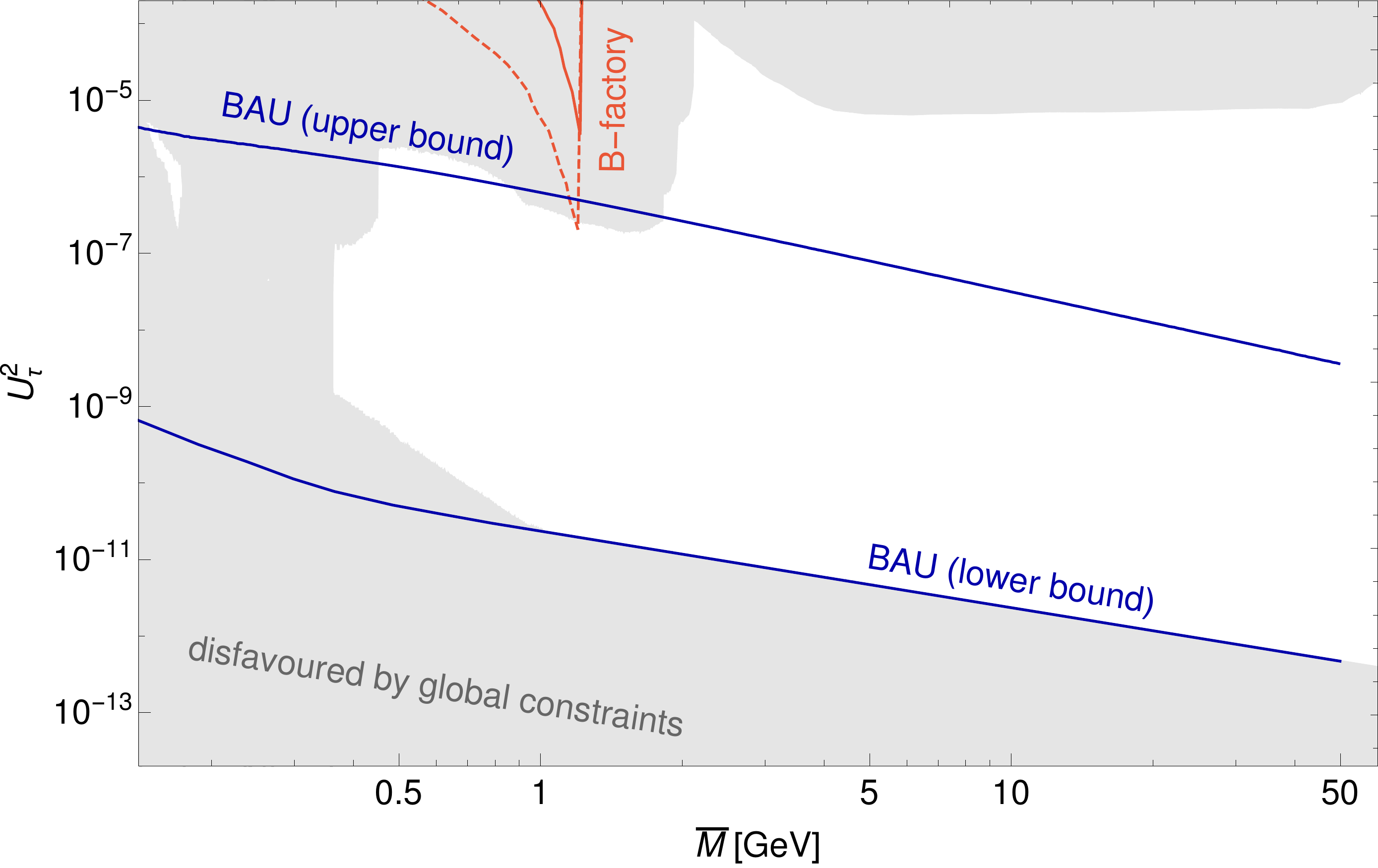}
	}	
	\quad
	\subfigure{
	\includegraphics[width=0.65\textwidth]{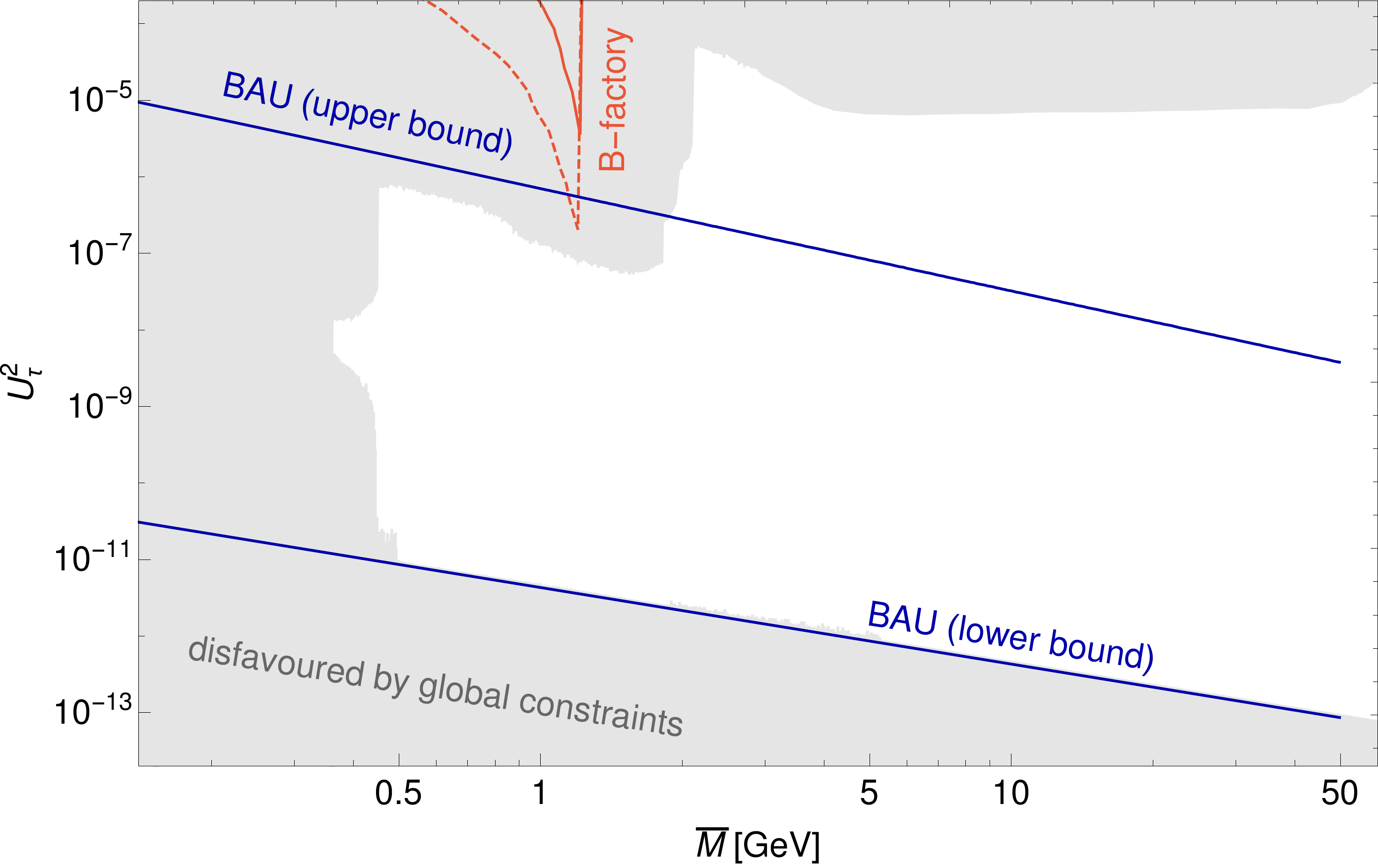}
	}
	\caption{Limits on $U_\tau^2$ as a function of $\bar{M}=(M_1+M_2)/2$ for normal hierarchy (top panel) and inverted hierarchy (bottom panel). The grey area corresponds to the parameter region that is disfavoured by the combined constraints discussed in subsection~\ref{other_constraints}. The dark blue lines are the upper and lower bound of $U_\tau^2$ consistent with neutrino oscillation data and the requirement to account for the observed BAU.  The conservative (red, solid) and most optimistic (red, dashed) $95\%$ c.l. limits on $U_\tau^2$ are shown for a kinematic analysis of $10^6$	$\tau^-\to \nu \pi^-\pi^+\pi^-$  decays at B-factories~\cite{Kobach:2014hea}.
Further, $U_\tau^2$ can e.g. be probed by SHiP \cite{Anelli:2015pba,Graverini:2214085}, LBNE/DUNE \cite{Anelli:2015pba,Graverini:2214085}, FCC-ee \cite{Blondel:2014bra,Blondel:2014bra}, CEPC \cite{Antusch:2016vyf} and ILC \cite{Antusch:2016vyf}.
However, no experimental sensitivities have been published for benchmark scenarios that would allow to extract the sensitivity to $U_\tau^2$ in a simple way.}
	\label{Fig:Flavor_Plots_tau}
\end{figure}

A full treatment of the overdamped regime is numerically demanding due to the large separation between the equilibration and oscillation
temperatures.
However using the semi-analytic approximations developed in ref.~\cite{Drewes:2016gmt}, one can remove the degrees of freedom with large equilibration rates
by fixing these to their quasi-static values. Then one may numerically solve the remaining set of differential equations,
which results in a significantly faster evaluation time.
For a subset of parameter choices used here, the semi-analytic solution was compared to the fully numerical result. The fully numerical approach
gives a systematically $\sim 50\%$ larger baryon asymmetry, therefore the bounds presented in figures
\ref{Fig:Flavor_Plots_e}-\ref{Fig:Flavor_Plots_tau}, as well as \ref{Fig:Lepto_NH}-\ref{Fig:Lepto_IH} may be relaxed accordingly.
We further facilitate the numerical analysis using the fact that, for $\epsilon \ll 1$ we can relate solutions for different parameter choices by
keeping the full time-dependence of the baryon asymmetry $B(z)$ and using the approximate rescaling property described in
table~\ref{table:rescale} taken from \cite{Drewes:2016gmt}.
\begin{table}
	\centering
	\begin{tabular}{r||ccc|cc}
		& \multicolumn{3}{c|}{Scaling of the parameters} & \multicolumn{2}{c}{Scaling of the observables}\\
		\hline
		\hline
		At the original scale & $M$ & $\Delta M^2$ & $\Im\, \omega$
		& $U_{a i}^2$ & $B(z=1)$\\
		\hline
		Rescaled & $\zeta M$ & $\eta \Delta M^2$
		& $\Im\, \omega + \log(\eta/\zeta^3)/6$
		& $\eta^{1/3} \zeta^{-2} U_{a i}^2$ & $B(\eta^{1/3}) \zeta \eta^{-1/3}$
	\end{tabular}
	\caption{The approximate scaling of the observables (right column) under a change of the fundamental parameters (left column).
	The $\Im \Omega$ is chosen such that the ratio of the oscillation and equilibration time scales remains constant. By finding a
	solution to the evolution equations $B(z)$ for a particular choice of parameters, we can obtain the approximate baryon asymmetry for a
	class of parameter choices related by this scaling.}
	\label{table:rescale}
\end{table}

We use these approximations to find the maximally allowed $U^2$ for a given $U_a^2/U^2$.
As choosing $\alpha_{1,2}$ and $\delta$ uniquely fixes the ratio $U_a^2/U^2$, we only need to scan over the remaining free parameters
$\Delta M$, $\omega$ and the discrete parameter $\xi$.
However we can use the rescaling property to determine the $\Delta M$ that gives us the correct BAU by finding
the $\eta$ that solves the equation
\begin{align}
	B(\eta^{1/3}) \eta^{-1/3} = B_\text{observed}\,,
	\label{rescaling:Mbfixed}
\end{align}
and rescale the mass splitting accordingly, which leaves us with optimising $U^2$ for $\omega$ and the discrete parameter $\xi$ only.
Note that when rescaling we choose $\zeta=1$ in this case to keep the absolute mass scale of the RH neutrinos unchanged.
By these means we find the upper bounds that are shown in figures~\ref{Fig:Lepto_NH}-\ref{Fig:Lepto_IH}.

To find the overall bounds for the mixing with individual flavours $U_a^2$ shown in figures~\ref{Fig:Flavor_Plots_e}-\ref{Fig:Flavor_Plots_tau},
we first rescale the mass splitting such that the BAU is maximal at the time of freezeout,
\begin{align}
	\frac{\mathrm{d} B(\eta^{1/3})}{\mathrm{d} \eta}=0\,, 
	\label{rescaling:etafix}
\end{align}
and then change the absolute mass scale to get agreement with the observed BAU,
\begin{align}
	\zeta =\frac{\eta^{1/3} B_\text{observed}}{ B(\eta^{1/3})}\,.
	\label{rescaling:zetafix}
\end{align}
However, as we do not fix $U_a^2/U^2$ in this case, we also need to scan over the parameters $\alpha_{1,2}$, $\delta$, $\Re \omega$ and $\xi$.

The smallest possible values of $U_a^2$ lie in a parametric region where neither of the analytic approximations for
the oscillatory or overdamped regime apply.
Decreasing $U_a^2$ also decreases the Yukawa couplings $Y_{ai}$ and therefore the baryon asymmetry, which requires for compensation a small mass splitting to
produce a resonant enhancement. However, as the oscillations need to happen before the sphaleron freezeout, there is also a lower limit on the mass splitting. This prevents a clear separation between the oscillation, equilibration, and sphaleron freezeout time scales that is required for either of the two approximations to be applicable.
Since both the equilibration and oscillation temperatures are comparable in size to the sphaleron freezeout temperature,
a fully numerical solution is a viable method for exploring this part of the parameter space, which gives us
the lower bounds in figures~\ref{Fig:Flavor_Plots_e}-\ref{Fig:Flavor_Plots_tau}.\\ 

\begin{figure}
	\centering
	\subfigure{
	\includegraphics[width=0.65\textwidth]{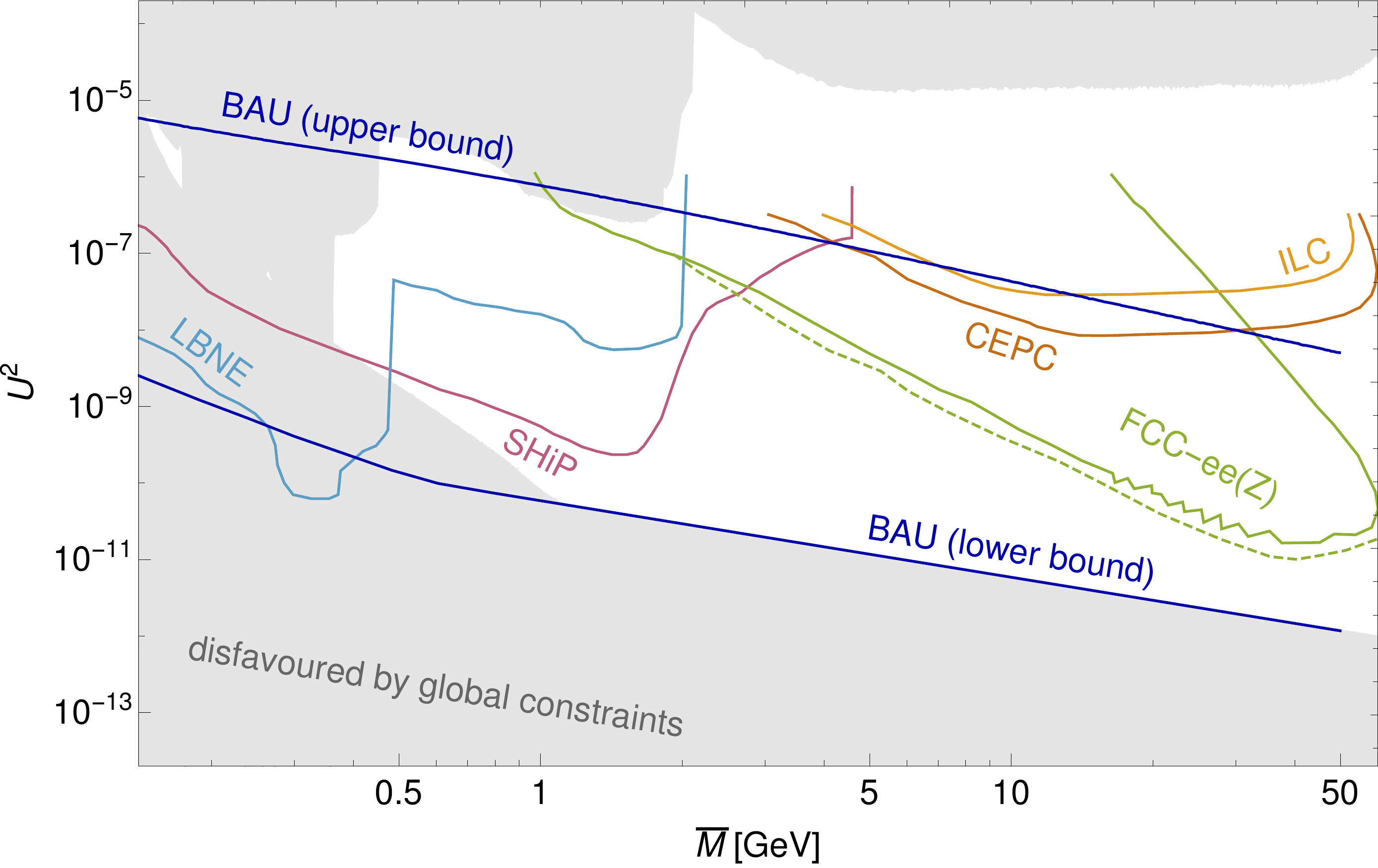}
	}	
	\quad
	\subfigure{
	\includegraphics[width=0.65\textwidth]{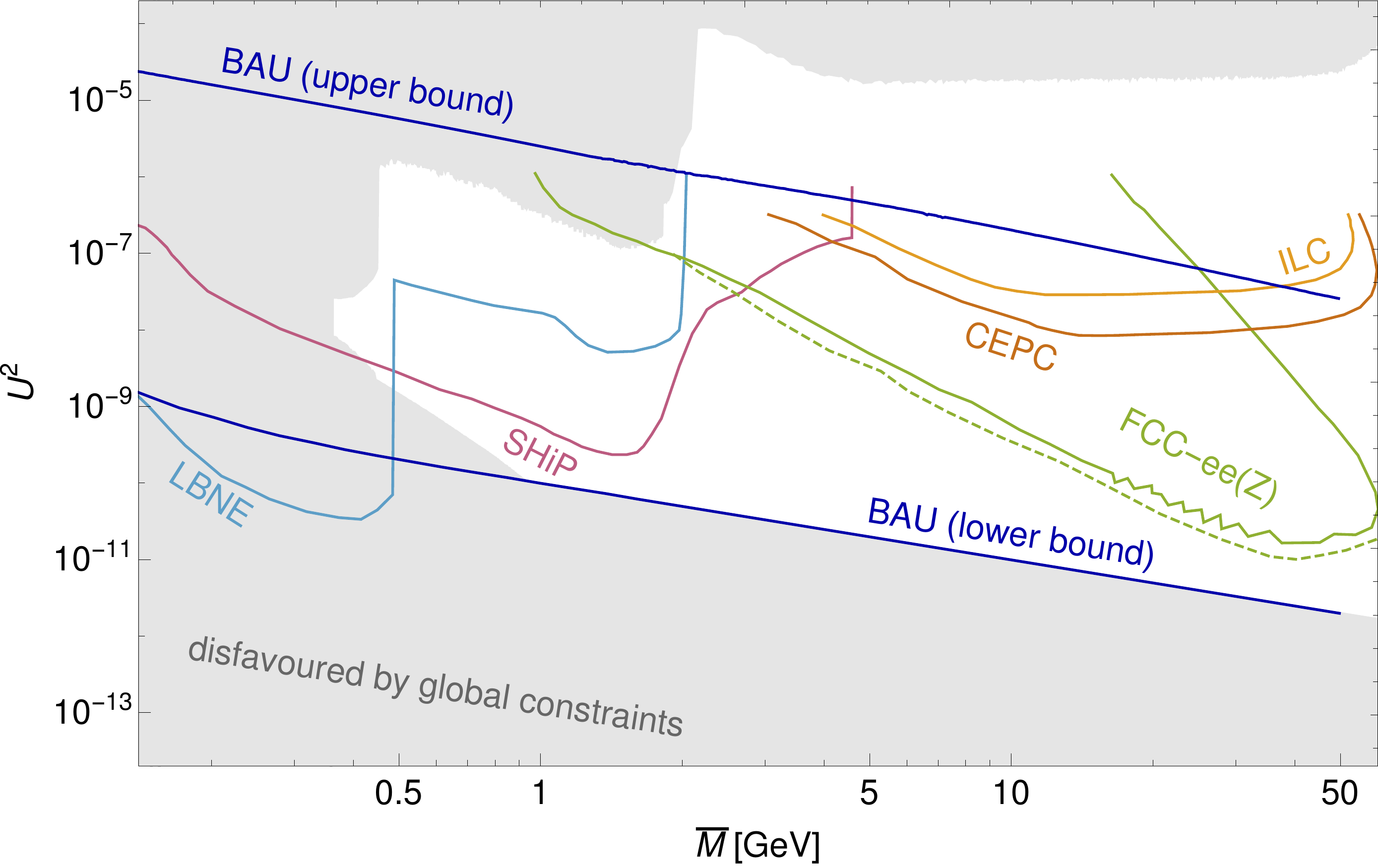}
	}
	\caption{Limits on the total $U^2$ as a function of $\bar{M}=(M_1+M_2)/2$ for normal hierarchy (top panel) and inverted hierarchy (bottom panel). The grey area corresponds to the parameter region that is disfavoured by the combined constraints discussed in subsection~\ref{other_constraints}. The dark blue lines are the upper and lower bound of $U_e^2$ consistent with neutrino oscillation data and the requirement to account for the observed BAU. These are compared to the sensitivity of future experiments: The SHiP lines (purple) show the $90\%$ c.l. upper limits assuming $0.1$ background events in  $2\times 10^{20}$ proton target collisions for a ratio of $U_e^2:U_\mu^2:U_\tau^2\sim 52:1:1$~\cite{Anelli:2015pba,Graverini:2214085}. The LBNE/DUNE sensitivity (light blue) is for the assumption of an exposure of $5\times 10^{21}$ protons on target for a detector length of $30\,{\rm m}$~\cite{Adams:2013qkq}. The solid FCC-ee(Z) lines (olive green) correspond to the expected reach of FCC-ee for $10^{12}$ $Z$ bosons with a displaced vertex at $10-100\,{\rm cm}$~\cite{Blondel:2014bra}. The expected sensitivities at $2\sigma$ for heavy neutrino searches via displaced vertices are presented for the FCC-ee(Z) (olive green, dashed), the CEPC (brown) and for the ILC (dark orange), each at the Z pole run for a centre of mass energy $m_{\rm cms}=m_Z$~\cite{Antusch:2016vyf}.
It is important to point out that the ILC can potentially do much better at higher centre of mass energies ~\cite{Antusch:2016vyf}. 
}
		\label{Fig:Flavor_Plots_tot}
\end{figure}

\subsection{Flavour mixing pattern predictions}
In the overdamped regime, the comparably large Yukawa couplings initially lead to the generation of a baryon asymmetry that tends to be much bigger than the observed value but at the same time is subject to a strong washout.
Since the damping rates $(\Gamma_N)_{ij}$ of the $N_i$ and the washout rates $\Gamma_a$ are governed by the same Yukawa couplings, the only way 
to avoid a complete washout of all lepton asymmetries (and therefore also $B$) in the overdamped regime is to realise a hierarchy in the magnitude of the washout rates $\Gamma_a$ for individual SM lepton flavours. Therefore we expect that leptogenesis with large $U^2$ is only feasible for specific flavour mixing patterns that realise such hierarchy. 
In the oscillatory regime, on the other hand, we do not expect significant constraints on the flavour mixing pattern from leptogenesis, at least not if all of the unknown parameters are allowed to vary freely.
In the following we address the question:

\begin{center}
\emph{
If one fixes $U^2$ to a specific value, for which range of $U_a^2/U^2$ can the observed BAU be generated?
}
\end{center}

\begin{landscape}
\begin{figure}
	\centering
	\subfigure{
	\includegraphics[width=0.45\textwidth]{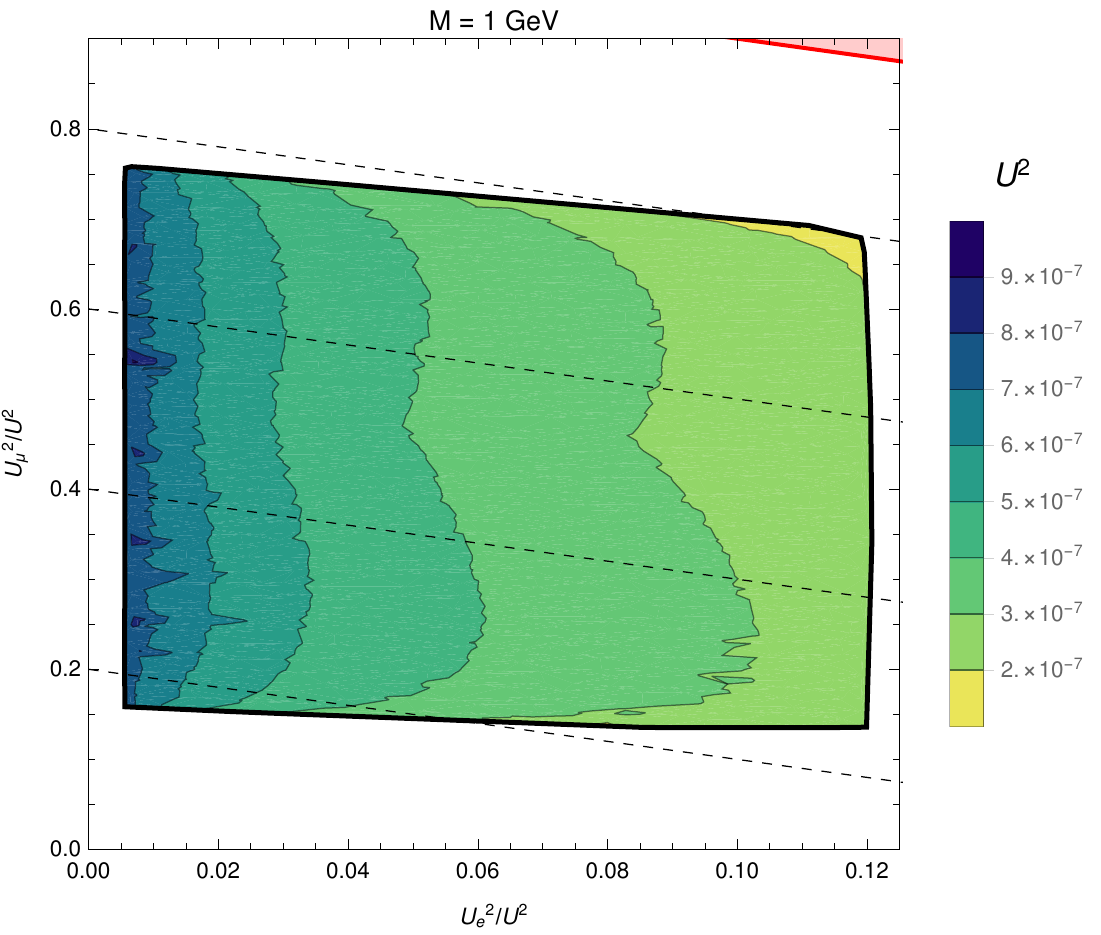}
	}	
	\quad
	\subfigure{
	\includegraphics[width=0.45\textwidth]{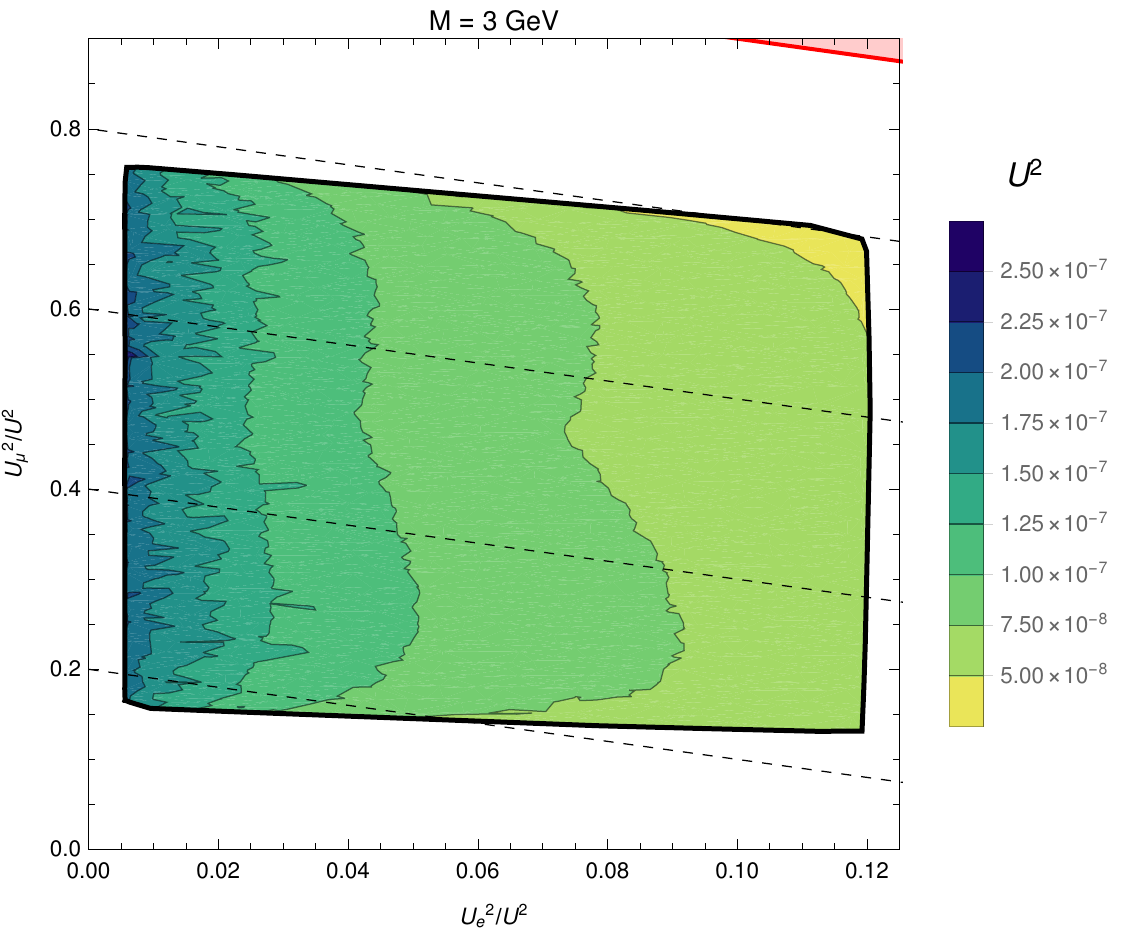}
	}
	\\
	\subfigure{
	\includegraphics[width=0.45\textwidth]{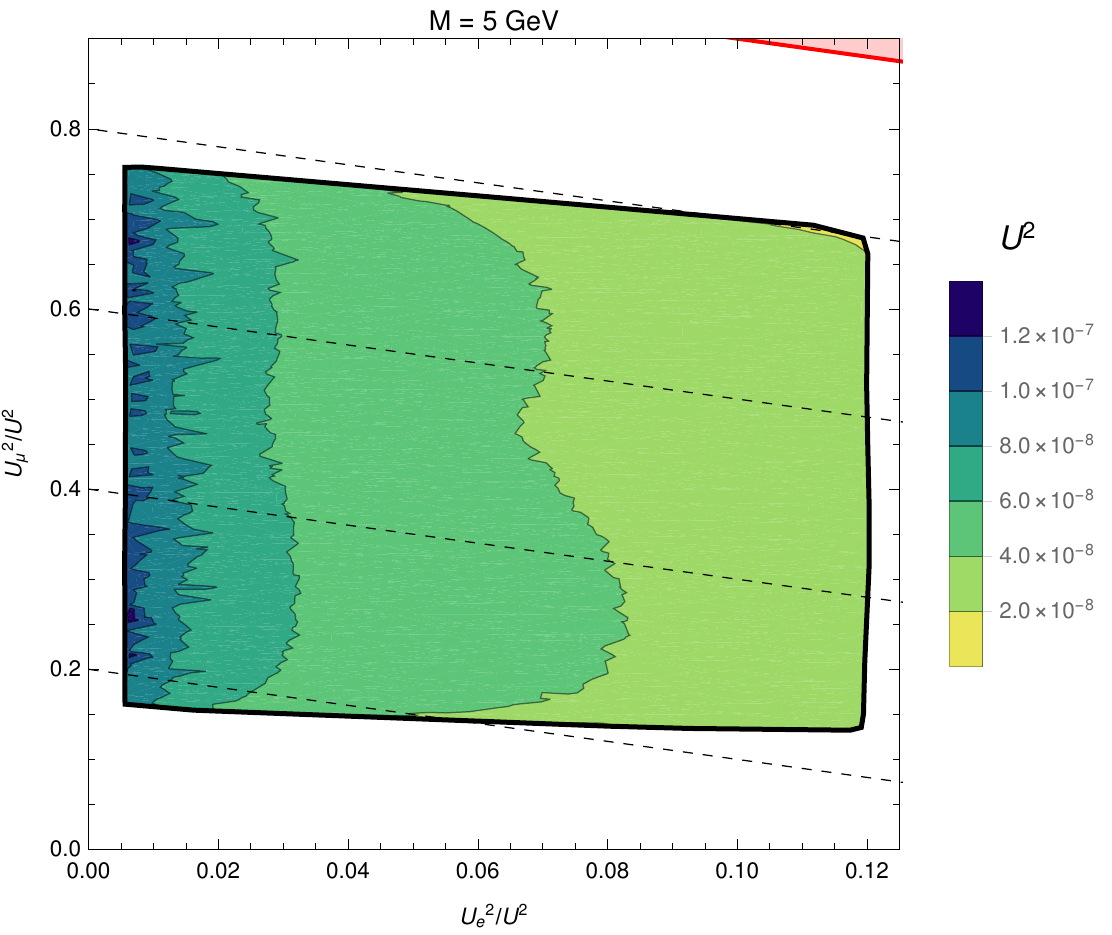}
	}
	\quad
	\subfigure{
	\includegraphics[width=0.45\textwidth]{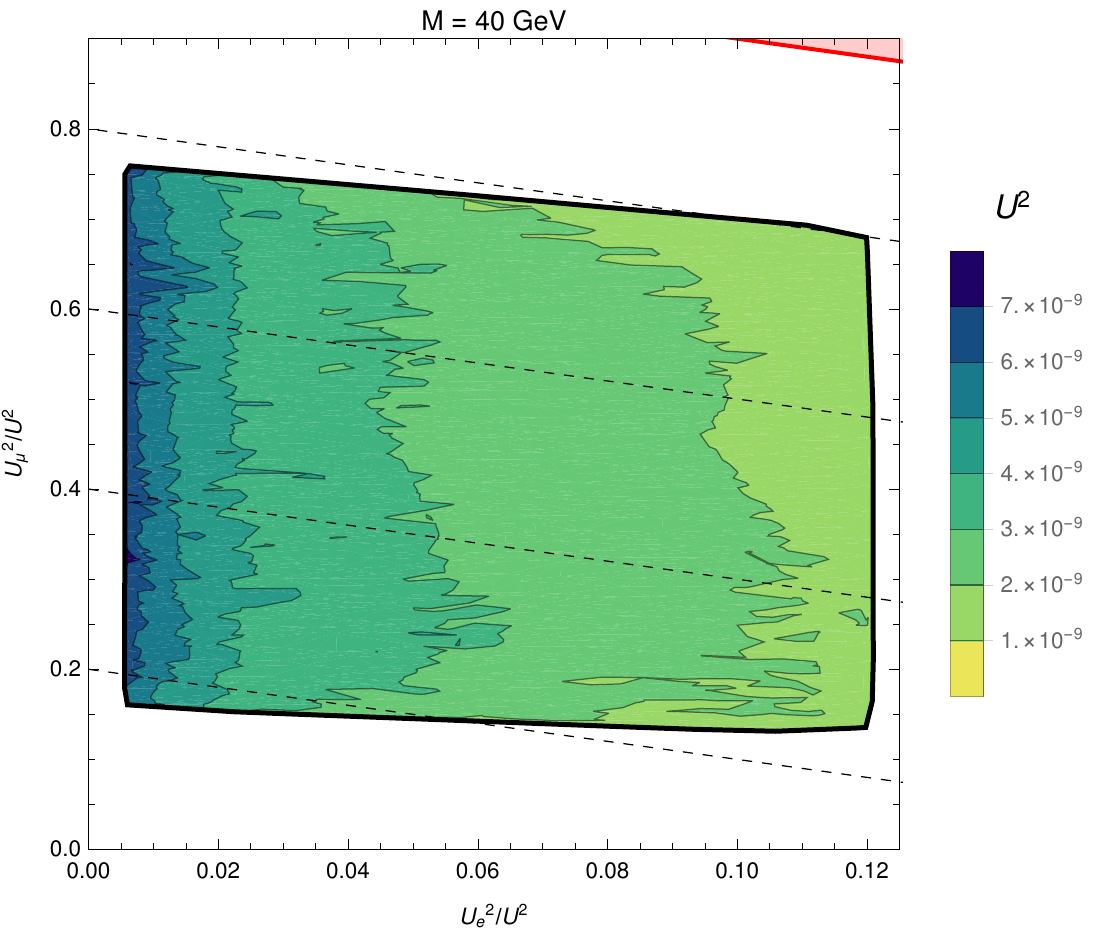}
	}
	\caption{Maximally allowed $U^2$ for given $U_a^2/U^2$ such that the
	 the observed BAU can be generated by leptogenesis.  The four plots are for different Majorana masses $M_i$: $1\,\GeV$ (top left), $3\,\GeV$ (top right), $5\,\GeV$ (bottom left) and $40\,\GeV$ (bottom right). These plots are shown in the $U_e^2/U^2$-$U_\mu^2/U^2$ plane for normal hierarchy.  The thick black lines display the region consistent with neutrino oscillation data, cf. the top panel of figure~\ref{fig:regions_NH.pdf}. $U_\tau^2/U^2$ is given by the requirement $\sum_a U_a^2/U^2=1$ and the iso-$U_\tau^2/U^2$ lines are given by the dashed black lines. The red area indicates the forbidden region $\sum_a U_a^2/U^2>1$.}
	\label{Fig:Lepto_NH}
\end{figure} 
\end{landscape}

\begin{landscape}
\begin{figure}
	\centering
	\subfigure{
	\includegraphics[width=0.45\textwidth]{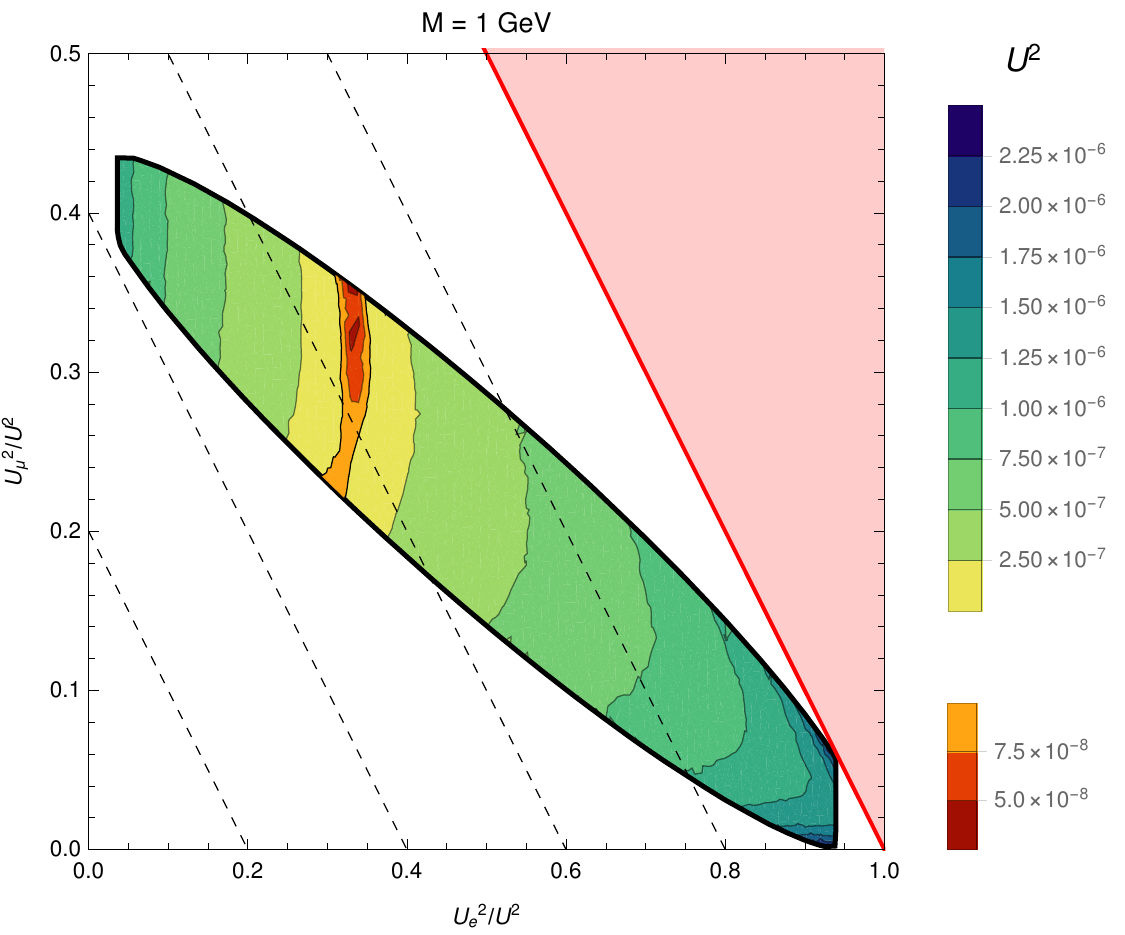}
	}	
	\quad
	\subfigure{
	\includegraphics[width=0.45\textwidth]{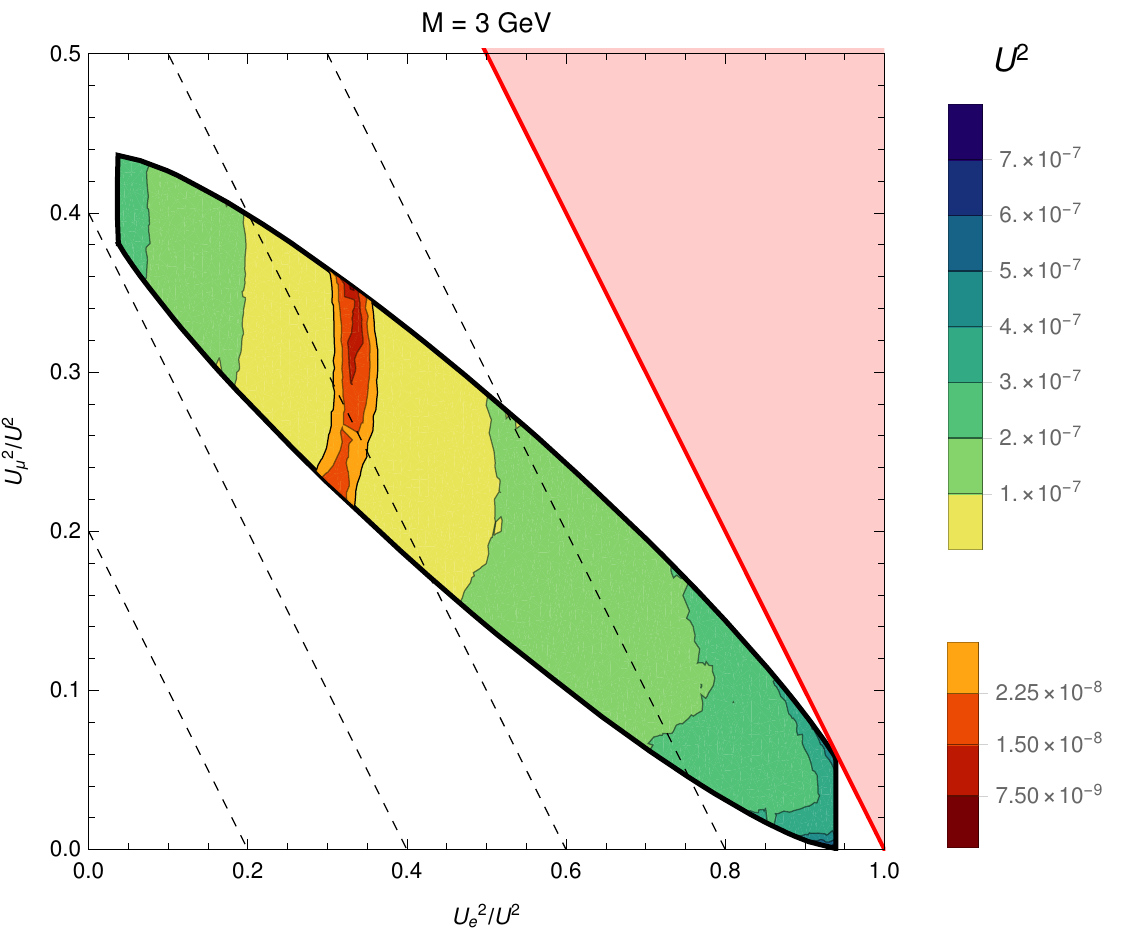}
	}
	\\
	\subfigure{
	\includegraphics[width=0.45\textwidth]{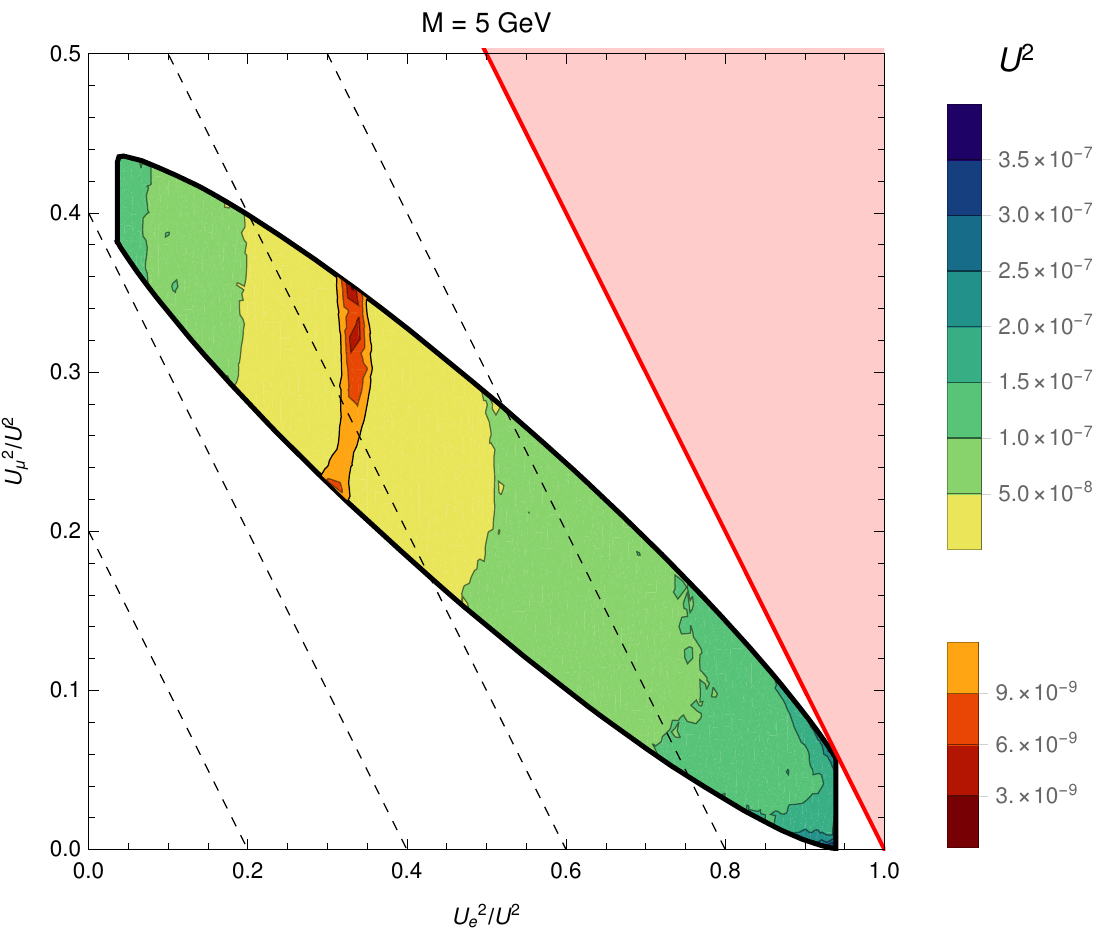}
	}
	\quad
	\subfigure{
	\includegraphics[width=0.45\textwidth]{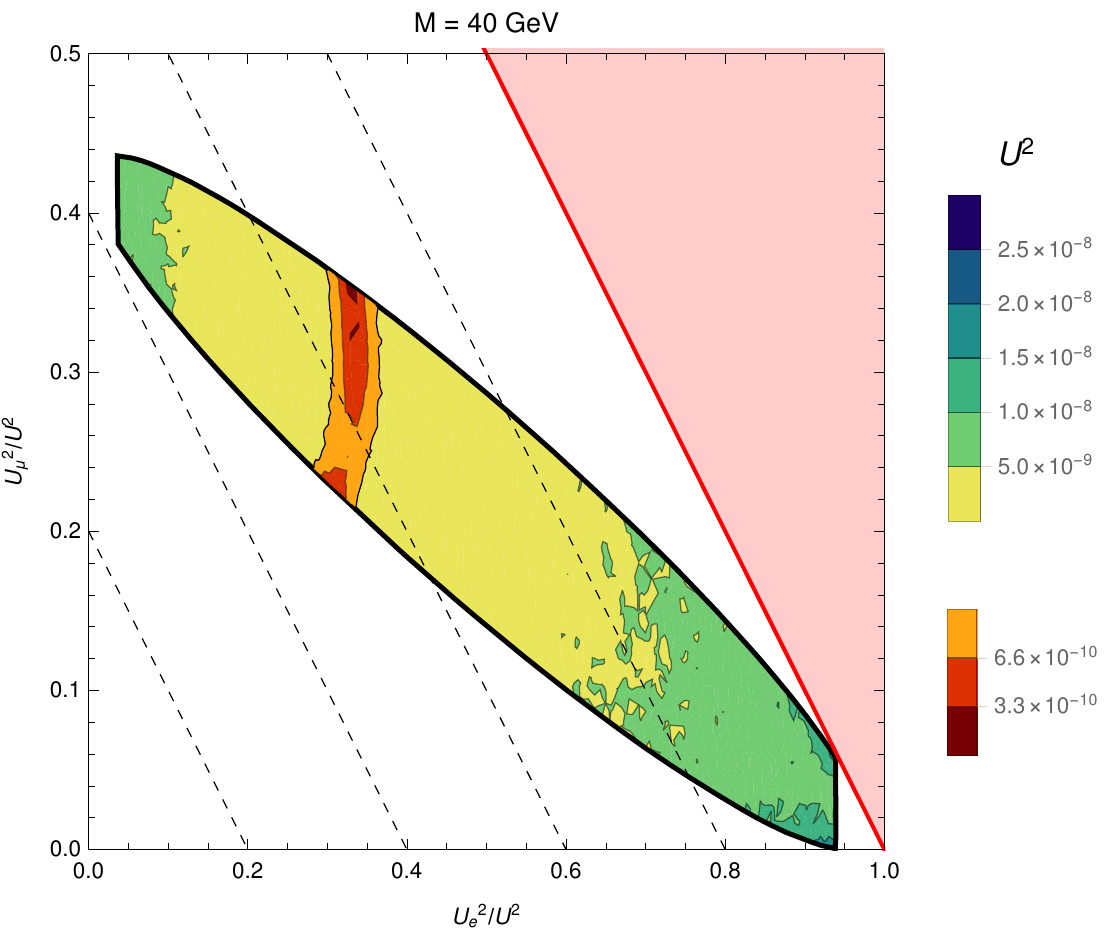}
	}
	\caption{
	Same as figure~\ref{Fig:Lepto_NH}, but for inverted hierarchy. (The thick black lines display the region consistent with neutrino oscillation data, cf. the bottom panel of figure~\ref{fig:regions_NH.pdf}.) 
	The region near the point $U_e^2=U_\mu^2=U_\tau^2$ is forbidden because it does not lead to a flavour asymmetric washout, which is required to generate a total $L\neq 0$ for $n_s=2$.
	\label{Fig:Lepto_IH}
	}
\end{figure}
\end{landscape}

\begin{landscape}
\begin{figure}
	\centering
	\subfigure{
	\includegraphics[width=0.45\textwidth]{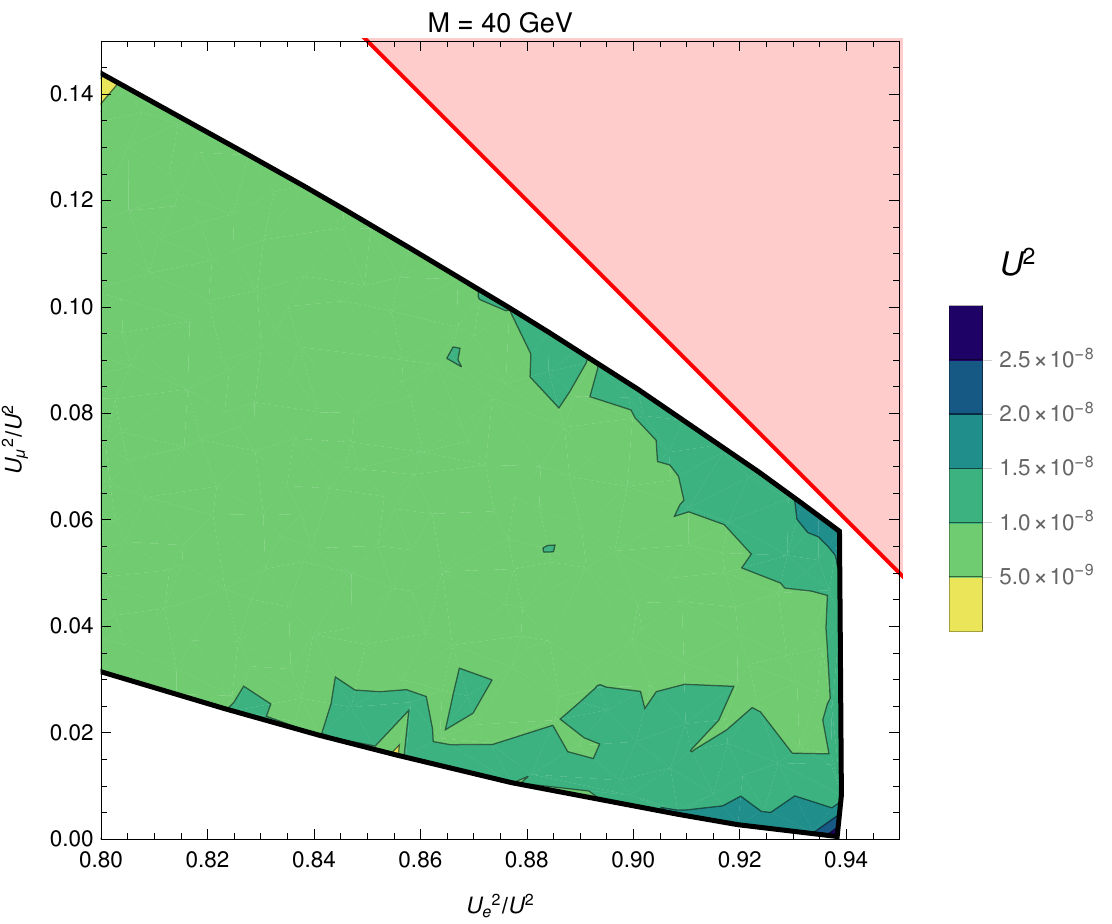}
	}	
	\quad
	\subfigure{
	\includegraphics[width=0.45\textwidth]{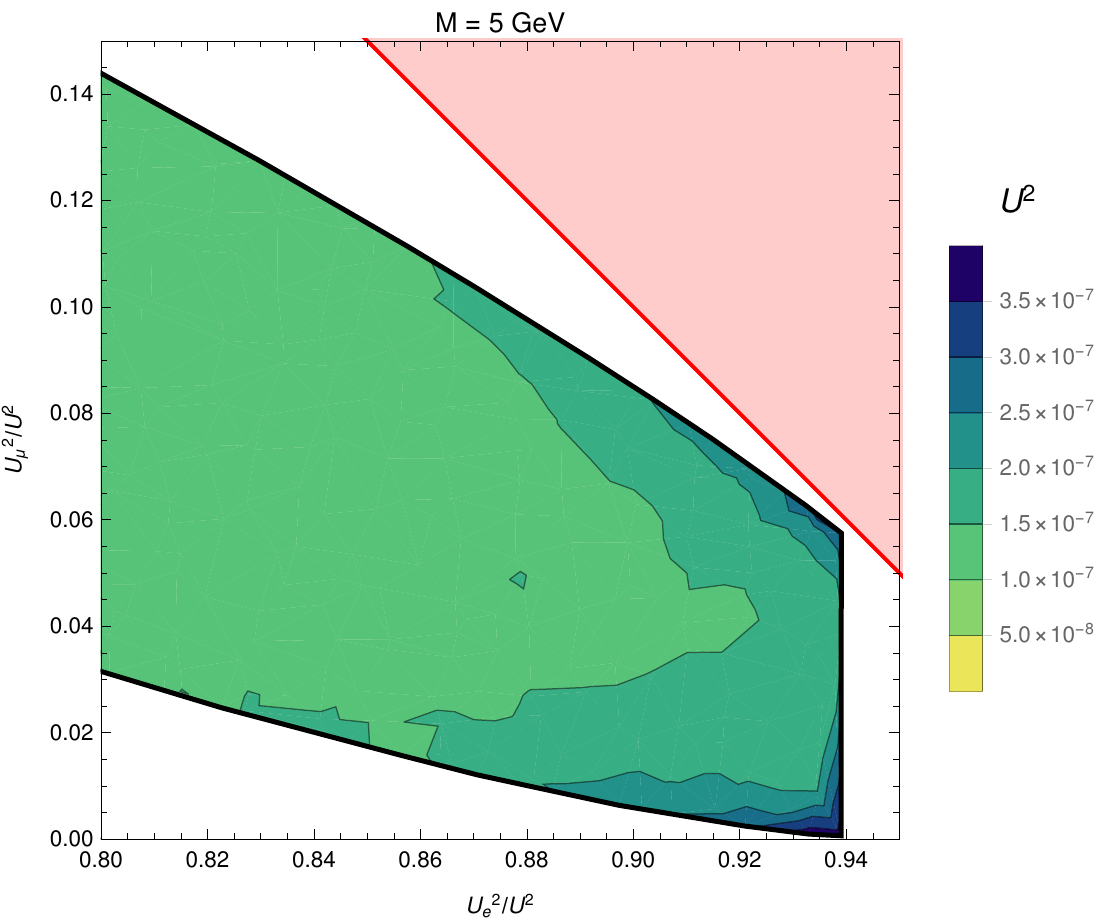}
	}
	\\
	\subfigure{
	\includegraphics[width=0.45\textwidth]{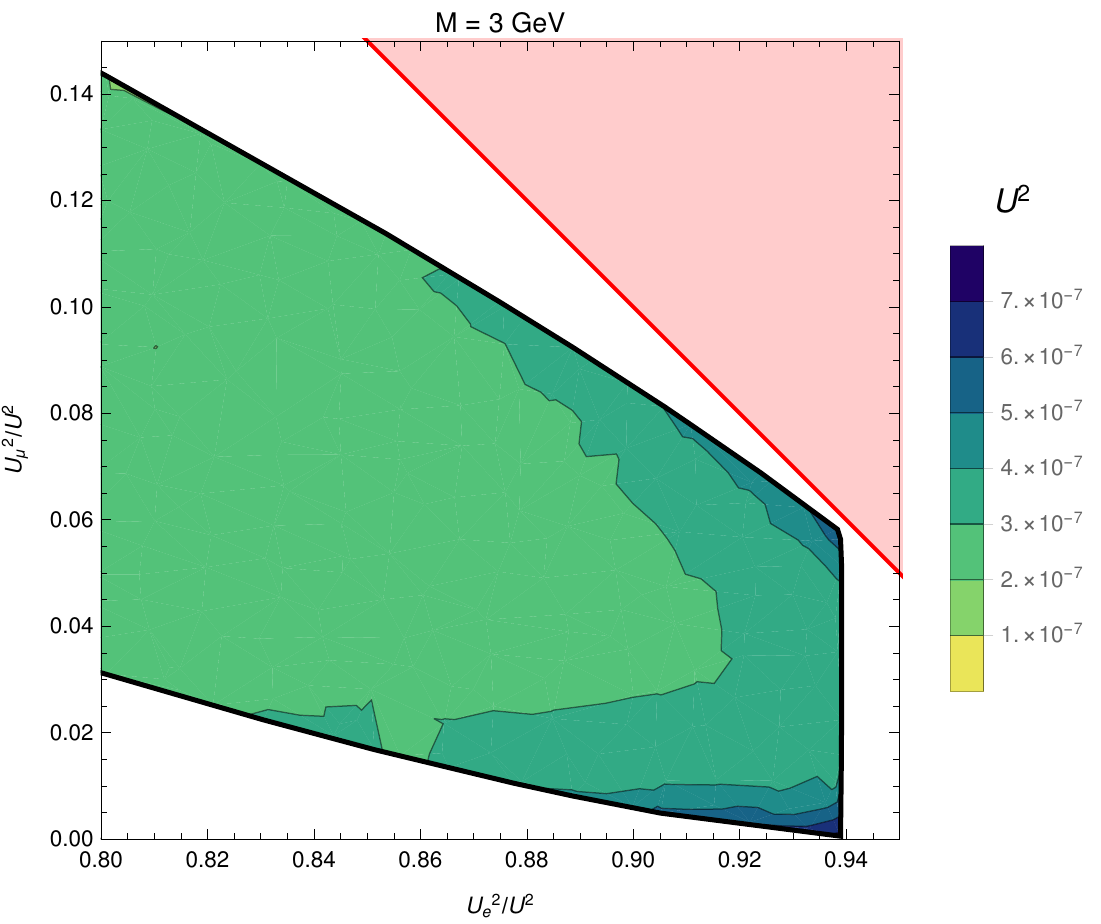}
	}
	\quad
	\subfigure{
	\includegraphics[width=0.45\textwidth]{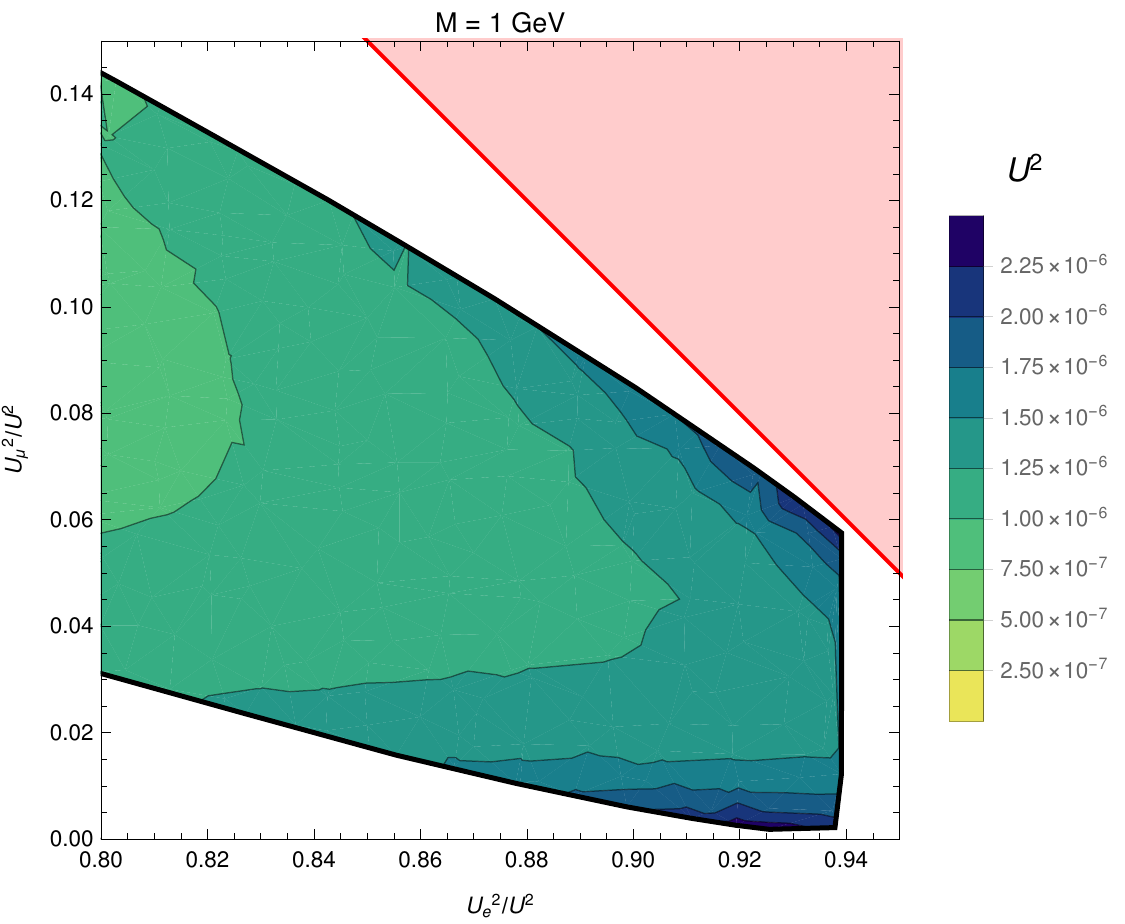}
	}
	\caption{Magnified view of figure~\ref{Fig:Lepto_IH} to show the corner of parameter space that yields maximal mixings $U^2$ separately for the Majorana masses $M_i$: $1\,\GeV$ (top left), $3\,\GeV$ (top right), $5\,\GeV$ (bottom left) and $40\,\GeV$ (bottom right).}
	\label{Fig:Lepto_IH_Zoom}
\end{figure}
\end{landscape}

We use the semi-analytic approach from ref.~\cite{Drewes:2016gmt}, as discussed in subsection~\ref{SubSec:Lepto}, to impose constraints on the flavour mixing pattern $U_a^2/U^2$ for $N_i$ with large $U^2$.
We require that the generated BAU lies within a $5\sigma$ range of the observed value $\eta_B=(8.06 - 9.11)\times 10^{-11}$ \cite{Ade:2015xua}.
The results are shown in figures~\ref{Fig:Lepto_NH}-\ref{Fig:Lepto_IH_Zoom}.
The red region is forbidden because it corresponds to $\sum_a U_a^2/U^2 > 1$.
Neutrino oscillation data can be explained within the solid black lines;
the dashed lines are iso-$U_\tau^2/U^2$ lines. Increasing the total $U^2$ makes the washout stronger such that the experimentally measured BAU can just be fulfilled when requiring a strongly flavour asymmetric washout. The coloured regions inside  figures~\ref{Fig:Lepto_NH}-\ref{Fig:Lepto_IH_Zoom} illustrate how the allowed region becomes smaller when increasing $U^2$. Furthermore, it is clearly visible that the maximally achievable $U^2$ requires a maximally asymmetric washout. For normal hierarchy, cf. figure~\ref{Fig:Lepto_NH}, this happens when the electron couples minimally, $U_e^2/U^2=0.0056$, what corresponds to $\alpha_2=-2\delta+\pi$. In case of inverted hierarchy,  cf. figures~\ref{Fig:Lepto_IH} and \ref{Fig:Lepto_IH_Zoom}, maximal mixing is achieved when the electron couples maximally, $U_e^2/U^2=0.94$, which corresponds to $\alpha_2-\alpha_1=\pi$. 
These  scenarios can are illustrated in figure~\ref{fig:Mixing_Angles_NH_IH.pdf} and discussed in appendix~\ref{App:Mixing} in more detail.

Note that the scenario with equal mixings $U_e^2=U_\mu^2=U_\tau^2$ does not allow for leptogenesis because it fails to produce the flavour-asymmetric washout that is required to produce a total $L\neq 0$.
For normal hierarchy, this scenario is in any case excluded by neutrino oscillation data, for inverted hierarchy the forbidden region is visible in red in figure~\ref{Fig:Lepto_IH}.

\section{Tests in the Future}\label{Sec:Future}
We assume that heavy neutral leptons will be found in a direct search experiment. This is the basic requirement for an experimental confirmation of the seesaw mechanism and leptogenesis.
Though circumstantial evidence may also come from a combination of indirect signatures, these can always be explained in the framework of effective field theory and are not unique to the seesaw model. 
In the following we discuss how much information about the model parameters can be extracted if the $N_i$ are discovered, and how much more  can be learnt if one in addition takes into account data from indirect searches. If there is no discovery, the same relations can of course be used to put stronger limits on the parameter space.

\subsection{Direct searches}
Heavy neutrinos that are lighter than the $W$ boson can be searched for at both, hadron
\cite{Helo:2013esa,Izaguirre:2015pga,Gago:2015vma,Dib:2015oka,
Dib:2016wge,Antusch:2016ejd,PhysRevLett.50.1427,Datta:1992qw,Datta:1993nm}
as well as lepton colliders
\cite{Abada:2014cca,Blondel:2014bra,Graverini:2015dka,Antusch:2015mia,Asaka:2015oia,Abada:2015zea,Antusch:2016vyf,Antusch:2016ejd}.
The reach of the LHC could further be extended with the MATHUSLA surface detector \cite{Chou:2016lxi}.
An overview of the signatures at different collider types has recently been presented in \cite{Antusch:2016ejd}. 
If the heavy neutrinos are lighter than the $B$ mesons, they can also be searched for at $B$ factories \cite{Kobach:2014hea,Canetti:2014dka,Shuve:2016muy,Milanes:2016rzr,Asaka:2016rwd} or fixed target experiments \cite{Gorbunov:2007ak}, including NA62 
\cite{Asaka:2012bb,Talk:Spadaro}, T2K \cite{Asaka:2012bb}, the SHiP experiment proposed at CERN \cite{Anelli:2015pba,Alekhin:2015byh,Graverini:2015dka} or a similar setup proposed at the DUNE beam at FNAL \cite{Akiri:2011dv,Adams:2013qkq,Rasmussen:2016njh}.

One can distinguish between two qualitatively different cases. 
If the mass resolution of the experiment is smaller than $\Delta M$, then $\bar{M}$ and $\Delta M$ can both be determined kinematically. That allows for independent measurements of the $U_{a 1}^2$ and $U_{a 2}^2$. 
If the mass resolution is not sufficient to resolve $\Delta M$, then this parameter cannot be directly determined. Moreover, the signatures of $N_1$ and $N_2$ cannot be distinguished, and branching ratios only allow to constrain the summed mixings $U_a^2$. 
This is the situation that most experiments will face because 
 $U^2$ that are much larger than the estimate (\ref{F0}) can only be made consistent with neutrino oscillation data if $\mu\ll1$.
It is straightforward to verify from the relations given in appendix \ref{App:Mixing} that, in the limit $\epsilon,\mu\rightarrow 0$, one can practically not obtain information about ${\rm Re}\omega$ from measurements of the $U_a^2$, and also the phases in $U_\nu$ cannot be uniquely reconstructed. As we have discussed in the previous sections, measurements of $\bar{M}$ and all of the $U_a^2$ 
nevertheless provide a strong test of the hypothesis that the heavy neutral leptons under investigation are part of the seesaw mechanism.
In the following we investigate how additional measurements can help to further constrain the parameter space in the limit $\epsilon, \mu\ll1$.
We come back to the case that $\Delta M$ can be resolved in section~\ref{Sec:Testability}.

\subsection{Heavy neutrino oscillations and $CP$ violation}

\paragraph{Measurements of the Dirac phase -}
A measurement of a non-vanishing Dirac phase $\delta$ would not only be a clear proof that $CP$ violation exists in the lepton sector, but also allow to make predictions about the heavy neutrino flavour mixing pattern $U_a^2/U^2$, see figure~\ref{fig:regions_NH.pdf}. 
If heavy neutrinos are found in collider or fixed target experiments, the consistency of the observed value of $\delta$ with the indirect constraints extracted from the active-sterile mixings $U_a^2$ provides a very strong test of the minimal seesaw mechanism with $n_s=2$ as the origin of light neutrino masses.
If $\Delta M$ is large enough to be  kinematically resolved, and if all $U_{ai}^2$ can be measured individually, then an independent measurement of $\delta$ allows to uniquely fix all model parameters and reconstruct the Lagrangian (\ref{eq:Lagrangian}), cf. section~\ref{Sec:Testability}.    
This would allow to predict the amount of unitarity violation in $V_\nu$, the rate of neutrinoless double $\beta$ decay and the outcome of future searches for violation of lepton number or lepton universality. Moreover, it would also allow to predict the value of the baryon asymmetry of the universe.

\paragraph{Measurement of the $CP$ violation in $N_i$-mediated meson decays}--
In refs.~\cite{Cvetic:2014nla,Cvetic:2015naa}, it is suggested
that the $CP$ violation could be measured in decays mediated by heavy neutrinos of charged pseudoscalar mesons $M$ such as $B^\pm$ and $D^\pm$ into lighter mesons $M'$ and leptons. The $CP$ asymmetry, characterised by the difference between the decay rates of mesons with opposite charge, can be approximated as
\begin{align}
\label{eq:A_cp}
\mathcal{A}_{CP}\equiv \frac{\sum_i \Gamma(M^-\to \ell_a^-\ell_b^- M'^+)-\Gamma(M^+\to \ell_a^+\ell_b^+ M'^-)}{\sum_i \Gamma(M^-\to \ell_a^-\ell_b^- M'^+)+\Gamma(M^+\to \ell_a^+\ell_b^+ M'^-)}\approx\frac{y}{y^2+1}\sin\left({\rm arg}\left[\frac{\theta_{a2}}{\theta_{a1}}\frac{\theta_{b2}}{\theta_{b1}}\right]\right)\,,
\end{align} 
with $y=\Delta M /\bar{\Gamma}_N$ the ratio of the mass splitting and the decay width of the
sterile neutrino
\begin{align}
	\bar{\Gamma}_{N} = \frac{G_F^2 \bar{M}^5}{96 \pi^3}\mathcal{N} U^2\,,
\end{align}
where $\mathcal{O}(\mathcal{N})=10$.
In the case of at least two almost mass-degenerate Majorana neutrinos the $CP$ violation can become large enough to be measurable if their mass splitting $\Delta M$ is comparable in size to
the decay width $\bar{\Gamma}_N$, assuming that the initial asymmetry of the produced mesons is under control.
As explained in ref.~\cite{Cvetic:2014nla,Cvetic:2015naa}  in more detail,
aside from the $CP$ asymmetry $\mathcal{A}_{CP}$, the observation of $CP$ violation also depends on an effective branching ratio, that can be approximated as
\begin{align}
{\rm Br}_{\rm eff}\sim 10^2 U^4\,.
\end{align}
Observing a possible $CP$ violation requires $|\mathcal{A}_{CP}|{\rm Br}_{\rm eff}$ to be larger than the inverse of the number of produced mesons in the experiment. Consequently the size of the $CP$ violating signal is determined by the absolute value of the active sterile mixing angle
and thus it is driven by $\Im \omega$. Note however that $\Im \omega$ can already be constrained by a measurement of $U_a^2$. 

In the minimal model with $n_s=2$ sterile neutrinos, it would be very challenging to measure the $CP$ violation in the parameter region where it can be responsible for the BAU.
For large mixing angles $U^2$, the relation $\theta_{a2}\simeq \ii\theta_{a1}$ forces the phase difference in eq.~(\ref{eq:A_cp}) to take values near $\pi/2$, cf. eq.~(\ref{pseudoDiractheta}). Therefore, in the limit $\epsilon \to 0$, $\mathcal{A}_{CP}$ vanishes, while the first non-vanishing contribution of $\mathcal{A}_{CP}$ is of order $\epsilon$.
This suppression has not been considered in refs.~\cite{Cvetic:2014nla,Cvetic:2015naa} where it is assumed that the sine function in eq.~(\ref{eq:A_cp}) is approximately one.  However, the perspectives may be better for scenarios with $n>2$, which contain more unconstrained $CP$ violating phases \cite{Zamora-Saa:2016qlk}.

\paragraph{Measurement of oscillations of the right-handed neutrinos}--
The mass splitting of heavy neutrinos may be observed as suggested in \cite{Cvetic:2015ura} where two almost degenerate $\GeV\,$ scale sterile neutrinos are considered. Detecting the width of possible semileptonic decays $B\to \mu e \pi$ can give information about the sterile mass splitting. In $\GeV\,$ seesaw models it is the size of the mass splitting that determines the time of the first oscillation of the sterile neutrinos. For $n_s=2$, highly degenerate sterile neutrinos are favoured for generating large enough mixings $U^2$ to be detected in future experiments. Consequently, a measurement of the mass splitting may rule out parts of the parameter space.  Note that there may be possible corrections if the Higgs term dominates the mass splitting, as discussed in section~\ref{Higgs}.

\begin{figure}
	\centering
	\includegraphics[scale=0.8]{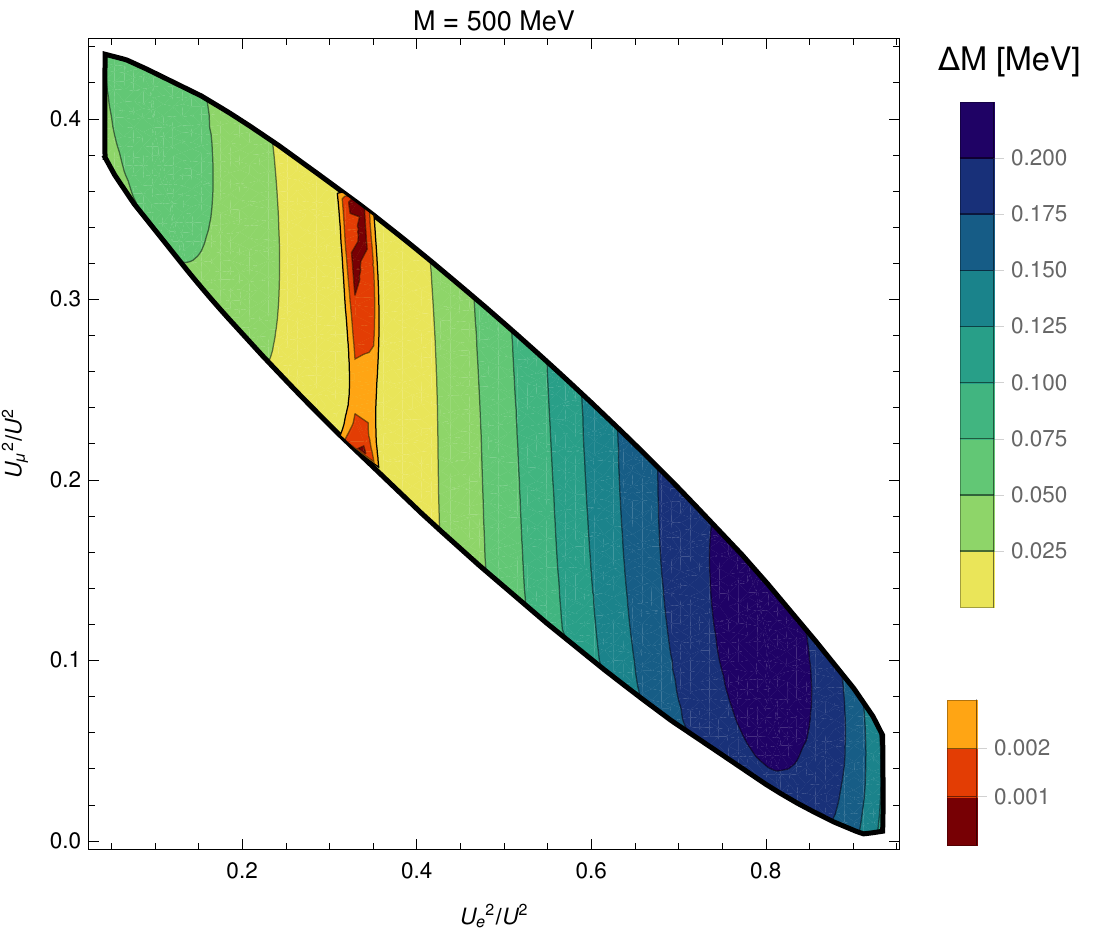}
	\caption{Maximally allowed mass splitting $\Delta M$ for given $U_a^2/U^2$ such that the observed BAU can be generated by leptogenesis. This is an example plot shown in the $U_e^2/U^2$-$U_\mu^2/U^2$ plane for inverted hierarchy with a fixed $\Im\omega=2$ (corresponding to $ U^2=5.38\times 10^{-9}$) and a mass $\bar{M}=0.5\,\GeV$. The thick black lines enclose the region within which neutrino oscillation data can be explained, cf. the bottom panel of figure~\ref{fig:regions_NH.pdf}. $U_\tau^2/U^2$ is given by the requirement $\sum_a U_a^2/U^2=1$.}\label{oscstuff}
\end{figure}

\paragraph{Flavour constraints in the oscillatory regime}--
In the oscillatory regime (or weak-washout regime) of leptogenesis the oscillations of the heavy neutrinos start to happen long before the first heavy neutrino reaches equilibrium as discussed in section \ref{Sec:Lepto}.
In this regime, the flavoured charge density for a given flavour $a$ are generated at very early times, long before the first heavy neutrino reaches equilibrium and can be approximated as~\cite{Drewes:2016gmt}
\begin{align}
\label{eq:osc_charge}
\frac{\Delta_a}{s}
&\approx
-\sum\limits_{\overset{i,j,c}{i\not=j}}
\frac{\Im[Y_{ai}^\dagger Y_{ic} Y_{cj}^\dagger Y_{ja}]}{{\rm sign}(M_{i}^2-M_{j}^2)}
\left(\frac{m_{\rm Pl}^2}{|M_{i}^2-M_{j}^2|}\right)^{\frac23}
\times 3.4
\times 10^{-4}\frac{\gamma_{\rm av}^2}{g_w}\,,
\end{align}
with $s$ the comoving entropy density $s=2\pi^2g_\star a_{\rm R}/45$.
This implies that the BAU effectively scales as $B\propto \Delta M^{-2/3}$. For given $U^2$, a BAU consistent with the observed value can be generated by adjusting the mass difference $\Delta M$.  
If all $U_a^2$ are measured, this allows to impose a constraint on the allowed $\Delta M$, as illustrated in figure~\ref{oscstuff}. Large values of $\Delta M$ are only consistent with specific flavour mixing patterns.
This can in part be understood from the fact that the production of a baryon asymmetry is suppressed in the case of a flavour symmetric washout,
requiring a bigger resonant enhancement.
If $\Delta M$ could be measured independently in some way, this can be used as a consistency check for leptogenesis.

\subsection{Neutrinoless double beta decay}\label{Sec:0nubb}
For heavy neutrino masses and mixings within the reach of fixed target experiments, the rate of $0\nu\beta\beta$ decay can significantly deviate from the standard prediction due to a sizeable contribution from heavy neutrino exchange \cite{Bezrukov:2005mx,Blennow:2010th,Ibarra:2010xw,Asaka:2011pb,Mitra:2011qr,LopezPavon:2012zg,Gorbunov:2014ypa,Drewes:2015iva}.
The rate of $0\nu\beta\beta$ decay can be expressed in terms of the \emph{effective Majorana mass}
\begin{eqnarray}\label{mee}
m_{\beta\beta}&=&\left|
\sum_a (U_\nu)_{ea}^2m_a + \sum_i \Theta_{ei}^2M_i f_A(M_i)
\right|\nonumber\\
&=&
\left|
[1-f_A(\bar{M})]m_{\beta\beta}^\nu
+\sum_i M_i\Theta_{e i}^2[f_A(M_i)-f_A(\bar{M})]
\right|\,.
\end{eqnarray}
The first term in the first line is the contribution from the exchange of light neutrinos,
\begin{equation}
m_{\beta\beta}^\nu=\sum_a (U_\nu)_{ea}^2m_a\,,
\end{equation}
whereas the second term is due to the exchange of heavy neutrinos. Since the typical momentum exchange of $0\nu\beta\beta$ is $\Lambda \sim 100\,\MeV$, masses $M_i$ larger than this scale make the exchanged heavy neutrinos virtual and lead to a contact interaction.
The resulting suppression is parametrised by the function $f_A$, for which we here use the \emph{Argonne model}  as discussed in ref.~\cite{Faessler:2014kka}, 
\begin{equation}
f_A(M)\simeq \frac{\Lambda^2}{\Lambda^2+M^2}\Big|_{\Lambda^2= (0.159\,\GeV)^2}\,.
\end{equation}
To first order in $\Delta M$, one can express eq.~(\ref{mee}) as
\begin{align}
m_{\beta\beta} \simeq\left| [1-f_A(\bar{M})]m_{\beta\beta}^\nu+2f_A^2(\bar{M})\frac{\bar{M}^2}{\Lambda^2}\Delta M(\Theta^2_{e1}-\Theta^2_{e2})\right|\,.
\end{align}
For $m_{\rm lightest}=0$ and $m_{\rm atm}^2 \gg m_{\rm sol}^2$, as well as in the limit $\epsilon \ll 1$, one finds $\Theta_{e2}\approx \theta_{e2}=\ii\theta_{e2}\approx \ii \Theta_{ea}$, such that $\Theta^2_{e1}-\Theta^2_{e2}\approx 2\theta_{e1}^2$. Consequently, the expressions for the two hierarchies can be recast in terms of the Casas-Ibarra parameters
\bse
\label{0nbb_dM}
\begin{align}
m_{\beta\beta}^{\rm NH}&\simeq \left| [1-f_A(\bar{M})]m_{\beta\beta}^\nu+f_A^2(\bar{M})\frac{\bar{M}^2}{\Lambda^2}\frac{\Delta M}{\bar{M}}|m_{\rm atm}|\ee^{2\Im\omega}\ee^{-2\ii(\Re\omega+\delta)}\right|\,,\\ \nonumber
m_{\beta\beta}^{\rm IH}&\simeq \bigg| [1-f_A(\bar{M})]m_{\beta\beta}^\nu+f_A^2(\bar{M})\frac{\bar{M}^2}{\Lambda^2}\frac{\Delta M}{\bar{M}}|m_{\rm atm}| \cos^2\uptheta_{13}\\
&\times \ee^{2\Im\omega}\ee^{-2\ii \Re\omega}(\xi \ee^{\ii \alpha_2/2}\sin \uptheta_{12}+\ii \ee^{\ii\alpha_1/2}\cos \uptheta_{12})^2
\bigg|\,.\label{oderplus}
\end{align}
\ese
In order to make predictions about the heavy neutrino parameters, the terms involving $f_A$ in eqs.~(\ref{0nbb_dM}) for each hierarchy need to be comparable to the experimental sensitivity, which implies that $\bar{M}$ is not too much larger than $\Lambda$.
For $\bar{M}\sim 1$ GeV and $\Delta M >10^{-4}$ GeV, an enhanced rate of $0\nu\beta\beta$ can be achieved while reproducing the correct value for the BAU \cite{Drewes:2016lqo,Hernandez:2016kel,Asaka:2016zib}. This does not only allow to constrain the  phases in $U_\nu$, $\Im\omega$ and $\Delta M$ but also the parameter $\Re\omega$, which is otherwise difficult to access. 
For small values of $\Delta M$, $m_{\beta\beta}$ is given by $m_{\beta\beta}\simeq \left| [1-f_A(\bar{M})]m_{\beta\beta}^\nu\right|$ \cite{Bezrukov:2005mx,Asaka:2011pb}. In this case it still provides a useful probe of the Majorana phase and $\bar{M}$, even if $U^2$ is too small for a direct measurement of $\bar{M}$. For inverted hierarchy $\Re\omega$ can in principle be extracted from $0\nu\beta\beta$ decay as all other parameters can be measured. However, the second term in eq.~(\ref{oderplus}) is increasingly suppressed when increasing $\bar{M}$, such that the sensitivity to $\Re\omega$ is effectively lost due to the error bars of the experiment. It has been previously argued in ref.~\cite{Hernandez:2016kel,Drewes:2016lqo,Asaka:2016zib} that a meaningful constraint on $\Re\omega$ can be inferred from $0\nu\beta\beta$ decay if $\bar{M}$ is smaller than a few $\GeV$.

\subsection{The $\mu\rightarrow e\gamma$ rare lepton decay}
The decay rate $\Gamma_{\ell_a\to\ell_b+\gamma}$ of radiative rare lepton decays $\ell_a\to\ell_b+\gamma$ is given by \cite{Cheng:1980tp,Bilenky:1977du}
\begin{align}
\label{eq:muegamma}
\Gamma_{\ell_a\to\ell_b+\gamma}=\frac{\alpha_{\rm em}m_a^5G_\mu^2}{2048\pi^4}|R_{ab}|^2\,,
\end{align}
with $G_\mu\simeq G_F(V_\nu V_\nu^\dagger)_{ee}(V_\nu V_\nu^\dagger)_{\mu\mu}
$ a modified Fermi constant $G_F$ that is  measured via the lifetime of the $\mu$
and affected by the existence of sterile neutrinos. Here
\begin{align}
\label{eq:R}
R_{ab}=\sum_c(V_\nu^*)_{b c}(V_\nu)_{a c}G\left(\frac{m_c^2}{M_W^2}\right)+\sum_i\Theta^*_{b i}\Theta_{a i}G\left(\frac{M_i^2}{M_W^2}\right)\,,
\end{align}
where the second term is due to an exchange of heavy neutrinos in the loop. The function taking account of the loop contributions is 
\begin{equation}\label{Gloop}
G(x)=\frac{10-43x+78 x^2-49 x^3 + 4 x^4 + 18 x^3 \log(x)}{3 (x - 1)^4}.
\end{equation}
The experimentally most constrained process is  $\mu\rightarrow e\gamma$ with a branching ratio of \cite{Adam:2013mnn}
\begin{align}
\label{eq:muegamma_exp}
B(\mu\rightarrow e\gamma)=\frac{\Gamma(\mu\rightarrow e\gamma)}{\Gamma(\mu \to e \nu_\mu \bar{\nu}_e)}<5.7\times 10^{-13}\,.
\end{align}
With eq.~(\ref{eq:muegamma}) one can express the branching ratio as follows
\begin{align}
B(\mu\rightarrow e\gamma)=\frac{\alpha_{\rm ew}}{32\pi}|R_{e\mu}|^2\,.
\end{align}
If all (heavy and light) neutrinos had the same mass, then the function $G$ would be a constant, and  $R_{e\mu}$ would vanish due to the unitarity of the full $5\times 5$ neutrino mixing matrix. 
This \emph{GIM suppression} is the reason why the rate of the decay $\mu\rightarrow e\gamma$ in the SM without heavy neutrinos is extremely small, as the deviation of $G$ from the constant value $G(0)=10/3$ is tiny ($\mathcal{O}[m_a^2/M_W^2]<10^{-23}$) 
such that we may approximate
\begin{align}
\label{eq:GIM_2}
|R_{e\mu}|\approx \sum_i\left|\theta_{ei}\theta^\dagger_{i\mu}\right|
\left[
G\left(\frac{M_i^2}{M_W^2}\right) - \frac{10}{3}
\right]\,.
\end{align}
One can express eq.~(\ref{eq:GIM_2}) in terms of the Casas-Ibarra parameters
\begin{align}
\label{eq:R_Casas}
|R_{e\mu}|\approx \frac{1}{2\bar{M}}
\left[
G\left(\frac{M_i^2}{M_W^2}\right) - \frac{10}{3}
\right]
m_{\rm atm}\ee^{2\Im\omega}\sin\uptheta_{13}\cos\uptheta_{13}\sin\uptheta_{23}\,,
\end{align}
which is true for both hierarchies and in the limits $\epsilon, \mu, m_{\rm sol}^2/m_{\rm atm}^2\ll 1$. 
While the branching ratio is thus $\propto 1/\epsilon$, it is nonetheless
bounded from above due to the direct upper bounds on the $U_a^2$.
For $\bar{M}\ll M_W$ future searches for lepton flavour violating decays are therefore unlikely to lead to additional constraints on the model parameters. In the regime $\bar{M}\sim M_W$ a discovery of $\mu \to e\gamma$ is possible because $G$ in eq.~(\ref{eq:R_Casas}) deviates
substantially from $10/3$ and the direct bounds on the $U_a^2$ are weaker.
The rate for $\mu\to e\gamma$ would then be sensitive to the same parameters $\bar{M}$, $\Im\omega$ as the observables $U_a^2$, cf. eq.~(\ref{eq:R_Casas}).

\subsection{Measurement of lepton universality violation}
A violation of lepton universality can be described by comparing decays of a particle $X$ (which can be a meson $M$ or a $\tau$-lepton) into leptons of different flavours $a\neq b$ \cite{Asaka:2014kia,Shrock:1980ct}:
\begin{align}
\label{eq:R_X}
R^M_{ab}=\frac{\Gamma(M^+\to \ell_a^+\nu_a)}{\Gamma(M^+\to \ell_b^+\nu_b)}\,,&&
R^\tau_{ab}=\frac{\Gamma(\tau^+\to \ell_a^+\nu_a\bar{\nu}_\tau)}{\Gamma(\tau^+\to \ell_b^+\nu_b\bar{\nu}_\tau)}\,.
\end{align}
A sensitive observable is the deviation from the SM prediction, that is given by~\cite{Shrock:1980ct}
\begin{align}
\Delta r^X_{ab}=\frac{R^X_{ab}}{R^X_{ab\,{\rm SM}}}-1\,.
\end{align}
A decomposition of eq.~(\ref{eq:R_X}) into mass eigenstates yields in case of the mesons
\begin{align}
R^M_{ab}=\frac{\sum_i \Gamma(M^+\to \ell_a^+\upnu_i)+\sum_i \Gamma(M^+\to \ell_a^+N_i)}{\sum_i \Gamma(M^+\to \ell_b^+\upnu_i)+\sum_i \Gamma(M^+\to \ell_b^+N_i)}\,,
\end{align}
with an analogous equation for the $\tau$.
We use the deviation
\begin{align}
\Delta r^X_{ab}=\frac{\sum_c|(V_\nu)_{ac}|^2+\sum_i U_{ai}^2G_{ai}}{\sum_c|(V_\nu)_{bc}|^2+\sum_i U_{ai}^2G_{bi}}-1=\frac{1+\sum_i U_{ai}^2[G_{ai}-1]}{1+\sum_i U_{bi}^2[G_{bi}-1]}-1\,,
\end{align}
where
\begin{align}
\label{eq:G}
G_{ai}=\vartheta(m_X-m_{\ell_a}-M_i)\frac{r_a+r_i-(r_a-r_i)^2}{r_a(1-r_a)^2}\sqrt{1-2(r_a+r_i)+(r_a-r_i)^2}\,,
\end{align}
with $r_a=m_{\ell_a}^2/m_X^2$ and $r_i=M_i^2/m_X^2$.
Note that, even if the decaying particle $X$ is lighter than the heavy neutrinos, such that the latter do not appear as final states and $G_{ai}=0$, the decay can be affected indirectly due to a unitarity violation of $V_\nu$ and yields
\begin{align}
\Delta r_{ab}^X=\frac{\sum_{c}|(V_\nu)_{ac}|^2}{\sum_{c}|(V_\nu)_{b c}|^2}-1=\frac{1-U_a^2}{1-U_b^2}-1 \simeq U^2_b-U^2_a\,.
\end{align}
For kaon decay the result from NA62 \cite{Lazzeroni:2012cx} compared to the SM prediction \cite{Cirigliano:2007xi} is
\bse
\begin{align}
R^K_{e\mu}&=(2.488\pm 0.010)\times 10^{-5}\,,\\
R^K_{e\mu\,\rm{SM}}&=(2.477\pm 0.001)\times 10^{-5}\,,
\end{align}
\ese
while from pion \cite{Czapek:1993kc} and tauon decay \cite{Pich:2009zza,Beringer:1900zz} we obtain
\bse
\begin{align}
R^\pi_{e\mu}&=(1.230\pm 0.004)\times 10^{-4}\,,\\
R^\pi_{e\mu\,\rm{SM}}&=(1.2354\pm 0.0002)\times 10^{-4}\,,
\end{align}
\ese
and
\bse
\begin{align}
R^\tau_{e\mu}&=0.9764\pm 0.0030\,,\\
R^\tau_{e\mu\,\rm{SM}}&=0.973\,,
\end{align}
\ese
with deviations $|\Delta r_{e\mu}^X|\simeq 5\times 10^{-3}$ for these three decay channels. In the following we make an estimate of the theory prediction for $|\Delta r_{e\mu}^X|$. For simplicity we neglect $G$ from eq.~(\ref{eq:G}) and note that this can lead to an $\mathcal{O}(1)$ error but does not affect our estimate given the necessary accuracy. $|\Delta r_{e\mu}^X|$ can then be expressed in terms of the Casas-Ibarra parameters
\begin{align}
|\Delta r_{e\mu}^X| 
\simeq |U^2_\mu-U^2_e|
\simeq
\begin{cases}
 \ee^{2\Im\omega}\frac{m_{\rm atm}}{2\bar{M}}\cos^2\uptheta_{13}\sin^2\uptheta_{23}
&\text{ for NH}
\\ \ee^{2\Im\omega}\frac{m_{\rm atm}}{4\bar{M}}
\left|1+3\xi \sin (\frac{\alpha_2-\alpha_1}{2})\sin 2\uptheta_{12}\right| &\text{ for IH}
\end{cases}
\,.
\end{align}
Here we have used eqs.~(\ref{eq:mixing_NHsimple}) and~(\ref{eq:mixing_IHsimple}) as well as the relative smallness of $\uptheta_{13}$ compared to $\uptheta_{23}$, $|\Im\omega|\gg 1$ and further limited ourselves to the $n_s=2$ mass degenerate case, where we have additionally neglected the mass splitting $\Delta M$. 
Note that the SM expectation $\Delta r_{e\mu\, \rm{SM}}^X=0$ is in slight tension with the experimentally measured values, however, it is within the $2\sigma$ range. We can further approximate the theory prediction $|\Delta r_{e\mu}^X| \simeq |U^2_\mu-U^2_e|<U^2$, that has to be smaller than $10^{-5}$ in order to be consistent with leptogenesis constraints~\cite{Drewes:2016gmt}. This also is in tension but again remains in the $2\sigma$ range of the measured value. Future experiments with increased sensitivity may yield constraints on $\Im\omega$, the Majorana mass $\bar{M}$ and the phase difference $\alpha_2-\alpha_1$.

\subsection{Full testability of the model}\label{Sec:Testability}
So far we have assumed that future experiments cannot resolve $\Delta M$, and therefore are unable measure $U_{a 1}^2$ and $U_{a 2}^2$ individually. As a consequence, the parameters $\Delta M$ and ${\rm Re}\omega$, which are crucial for leptogenesis, remain largely unknown (apart from the constraints discussed in section~\ref{Sec:0nubb}, which can be obtained if the $N_i$ are lighter than a few GeV and $0\nu\beta\beta$ decay is observed).
This assumption may be too pessimistic for the SHiP experiment, which has an estimated mass resolution of $\sim 10$ MeV and could detect decays with all three SM flavours in the final state.
The FCC-ee is expected to probe mixings $U_{a i}^2<10^{-9}$ \cite{Blondel:2014bra,Antusch:2016vyf} and will have an energy resolution in the sub-MeV range \cite{Gomez-Ceballos:2013zzn}. 
Also the ILC is expected to have an excellent mass resolution. 
Consequently, the mass resolution in the search of RH neutrinos will be improved, potentially allowing for a measurement of all individual  $U_{a i}^2$ in the viable parameter space for leptogenesis. 
If complemented with an independent measurement of the Dirac phase $\delta$ in future neutrino oscillation experiments, such as DUNE or NOvA, this would in principle allow to uniquely determine all fundamental parameters in the seesaw Lagrangian.
In practice, the finite experimental resolution might lead to considerable error bars on some parameters, in particular ${\rm Re}\omega$.

At first sight, inserting eq.~(\ref{CasasIbarraDef}) into eq.~(\ref{eq:U_theta}) suggests that there exist multiple choices of the phases that lead to the same $U_{ai}^2$. 
However, most of these are physically equivalent and owed to the multiple coverage of the paramater space in the Casas Ibarra parametrisation: Shifting $\Re\omega \to \Re\omega +\pi$ leads to a sign change of $Y^\dagger_{ai}$, which can always be compensated by  field redefinitions $N_i\rightarrow -N_i$. 
Swapping the sign of $\xi$ and shifting $\alpha \to \alpha +2\pi$ also only changes the sign of $Y_{ai}$ for normal hierarchy, and even leaves $Y_{ai}$ invariant for inverted hierarchy. 
Changing simultaneously the sign of $\xi$, $\Im\omega$ and $\Delta M$ while shifting $\Re\omega \to \pi-\Re\omega$ results in a swap of the labels $N_1$ and $N_2$, without any physical consequences.
The only remaining choices of the phases that are physically inequivalent and lead to the same $U_{ai}^2$ are related by the transformation $(\delta, \alpha, \Re\omega) \to (-\delta, 2\pi-\alpha, -\Re\omega)$. This degeneracy can be broken by an independent measurement of $\delta$.\footnote{
In the first version of this manuscript we oversaw this transformation and incorrectly claimed that all model parameters can be extracted from measurements of the $U_{a i}^2$ alone. This error has previously been pointed out in the conference proceedings \cite{Drewes:2016blc}, which are based on the work presented here. 
}
Since $M$ and $\Delta M$ can be extracted kinematically, a measurement of all $U_{ai}^2$ and $\delta$ therefore would allow to fully reconstruct the Lagrangian (\ref{eq:Lagrangian}).

In the case of a mass resolution worse than $\Delta M$ one cannot gain information about the mixings $U_{a1}^2$ and $U_{a2}^2$ independently, but could only measure the sum $U_a^2=\sum_i U_{ai}^2$. On the one hand, $U_a^2$ is invariant under an additional transformation of the phases that has no simple analytic form, but requires solving the trigonometric equations in appendix  \ref{App:Mixing}.
On the other hand, the $U_a^2$ do not depend on $\Re\omega$ in the limit $\mu\to 0$. Consequently neither $\Delta M$ nor $\Re\omega$ can be constrained in this case. Nevertheless, this case would still allow for a powerful test of the hypothesis of the sterile neutrinos being the origin of the neutrino masses and the BAU: Measuring all $U_a^2$ would fix $\Im\omega$ uniquely for $\mu=0$, as $U^2$ is independent of $\xi$, cf. eqs.~(\ref{U2NH}) and (\ref{U2IH}) . In the case of small $\epsilon\ll 1$ the $U_a^2/U^2$ are completely determined by the phases $(\alpha, \delta)$, as these ratios do not depend on $\Delta M$ and $\Re\omega$ in a good approximation, cf. figure~\ref{fig:regions_NH.pdf}. 
Consistency of the measured $U_a^2$ with these relations and an independent measurement of $\delta$ would clearly indicate that the $N_i$ generate the light neutrino masses via the seesaw mechanism. 
Comparing the observed $(\alpha, \delta)$ with figure~\ref{Fig:Lepto_NH} and \ref{Fig:Lepto_IH} would allow to test the hypothesis that the $N_I$ also generated the BAU.

\section{Conclusions}
The extension of the SM by two RH neutrinos with masses below the electroweak scale can simultaneously explain the observed neutrino masses via the seesaw mechanism and the BAU via low scale leptogenesis. 
An attractive feature of this scenario is the possibility to discover heavy neutrinos $N_i$ in direct search experiments, such as SHiP, NA62, similar facilities at LBNF or T2K, LHCb, BELLE II, ATLAS, CMS or a future lepton collider.
In addition to the light neutrino masses $m_a$ and the angles and phases in the mixing matrix $U_\nu$, this model adds only four new parameters to the SM. These can be chosen as the masses $M_1$ and $M_2$ of the heavy neutrinos and the real and imaginary part of a complex angle $\omega$.  $\omega$ determines both, the misalignment between the heavy neutrino mass and interaction basis as well as the overall strength $U^2=\sum_{a=e,\mu,\tau} U_a^2$ of the mixing of the heavy neutrinos with the SM neutrinos. Here $U_a^2=\sum_i U_{ai}^2$ is the sum of the mixings of both heavy neutrino states $N_i$ with the SM flavour $a$. 
An experimental discovery is most likely in the regime where $U^2$ is large. 
Large $U^2$ can be made consistent with small $m_a$ if the heavy neutrino masses are quasi-degenerate.
In addition, leptogenesis with two heavy neutrinos also requires a mass degeneracy.
From a model building viewpoint, the region with large $U^2$ and small $\Delta M$ can be motivated by a approximate $U(1)_{B-L}$ symmetry.
Because of this, it is convenient to use the average mass $\bar{M}$ and mass splitting $\Delta M$ of the two heavy neutrinos instead of their individual masses $M_1$ and $M_2$.

In most of the parameter region where the observed BAU can be reproduced by two heavy neutrinos, their masses are required to be too degenerate to be resolved experimentally. 
This does not only make a direct determination of $\Delta M$ impossible, but also implies that the mixings $U_{a1}^2$ and $U_{a2}^2$  of the two heavy neutrinos with the different SM generations cannot be measured individually. Instead, experiments can only constrain their sum $U_a^2$.
Moreover, values of $U^2$ that lie within reach of existing experiments 
require comparably large values of ${\rm Im}\omega$. 
In the parameter region with $\Delta M/\bar{M}\ll 1$ and ${\rm Im}\omega>1$, the sensitivity of $U^2$ to ${\rm Re}\omega$ is practically lost. The two parameters $\Delta M$ and ${\rm Re}\omega$, which are crucial for leptogenesis, cannot be determined by neutrino oscillation data and direct searches for the $N_i$ alone. Hence, leptogenesis in this scenario is not fully testable even if the $N_i$ are discovered and the Dirac phase $\delta$ in $U_\nu$ is measured
(testable in the sense that all observables, including the BAU, can be calculated uniquely from measured quantities).
However, our analysis shows that the parameter space of the model can be severely constrained, and testable predictions can be made based on the requirement to explain light neutrino oscillation data and the observed BAU.
\begin{itemize}
\item In the limit $\Delta M/\bar{M}\ll 1$ and ${\rm Im}\omega\gg 1$, the flavour mixing pattern $U_a^2/U^2$ of the heavy neutrinos is determined by light neutrino parameters alone. With the exception of one Majorana phase, all of these may be measured in foreseeable time. The precise relations are given in appendix \ref{App:Mixing}. This point has recently also been made in ref.~\cite{Hernandez:2016kel}.

\item By combining the negative results of various direct and indirect search experiments with lifetime constraints from the requirement not to disturb BBN, one can impose considerably stronger bounds on the $U_a^2$ than by simply superimposing them. We present the results of 
the first global analysis of this kind in figures~\ref{Fig:Flavor_Plots_NH} and \ref{Fig:Flavor_Plots_IH}.
\begin{itemize}
\item  The combined constraints rule out most of the parameter region below $\bar{M}\simeq350$ MeV. Moreover, they impose a much stronger lower bound on $U_\mu^2$ and $U_\tau^2$ than BBN and the seesaw relation alone for $\bar{M}<m_K$.
However, before completely disregarding the mass region $\bar{M}<350$ MeV, a careful re-analysis of the lower bounds on $U_a^2$ from BBN should be performed in the regime $\bar{M}>m_\pi$.\footnote{For masses $\bar{M}<m_\pi$ a revised calculation has recently been presented in ref.~\cite{Ruchayskiy:2012si}, where the authors also perform a comparison to earlier results in the literature.}
In this work, we used the simple estimate that the heavy neutrino lifetime should be shorter than $0.1$s, which may be too simplified. 
Moreover, there exist different interpretations of the results of the PS191 experiment in the literature \cite{Bernardi:1987ek,Ruchayskiy:2011aa,Artamonov:2014urb}, and the combined constraints strongly depend on these differences. In our analysis we have chosen to use the results of ref.~\cite{Ruchayskiy:2011aa}.
Finally, we have fixed the known neutrino masses and mixing angles to their best fit values in figures~\ref{Fig:Flavor_Plots_NH} and \ref{Fig:Flavor_Plots_IH}. The error bars on these quantities should slightly relax the exclusion bounds presented there.
\item In the regime $m_K<\bar{M}<m_D$ the combined constraints impose a much stronger upper bound on $U_\tau^2$ than those from direct searches alone. For normal hierarchy, they impose a stronger upper bound on $U_e^2$.
\item Any improvement in the measurement of light neutrino properties will tighten the combined constraints. In particular, a measurement of the Dirac phase $\delta$ and the light neutrino mass hierarchy would be highly desirable.
\end{itemize}
It should be pointed out that the combined constraints are  specific to the minimal scenario with two mass degenerate heavy neutrinos or models that can effectively be described in this way (e.g.\ the $\nu$MSM). They are considerably weaker if more heavy neutrinos exist in the mass range we consider or the masses are not degenerate \cite{Drewes:2015iva}.

\item As shown in section~\ref{Sec:Future}, future experiments can either discover the $N_i$ or impose much stronger constraints on their properties.
For $\bar{M}$ below $1-2$ GeV, a measurement of neutrinoless double $\beta$ decay can provide information about ${\rm Re}\omega$. Moreover, various processes that are sensitive to the oscillations of heavy neutrinos in the laboratory and the degree of lepton number violation they mediate can be used to extract information about $\Delta M$. However, as pointed out in section~\ref{Higgs}, a quantitative analysis requires careful inclusion of $\mathcal{O}[U^2]$ corrections to the heavy neutrino mass spectrum at tree level as well as loop corrections.

\item In figures~\ref{Fig:Flavor_Plots_tot}, \ref{Fig:Flavor_Plots_e}, \ref{Fig:Flavor_Plots_mu} and \ref{Fig:Flavor_Plots_tau} we present estimates for the range of allowed values for the mixings $U_a^2$ of heavy neutrinos with individual SM flavours that are consistent with successful leptogenesis, and compare these to the expected sensitivity of planned experiments. 
These results may be used to estimate the discovery potential of experiments that are only sensitive to one or two SM flavours, or which have very different sensitivities to the individual flavours.
\item Our results shown ~\ref{Fig:Flavor_Plots_tot}, \ref{Fig:Flavor_Plots_e}, \ref{Fig:Flavor_Plots_mu} and \ref{Fig:Flavor_Plots_tau} may be compared to those found in ref.~\cite{Hernandez:2016kel}, which appear much more pessimistic. The reason for the apparent discrepancy may lie in the fact that the authors of that study performed a Bayesian analysis to estimate the likelihood that heavy neutrinos with given parameters are responsible for the BAU. While such analysis in principle allows to different information about the parameter space, this additional information in the present case clearly depends on the choice of the choice of parametrisation and priors. 
In contrast, our analysis aims at identifying the largest and smallest $U_a^2$ that are consistent with neutrino oscillation data and successful leptogenesis, without imposing any theoretical prejudice about the model parameters.
In this sense, the two analyses should be viewed as complementary.

\item As discussed in section~\ref{Sec:Testability}, the model is fully testable if the experimental resolution is sufficient to determine $\Delta M$. 
Independent measurement of all $U_{ai}^2$ would, in combination with a determination of the Dirac phase $\delta$ in the light neutrino mixing matrix $U_\nu$, in principle allow to determine all model parameters and fully reconstruct the Lagrangian (\ref{eq:Lagrangian}). This would allow to predict the BAU and the outcome of other future experiments, such as searches for neutrinoless double $\beta$ decay. In practice, the finite experimental sensitivity would of course imply that the error bars on some quantities (in particular ${\rm Re}\omega$) will be sizeable in foreseeable time.

\item Even in the case $\Delta M/\bar{M}\ll 1$ and ${\rm Im}\omega\gg 1$, in which the values of $\Delta M$ and ${\rm Re}\omega$ cannot be extracted from measurements of the $U_{a i}^2$, a measurement of the individual $U_a^2$ still provides a powerful test of the minimal low scale leptogenesis scenario. 
As shown in section~\ref{Sec:Lepto}, the requirement to explain the observed BAU allows to make predictions for the flavour mixing pattern $U_a^2/U^2$ for given $\bar{M}$ and $U^2$, cf. figures \ref{Fig:Lepto_NH} and \ref{Fig:Lepto_IH}. Finding heavy neutral leptons with flavour mixing patterns within these regions will provide circumstantial evidence for leptogenesis.

\end{itemize}

The low scale seesaw and leptogenesis provide a simple explanation for at least two of the most studied questions in fundamental physics, the origin of neutrino masses and the origin of the baryonic matter in the universe. A significant fraction of the parameter space can be tested in existing or proposed experiments, and in a part of this region, the model is fully testable.
The global constrains and estimates of the cosmologically relevant parameter space presented here are state of the art, but still suffer from order one uncertainties. In the wake of upcoming experiments, further theoretical work will be necessary to reduce these to a level that would be required for a comparison with experimental data if any heavy neutral leptons are discovered.

\subsection*{Acknowledgements}
We would like to thank Elena Graverini, Nicola Serra, Walter Bonivento, Gaia Lanfranchi, Oliver Fischer and Jilberto Antonio Zamora Saa for helpful discussions about experimental aspects. 
We would also like to thank Dmitry Gorbunov and Stefan Antusch for pointing out errors in the experimental sensitivities shown in the first version of this manuscript and Mikhail Shaposhnikov for general comments.
This research was supported by the DFG cluster of excellence 'Origin and Structure of the Universe' (www.universe-cluster.de).

\appendix
\section{Heavy neutrino mixing angles}\label{App:Mixing}
In the following we write down analytic expressions for the six different combinations of the flavour mixings $U_{ai}^2$ for both the normal and the inverted hierarchy in the case of $n_s=2$ sterile neutrinos. This requires the lightest SM neutrino mass to be zero. The other two masses are denoted by $m_a$, while $s_{ab}$ and $c_{ab}$ are shorthand notations for $\sin\uptheta_{ab}$ and $\cos\uptheta_{ab}$, respectively, with the neutrino mixing angles $\uptheta_{ab}$ for $a,b=1,2,3$.\footnote{We assume $s_{ab}$ and $c_{ab}$ to be the positive real roots of $s_{ab}^2$ and $c_{ab}^2$ from table~\ref{tab:active_bounds}.} Experimental values are given in table~\ref{tab:active_bounds}. As this paper mainly focuses on large $\Im\omega$ it is worth mentioning the limits: 
\begin{align}
\label{eq:limits}
\lim_{\Im\omega\gg 1}\tanh(2\Im\omega)= 1\,,\quad \lim_{\Im\omega\gg 1}\cosh(2\Im\omega) =\frac12 {\rm exp}(2\Im\omega)\,.
\end{align}
Nevertheless, the following expressions for $U^2_{ai}$ are given in terms of the Casas-Ibarra parameters for any $\Im\omega$, cf. subsection~\ref{SubSec:model}.

\paragraph{Normal hierarchy:}
\bse
\begin{align}
\frac{M_1 U^2_{e1}+M_2U^2_{e2}}{\cosh(2\Im\omega)}&= a_1^+ -a_2\sin\left(\frac{\alpha_2}{2}+\delta\right)\tanh(2\Im\omega)\,,\\
\frac{M_1 U^2_{\mu 1}+M_2U^2_{\mu 2}}{\cosh(2\Im\omega)}&=a_3^+ -a_4\cos(\delta)-\left[a_5\sin\left(\frac{\alpha_2}{2}\right)-a_6\sin\left(\frac{\alpha_2}{2}+\delta\right)\right]\tanh(2\Im\omega)\,,\\
\frac{M_1 U^2_{\tau 1}+M_2U^2_{\tau 2}}{\cosh(2\Im\omega)}&=a_7^+ +a_4\cos(\delta)+\left[a_5\sin\left(\frac{\alpha_2}{2}\right)+a_{8}\sin\left(\frac{\alpha_2}{2}+\delta\right)\right]\tanh(2\Im\omega)
\,,
\end{align}
\label{eq:mixing_NH}
\ese

\bse
\begin{align}
\frac{M_1 U^2_{e1}-M_2U^2_{e2}}{\cos(2\Re\omega)}&= a_1^- -a_2\cos\left(\frac{\alpha_2}{2}+\delta\right)\tan(2\Re\omega)\,,\\
\frac{M_1 U^2_{\mu 1}-M_2U^2_{\mu 2}}{\cos(2\Re\omega)}&=-a_3^- -a_4\cos(\delta)-\left[a_5\cos\left(\frac{\alpha_2}{2}\right)-a_6\cos\left(\frac{\alpha_2}{2}+\delta\right)\right]\tan(2\Re\omega)\,,\\
\frac{M_1 U^2_{\tau 1}-M_2U^2_{\tau 2}}{\cos(2\Re\omega)}&=-a_7^-+a_4\cos(\delta)+\left[a_5\cos\left(\frac{\alpha_2}{2}\right)+a_{8}\cos\left(\frac{\alpha_2}{2}+\delta\right)\right]\tan(2\Re\omega)
\,,
\end{align}
\label{eq:mixing_NH_minus}
\ese

with $a_1$ to $a_8$ positive real values given by active neutrino masses and their mixing angles:
\begin{align}
a_1^\pm &= m_2c_{13}^2s_{12}^2\pm m_3s_{13}^2\,,\notag\\
a_2&=2\sqrt{m_2m_3}c_{13}s_{12}s_{13}\xi\,,\notag\\
a_3^\pm &=\pm m_2(c_{12}^2c_{23}^2+s_{12}^2s_{13}^2s_{23}^2)
	+ m_3c_{13}^2s_{23}^2\,,\notag\\
a_4&=2m_2c_{12}c_{23}s_{12}s_{13}s_{23}\,,\notag\\
a_5&=2\sqrt{m_2m_3}c_{12}c_{13}c_{23}s_{23}\xi\,,\notag\\
a_6&=2\sqrt{m_2m_3}c_{13}s_{12}s_{13}s_{23}^2\xi\,,\notag\\
a_7^\pm &=\pm m_2(c_{23}^2s_{12}^2s_{13}^2+c_{12}^2s_{23}^2)
+ m_3c_{13}^2c_{23}^2\,,\notag\\
a_{8}&=2\sqrt{m_2m_3}c_{13}c_{23}^2s_{12}s_{13}\xi
\,.
\end{align} 
Due to the fact that for normal hierarchy we have $m_1=0$, eq.~(\ref{eq:mixing_NH}) and thus the mixings do not depend on $\alpha_1$, which is why we can use $\alpha_1=0$.\\
For the following discussion we limit ourselves to the parametric region where leptogenesis makes flavour predictions for the active-sterile mixing, namely the overdamped regime. In this case, in order to fulfil leptogenesis constraints, it is required to have large $\im\omega$ and therefore high degeneracies, i.e. we neglect $\mu$, such that $M_1\simeq M_2\simeq \bar{M}$, and only keep terms of order $1/\epsilon$, cf. eq.~(\ref{mu_eps}). In this case experiments might not distinguish the single sterile neutrinos and will thus be only sensitive to the sum $U_a^2=\sum_{i=1,2}U_{ai}^2$. This sum can be obtained from eq.~(\ref{eq:mixing_NH}) by using the limits~(\ref{eq:limits}). 

\begin{figure}
	\centering
	\includegraphics[scale=0.8]{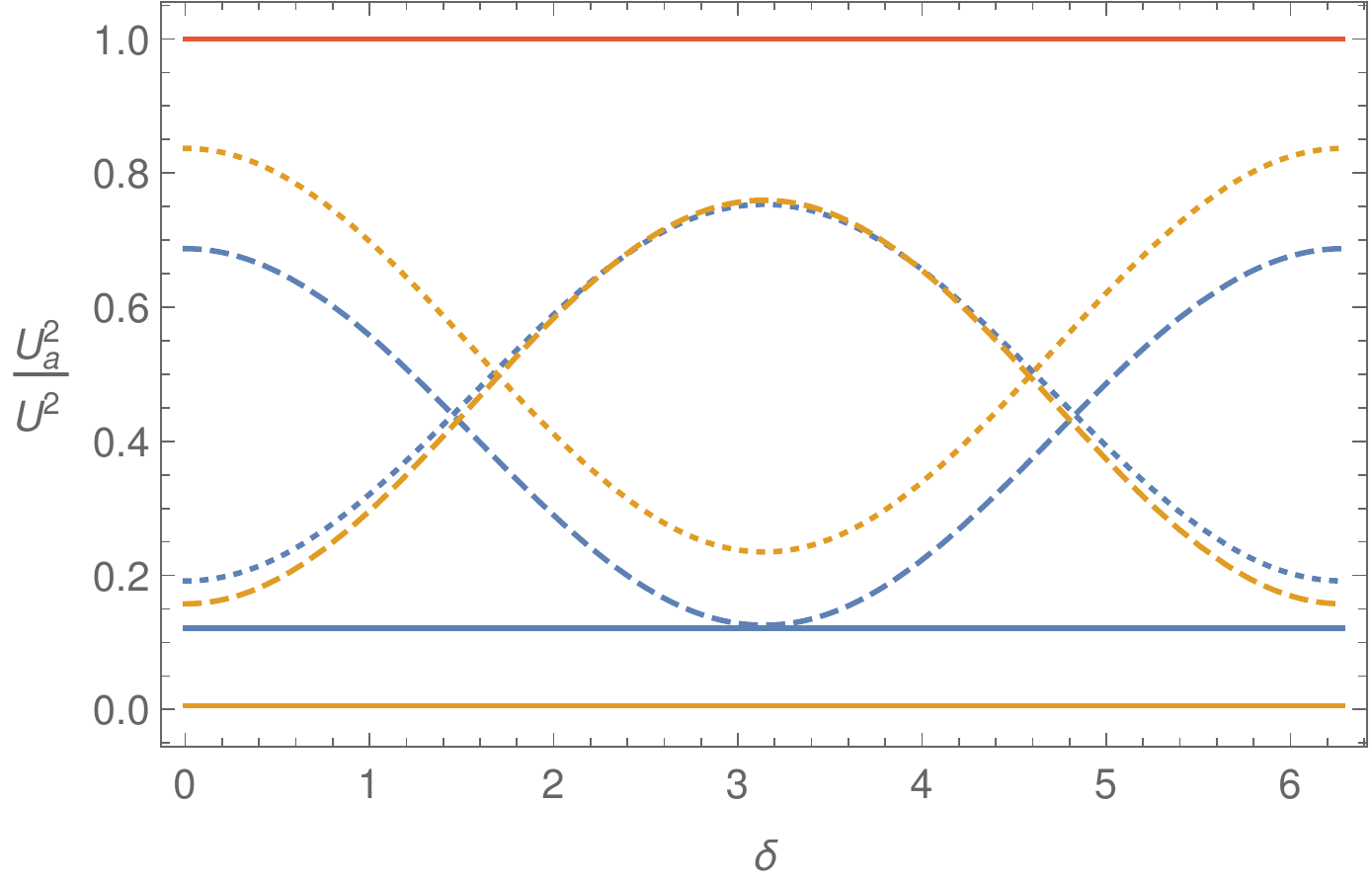}
	\includegraphics[scale=0.8]{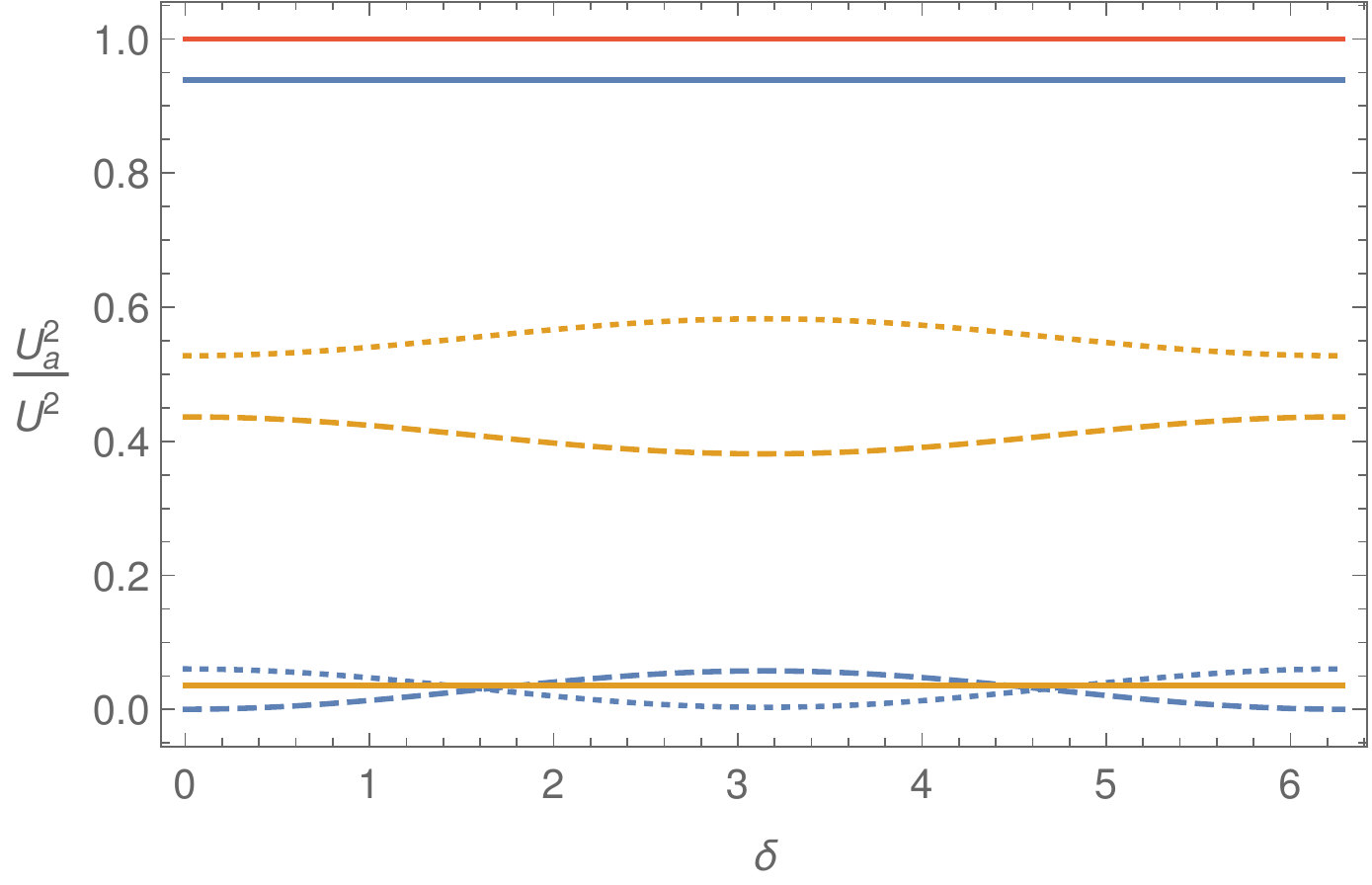}
	\caption{Relative mixing $U_a^2/U^2$ over $\delta$ with $a=e,\mu,\tau$ for $n_s=2$ sterile flavours in the limit of large $\Im\omega$, cf. eq.~(\ref{eq:R_exp}). The electron mixings are given by solid lines, the muon mixings by dashed and the tauon mixings by dotted lines. The red line corresponds to the total mixing $U^2=\sum_a U_a^2$. The upper panel assumes normal hierarchy, where the orange and blue curves indicate the cases $\alpha_2=-2\delta+\pi$ and $\alpha_2=-2\delta+3\pi$, respectively. The lower panel corresponds to inverted hierarchy, where the blue and orange curves indicate the cases $\alpha_2=\pi$ and $\alpha_2=-\pi$, respectively.}
	\label{fig:Mixing_Angles_NH_IH.pdf}
\end{figure}

On the one hand, the minimal electron (relative) mixing angle $U^2_e/U^2=0.00557$, which turns out to maximise $U^2$ under the condition of fulfilling leptogenesis constraints, is obtained by the choice $\alpha_2=-2\delta+\pi$. The maximal $\mu$ and  $\tau$ (relative) mixing $U^2_\mu/U^2$ and $U^2_\tau/U^2$ are given by $\delta=\pi$ and $\delta=0$ with this particular value of $\alpha_2$, respectively.\footnote{In fact, the maximal possible total mixing $U^2$ in NH is obtained for the minimal electron relative mixing $U^2_e/U^2$. However, this is not necessarily true for a given mass splitting $\Delta M$, as it is the case for $\Delta M \lesssim 10^{-10}$ with $M=1 \,\GeV$.}
On the other hand it is worth mentioning that the maximal electron (relative) mixing occurs for $\alpha_2=-2\delta+3\pi$, while the maximal $\mu$ and  $\tau$ (relative) mixing $U^2_\mu/U^2$ and $U^2_\tau/U^2$ for $\delta=0$ and $\delta=\pi$, respectively, with this value of $\alpha_2$. Note that maximal muon mixing corresponds to minimal tauon mixing and vice versa as the plot for the relative mixings for normal hierarchy in the upper left panel of figure~\ref{fig:Mixing_Angles_NH_IH.pdf} demonstrates.

\paragraph{Inverted hierarchy:}
\bse
\begin{align}
\frac{M_1 U^2_{e1}+M_2U^2_{e2}}{\cosh(2\Im\omega)}&=b_1^+ +b_2\sin\left(\frac{\tilde{\alpha}}{2}\right)\tanh(2\Im\omega)\,,\\
\frac{M_1 U^2_{\mu 1}+M_2U^2_{\mu 2}}{\cosh(2\Im\omega)}&=b_3^+ -b_4^+\cos(\delta)\notag\\
&-\left[b_5\sin\left(\frac{\tilde{\alpha}}{2}\right)+b_6\sin\left(\frac{\tilde{\alpha}}{2}-\delta\right)-b_7\sin\left(\frac{\tilde{\alpha}}{2}+\delta\right)\right]\tanh(2\Im\omega)\,,\\
\frac{M_1 U^2_{\tau 1}+M_2U^2_{\tau 2}}{\cosh(2\Im\omega)}&=b_8^+ +b_4^+ \cos(\delta)\notag\\
&-\left[b_{9}\sin\left(\frac{\tilde{\alpha}}{2}\right) -b_{6}\sin\left(\frac{\tilde{\alpha}}{2}-\delta\right)+b_{7}\sin\left(\frac{\tilde{\alpha}}{2}+\delta\right)\right]\tanh(2\Im\omega)\,,
\end{align}
\label{eq:mixing_IH}
\ese

\bse
\begin{align}
\frac{M_1 U^2_{e1}-M_2U^2_{e2}}{\cos(2\Re\omega)}&=b_1^- -b_2\cos\left(\frac{\tilde{\alpha}}{2}\right)\tan(2\Re\omega)\,,\\
\frac{M_1 U^2_{\mu 1}-M_2U^2_{\mu 2}}{\cos(2\Re\omega)}&=-b_3^- -b_4^-\cos(\delta)\notag\\
&+\left[b_5\cos\left(\frac{\tilde{\alpha}}{2}\right)+b_6\cos\left(\frac{\tilde{\alpha}}{2}-\delta\right)-b_7\cos\left(\frac{\tilde{\alpha}}{2}+\delta\right)\right]\tan(2\Re\omega)\,,\\
\frac{M_1 U^2_{\tau 1}-M_2U^2_{\tau 2}}{\cos(2\Re\omega)}&=-b_8^-+b_4^-\cos(\delta)\notag\\
&+\left[b_{9}\cos\left(\frac{\tilde{\alpha}}{2}\right) -b_{6}\cos\left(\frac{\tilde{\alpha}}{2}-\delta\right)+b_{7}\cos\left(\frac{\tilde{\alpha}}{2}+\delta\right)\right]\tan(2\Re\omega)\,,
\end{align}
\label{eq:mixing_IH_minus}
\ese
with $b_1$ to $b_{9}$ positive real numbers given by the active neutrino masses and mixings:

\begin{align}
b_1^\pm&= m_1c_{12}^2c_{13}^2\pm m_2s_{12}^2c_{13}^2\,,\notag\\
b_2&=2\sqrt{m_1m_2}c_{12}s_{12}\xi\,,\notag\\
b_3^\pm&=\pm m_1(c_{23}^2s_{12}^2+c_{12}^2s_{13}^2s_{23}^2)
	+ m_2(c_{12}^2c_{23}^2+s_{12}^2s_{13}^2s_{23}^2)\,,\notag\\
b_4^\pm &=2(\pm m_2 -m_1)c_{12}c_{23}s_{12}s_{13}s_{23}\,,\notag\\
b_5&=2\sqrt{m_1m_2}(c_{12}c_{23}^2s_{12}-c_{12}s_{12}s_{13}^2s_{23}^2)\xi\,,\notag\\
b_6&=2\sqrt{m_1m_2}c_{12}^2c_{23}s_{13}s_{23}\xi\,,\notag\\
b_7&=2\sqrt{m_1m_2}s_{12}^2c_{23}s_{13}s_{23}\xi\,,\notag\\
b_8^\pm&=\pm m_1(c_{12}^2 c_{23}^2 s_{13}^2 +s_{12}^2 s_{23}^2) 
+ m_2(c_{23}^2s_{12}^2 s_{13}^2 + c_{12}^2  s_{23}^2 )\,,\notag\\
b_{9}&=2\sqrt{m_1m_2}c_{12}s_{12}(s_{23}^2-c_{23}^2s_{13}^2)\xi\,,
\end{align} 
and $\tilde{\alpha}=\alpha_2-\alpha_1$. This implies that for inverted hierarchy the Yukawa matrices $Y$ do only depend on the difference $\alpha_2-\alpha_1$, which allows us to set $\alpha_1=0$ and take $\alpha_2$ as the running parameter. 

Analogously to the discussion for normal hierarchy, we limit ourselves to the large $\im\omega$ and small $\mu$ limit, which is why experiments might only be sensitive to the sum $U_a^2=\sum_{i=1,2}U_{ai}^2$, that is obtained from eq.~(\ref{eq:mixing_IH}) by using the limits~(\ref{eq:limits}). 

In case of inverted hierarchy it is the maximal electron (relative) mixing $U^2_e/U^2=0.939$, given for $\alpha_2=\pi$, that maximises $U^2$ given the leptogenesis constraints. The maximal  $U_\mu^2/U^2$ and $U_\tau^2/U^2$ for this value of $\alpha_2$ are given for $\delta=\pi$ and $\delta=0$, respectively. The bottom panel of figure~\ref{fig:Mixing_Angles_NH_IH.pdf} shows the relative mixings for inverted hierarchy. There we can see that maximal muon mixing corresponds to minimal tauon mixing and vice versa. The results from eqs. (\ref{eq:mixing_NH}) and (\ref{eq:mixing_IH}) are consistent with previous results from ref.~\cite{Hernandez:2016kel} and lead to the approximated expressions~(\ref{eq:mixing_total})--(\ref{eq:mixing_IHsimple}).

\bibliographystyle{JHEP}
\bibliography{all}
\end{document}